\documentclass{osa-article}

\journal{aop}

\articletype{Tutorial}

\usepackage{tikz}
\usetikzlibrary{%
    decorations.pathreplacing,%
    decorations.pathmorphing,%
    arrows
}
\usetikzlibrary{positioning, calc,intersections,shapes,matrix,fit}

\definecolor{huntergreen}{rgb}{0.21, 0.37, 0.23}
\definecolor{lavenderpurple}{rgb}{0.59, 0.48, 0.71}
\definecolor{coquelicot}{rgb}{1.0, 0.22, 0.0}
\definecolor{crimsonglory}{rgb}{0.75, 0.0, 0.2}
\definecolor{deeppink}{rgb}{1.0, 0.08, 0.58}
\definecolor{electricviolet}{rgb}{0.56, 0.0, 1.0}
\definecolor{electricgreen}{rgb}{0.0, 1.0, 0.0}
\definecolor{mint}{rgb}{0.24, 0.71, 0.54}
\definecolor{dodgerblue}{rgb}{0.12, 0.56, 1.0}
\definecolor{lincolngreen}{rgb}{0.11, 0.35, 0.02}
\definecolor{persianblue}{rgb}{0.11, 0.22, 0.73}
\definecolor{patriarch}{rgb}{0.5, 0.0, 0.5}
\definecolor{amaranth}{rgb}{0.9, 0.17, 0.31}

\begin{document}

\title{Entanglement of a pair of quantum emitters via continuous fluorescence measurements: \\ a tutorial}

\author{Philippe Lewalle,\authormark{1,*} Cyril Elouard, \authormark{1,2} Sreenath K.~Manikandan,\authormark{1} Xiao-Feng Qian,\authormark{3} Joseph H.~Eberly,\authormark{1} and Andrew N.~Jordan\authormark{1,4}}

\address{\authormark{1}Department of Physics and Astronomy, and Center for Coherence and Quantum Optics, University of Rochester, Rochester, NY 14627, USA \\
\authormark{2}QUANTIC lab, INRIA Paris. 2 rue Simone Iff, 75012 Paris, France\\
\authormark{3}Department of Physics and Center for Quantum Science and Engineering, Stevens Institute of Technology, Hoboken, NJ 07030, USA \\
\authormark{4}Institute for Quantum Studies, Chapman University, Orange, CA 92866, USA }

\email{\authormark{*}plewalle.quantum@gmail.com} 

\def\la{\langle}
\def\ra{\rangle}
\def\cur{\mathcal{I}}
\def\be{\begin{equation}}
\def\ee{\end{equation}}
\newcommand{\partl}[3]{ \frac{\partial^{#3}#1}{ \partial #2^{#3}} }	
\newcommand{\limit}[2]{\underset{#1 \rightarrow #2}{\text{lim}} \;}
\newcommand{\ket}[1]{\left\vert #1 \right\rangle}
\newcommand{\bra}[1]{\left\langle #1 \right\vert}
\newcommand{\ip}[2]{\langle #1 \vert #2 \rangle}
\newcommand{\exval}[3]{\langle #1 \vert #2 \vert #3 \rangle}
\newcommand{\exvalbig}[3]{\big\langle #1 \big\vert #2 \big\vert #3 \big\rangle}
\newcommand{\exvalbigg}[3]{\bigg\langle #1 \bigg\vert #2 \bigg\vert #3 \bigg\rangle}
\newcommand{\abs}[1]{\left| #1 \right|}


\begin{abstract}
We discuss recent developments in measurement protocols that generate quantum entanglement between two remote qubits, focusing on {the theory of} joint continuous detection of their spontaneous emission.
We consider a device geometry similar to that used in well--known Bell--state measurements, {which we analyze using a conceptually transparent model of} stochastic quantum trajectories; 
we use this to review photodetection, the most straightforward case, and then generalize to the diffusive trajectories from homodyne and heterodyne detection as well. 
Such quadrature measurement schemes are a realistic two--qubit extension of existing circuit--QED experiments which obtain quantum trajectories by homodyning or heterodyning a superconducting qubit's spontaneous emission, or an adaptation of existing optical measurement schemes to obtain jump trajectories from emitters. 
We mention key results, presented from within a single theoretical framework, and draw connections to concepts in the wider literature on entanglement generation by measurement {(such as path information erasure and entanglement swapping)}. 
The photon which--path information acquisition, and therefore the two--qubit entanglement yield, is tunable under the homodyne detection scheme we {discuss}, at best generating equivalent average entanglement dynamics as in the comparable photodetection case. 
{In addition to deriving this known equivalence, we extend past analyses in our characterization of the measurement dynamics: We include derivations of bounds on the fastest possible evolution towards a Bell state under joint homodyne measurement dynamics, and characterize the maximal entanglement yield possible using inefficient (lossy) measurements. }
\end{abstract}

\setcounter{tocdepth}{1}
\tableofcontents


\section{Introduction}

The present work draws on three distinct areas of research; entanglement, quantum trajectories arising from continuous measurement, and spontaneous emission all play a role in what follows. Entanglement has interested and confounded the physics community since it was first noted \cite{Schrodinger1935, Schrodinger_1936, EPR1935}.
It has since been shown to be a valuable resource for applications to cryptography or metrology, 
and protocols have been found to generate and preserve it. 
Spontaneous fluorescence is a long--studied and fundamental example of behavior arising from the interaction between a physical system and its optical environment.
When left unmonitored, spontaneous emission becomes a source of disentanglement \cite{YuEberly_2004, Mintert2005} and/or decoherence; when collected and measured, however, this need not be the case. 
Recent efforts to track the impact of environment--induced dynamics, such as a qubit's fluorescence, have led to stochastic quantum trajectories (SQTs) \cite{BookCarmichael, Wiseman1996Review, Plenio_Knight_RMP98, BookWiseman, BookBarchielli, BookJacobs, Brun2001Teach, Jacobs2006, Gross_2018}, which arise from time--continuous measurements of a system. 
{These stochastic dynamics are those which occur \emph{conditioned} the outcomes of measurements on the system's environment, in contrast with the dissipative dynamics arising in un-monitored open quantum systems.}

\par We find our present topic at the intersection of these three areas. 
Specifically, we {take} recent theoretical \cite{Santos2011, bolund2014stochastic, Jordan2015flor, FlorTeach2019,PL-Dissertation} and experimental \cite{PCI-2013, Campagne-Ibarcq2016, PCI-2016-2, Naghiloo2016flor, Mahdi2016, Tan2017, Ficheux2018, Cottet_Thesis, FicheuxThesis, MahdiTeachThesis, Mahdi2017Qtherm} developments in the continuous monitoring of a single qubit's spontaneous emission {as our point of departure}. 
{While our work is inspired by the recent progress in experiments using superconducting qubits, our analysis is not restricted to this specific platform.}
{From there, we} consider how {two qubit generalizations of such measurement protocols} might be used to entangle the continuously--monitored quantum emitters. 
Specifically, we investigate a device, illustrated in Fig.~\ref{fig-splitterphase}, in which two remote identical qubits' spontaneous emission is mixed and monitored continuously. 
We review and expand on key results in this area \cite{Viviescas2010, Mascarenhas2011, Santos2012}, using a straightforward and conceptually transparent model, 
drawing connections to the wider literature as we go.

{The overall aim of this paper is to illustrate how an observer may infer the presence or absence of coherent correlations (entanglement) between remote systems, using a quantum state evolution that is \emph{conditioned} on the real--time outcomes of measurements tracking the two--qubit state. 
We begin by establishing some straightforward principles (based on standard ideas for open quantum systems and Bayesian inference) in Sec.~\ref{sec-model}. Some general conceptual discussion there coalesces into a measurement model which underpins the rest of the paper.}
{Sec.~\ref{sec-erasure} applies the model to answer the question: What joint measurements and measurement settings are suitable for generating correlations between subsystems? This discussion necessarily emphasizes the erasure of which--path information, drawing connections between our model and the wider literature on the subject.}
{In Secs.~\ref{sec-photodetect} and \ref{sec-homodyne}, we perform detailed analysis of the dynamical changes in entanglement due to time--continuous tracking of the two qubits, emphasizing the correlations between them that an observer may infer by acquiring different forms of quantum--mechanically \emph{complete} information, using ideal measurement devices.}
{This may be contrasted with the case in which an observer attempts to draw inferences based on \emph{incomplete} information obtained via inefficient measurements, which we discuss in Sec.~\ref{sec-inefficient}.}
{Some conclusions and further discussion of the literature context into which these ideas fit are included in Sec.~\ref{sec-conclusions}. Finally, some technical details and additional pedagogical points are covered in the appendices.}

\section{Continuously Monitoring Two--Qubit Fluorescence: Formalism \label{sec-model}}

Continuous monitoring of a quantum system, given its initial state, leads to a real--time estimate of the system's evolution, conditioned on the measurement record. 
Such an evolution may generically include both unitary dynamics, and the stochastic effects of measurement backaction. 

{The stochasticity of the dynamics is a direct outgrowth of the randomness implicit in quantum measurement. 
Whereas the evolution described by the Schr\"{o}dinger equation alone pertains to closed system dynamics, in which measurements cannot be directly accounted for due to the system's isolation from any possible probe, we here consider an example of open system evolution. 
More specifically, we will consider that quantum degrees of freedom in our apparatus illustrated in Fig.~\ref{fig-splitterphase}, may be divided into a primary system (two qubits), and their environment (optical degrees of freedom which may contain some emitted photons). 
The optical degrees of freedom may be monitored, and then inferences can be drawn to update the two--qubit state in real time, conditioned on the outcomes of the optical measurements at each time step.
Generically, such conditional stochastic evolution of the system (monitored via generalized measurements) is called a stochastic quantum trajectory (SQT). }

\subsection{Open Systems and (Un)Conditional Evolution}

{We may formalize notions of system--environment interaction, and the resulting dissipative or conditional evolution, by reviewing the Kraus representation \cite{jordan1961dynamical, kraus1971general, BookKraus} (or operator--sum representation \cite{BookNielsen}). 
We suppose we have a quantum system in the Hilbert space $\mathcal{H}_{S}$, with free dynamics governed by the Hamiltonian $\hat{H}_{S}$. 
The system is also interacting with one (or more) environmental degree(s) of freedom in the Hilbert space $\mathcal{H}_{E}$, and characterized by the free Hamiltonian $\hat{H}_{E}$.}
{A generic interaction Hamiltonian $\hat{H}_{int}$ describes possible interactions between the system and the environment. 
The evolution of the system and environment together is assumed unitary in the extended Hilbert space $\mathcal{H}_{SE}=\mathcal{H}_{S}\otimes \mathcal{H}_{E}$, described by the Hamiltonian $\hat{H}_{SE}=\hat{H}_{S}+\hat{H}_{E}+\hat{H}_{int}$.}
{
This formulation implies that our system of interest is completely isolated, except for its interaction with the environmental degree(s) of freedom that has(have) been specified explicitly; these environmental degrees of freedom will mediate all possibility of measurement going forward.}

{
We consider the scenario where the system and its environment are initially uncorrelated, as per
\begin{equation} \label{ic-se}
    \rho_{SE}(0)=\rho(0)\otimes\varrho,
\end{equation}
where $\rho$ denotes the state of the system, and $\varrho$ the state of the environment.}
{
The combined state of the system and environment evolves as
 \begin{equation}
     \rho_{SE}(dt)=\hat{U}(dt)\rho_{SE}(0)\hat{U}^{\dagger}(dt),\quad\text{for}\quad \hat{U}(dt)=\exp[-i\,dt\,\hat{H}_{SE}]
 \end{equation}
(in units $\hbar \rightarrow 1$). 
We may further choose a basis in which $\varrho=\sum_{k}\wp_{k}|k\rangle\langle k|$ is diagonal. 
The reduced evolution of the system of interest is obtained by tracing over the reservoir degrees of freedom, i.e.
 \begin{eqnarray}
     \rho(t+dt)&=& \mathrm{tr}_{E}\{\rho_{SE}(t+dt)\} = \sum_{j}\langle j|\hat{U}(dt)\rho(t)\otimes\bigg(\sum_{k}\wp_{k}|k\rangle\langle k|\bigg)\hat{U}^{\dagger}(dt)|j \rangle\nonumber\\ &=& \sum_{jk}\hat{\mathcal{M}}_{jk}(dt)\rho(t)\hat{\mathcal{M}}_{jk}^{\dagger}(dt).\label{dynmap}
 \end{eqnarray}
 The operators $\hat{\mathcal{M}}_{jk}(dt)=\sqrt{\wp_{k}}\langle j|\hat{U}(dt)|k\rangle$ are \emph{Kraus operators}.
The map \eqref{dynmap} is trace preserving, and the Kraus operators obey the completeness relation  $\sum_{jk}\hat{\mathcal{M}}_{jk}^{\dagger}(t)\hat{\mathcal{M}}_{jk}(t)=1\!\!1$, such that they form elements of a positive operator valued measure (POVM).}
{This description will be useful going forward, not only because the sum over $j$ and $k$ allows us to describe the dynamics of an open system, but because the individual $\hat{\mathcal{M}}_{jk}$ will allow us to describe the evolution of a quantum system \emph{conditioned} on its environment evolving from $\ket{k}$ to $\ket{j}$.}
{Initial conditions of the form \eqref{ic-se} adequately describe the case when the environment is probed directly with some detection apparatus in every time interval, possibly exchanging energy, entropy and information through the coupling with the measurement apparatus.

{The dissipative dynamics due to a quantum system's interaction with an unmonitored environment are often formalized via the Lindblad master equation \cite{Lind1976}. The latter approach is effectively a time--continuum description of the dynamics above, and is otherwise equivalent.}
{Specifically, expansion of the dynamical rule \eqref{dynmap} to $O(dt)$ leads to\footnote{There exist a large class of environments whose effect of the system can be described as a Markovian master equation---the Lindblad equation. This generically happens when the environment has a sufficiently short memory of its interaction with the system, which in turn requires that the coupling to the system is sufficiently weak and does not depend too quickly on the frequency \cite{BookBreuer}. 
Due to the short memory time of the environment, the evolution of the system can be computed from an uncorrelated approximation of the system-environment state at time $t$, i.e.~$\rho(t)\otimes\varrho$ as in eq.~\eqref{ic-se}, where the environment is assumed to be in its equilibrium state. See e.g.~Ref.~\cite{BookCohenTannoudji} for an accessible derivation.}}
\be \label{LindME}
\dot{\rho} = i[\rho,\hat{H}_S] + \sum_c \left\lbrace \hat{L}_c\,\rho\,\hat{L}_c^\dag - \tfrac{1}{2} \left( \hat{L}_c^\dag\hat{L}_c\,\rho + \rho\,\hat{L}_c^\dag\hat{L}_c\right)\right\rbrace \approx \frac{\rho(t+dt) - \rho(t)}{dt}.
\ee
The density matrix $\rho$ again describes the state of the system, while the Lindblad operators $\hat{L}$ describe the effects of different channels $c$ by which information may ``leak'' into environmental degrees of freedom. 
Closed system evolution alone may be described by the first term $\dot{\rho} = i[\rho,\hat{H}_S]$ (this is equivalent to the Schr\"{o}dinger equation).}
{The remainder of the master equation \eqref{LindME}, concerning interaction via open channels to the environment, and causes the state purity to degrade over time. 
In other words, the accumulated loss of information to the environment prevents the system state from being described as a pure state (coherent) superposition, as $\rho = \ket{\psi}\bra{\psi}$; lost information leads instead to (incoherent) mixed  states $\rho = \sum_i \wp_i \ket{\psi_i}\bra{\psi_i}$ (where $\wp_i$ is the probability to draw a pure state $\ket{\psi_i}$ from an ensemble).}
{In this sense, $\mathrm{tr}_E$ in \eqref{dynmap} may be regarded as an average over all of the measurement outcomes that \emph{could have been} obtained if a measurement had been performed on the environment. } 

{We begin developing these ideas in our specific problem of interest, by considering the spontaneous emission of one qubit. }
{The typical decay statistics of spontaneous emission may be derived from an interaction 
\be 
\hat{H}_{int} = \sum_j g_j(\hat{a}_j\,\hat{\sigma}_+ + \,\hat{a}_j^\dag\,\hat{\sigma}_-),
\ee
between a qubit with raising and lowering operators $\hat{\sigma}_-^+$, and many modes $j$ of its electromagnetic environment with frequencies $\omega_j$, and photon creation and annihilation operators $\hat{a}_j^\dag$ and $\hat{a}_j$. 
The coupling between the qubit and each field mode is characterized by the constant $g_j$. }
{The resulting dynamics of the qubit may be captured in the Lindblad form \eqref{LindME}, however (see e.g.~\cite{Jacobs2006} for an accessible derivation).  
Specifically, a qubit (two level quantum system) decaying into its optical environment may be described by 
\be \label{Lind1QF} 
\dot{\rho} = \gamma\,\hat{\sigma}_-\,\rho\,\hat{\sigma}^+ - \tfrac{1}{2}\hat{\sigma}^+\hat{\sigma}_-\, \rho - \tfrac{1}{2} \rho \,\hat{\sigma}^+\hat{\sigma}_- \quad\leftrightarrow\quad \left(\begin{array}{cc} 
\dot{\rho}_{ee} & \dot{\rho}_{eg} \\ \dot{\rho}_{ge} & \dot{\rho}_{gg}
\end{array}\right) = \left(\begin{array}{cc} 
-\gamma\,\rho_{ee} & -\tfrac{\gamma}{2}\,\rho_{eg} \\
-\tfrac{\gamma}{2}\,\rho_{ge} & \gamma\,\rho_{gg}
\end{array}\right),
\ee
where $\rho$ is the qubit density matrix, $\hat{L} = \sqrt{\gamma}\,\hat{\sigma}_-$ describes decay at rate $\gamma = 1/T_1$, and any unitary contributions to the dynamics have been eliminated by going into a rotating frame (see e.g.~\cite{Jacobs2006, FlorTeach2019}). 
The equation \eqref{Lind1QF} describes exponential decay of the qubit's excited state population, i.e.~$\rho_{ee}(t) = \rho_{ee}(0)e^{-\gamma\,t}$.}

We are here interested in developing the case of \emph{monitored} spontaneous emission, in which case we may discuss not only the \emph{unconditioned} evolution above, but also evolution that is \emph{conditioned} on the outcomes of measurements performed on the emitted photons.
Tracking of a single qubit in this way has been acheived in recent experiments \cite{PCI-2013, Campagne-Ibarcq2016, PCI-2016-2, Naghiloo2016flor, Mahdi2016, Tan2017, Ficheux2018, Cottet_Thesis, FicheuxThesis, MahdiTeachThesis, Mahdi2017Qtherm}. 
Following such implementations, we should imagine that spontaneously emitted photons are captured, and routed to a detector through a transmission line. 
Connecting these ideas more formally to the description above, the measurement is performed on a temporal mode of the line, i.e.~on traveling modes propagating from the qubit to the detector \cite{Gardiner85, BookWiseman, Ciccarello17}.\footnote{The travelling mode interacting with the qubit between times $t$ and $t+dt$ is related to the eigenmodes of the line (associated with operators $\hat{a}_j$ and  frequencies $\omega_j$) according to $\hat{a}_\text{in}(t) = (\gamma)^{-1/2}\sum_j g_j \hat{a}_j e^{-i(\omega_j-\omega_0)t}$, where the traveling mode operators obey the canonical commutation $[\hat{a}_\text{in}(t),\hat{a}_\text{in}(t')^\dagger] = \delta(t-t')$. The notion of sequential interactions between qubit and environment may be retrieved considering the interaction picture Hamiltonian 
$$
    \hat{H}_{int}(t) = \sum_j g_j \left(\hat{a}_j \hat{\sigma}^+e^{-i(\omega_j-\omega_0)t} +\hat{a}_j^\dag \hat{\sigma}_-e^{i(\omega_j-\omega_0)t}\right) = \sqrt{\gamma}\left(\hat{a}_\text{in}(t)\hat{\sigma}^+ +\hat{a}_\text{in}^\dagger(t)\hat{\sigma}_-\right).
$$
See e.g.~Ref.~\cite{Ciccarello17} for further context and details. 
\label{fn-io}}
In each timestep, a new mode interacts with the qubit, and will then be measured; we formalize this below in a way that closely follows collision models of quantum optics \cite{Ciccarello17}.
We may then describe a single timestep as an interaction between the qubit and single field mode, initialized in vacuum.\footnote{One may also imagine that the qubit is placed in a bad cavity coupled to the transmission line (i.e.~a cavity which decays quickly into the line). 
In this case, the cavity captures spontaneously emitted photons and then quickly releases them to the line which carries them to detectors. 
The photon number states we write in our state updates from eq.~\eqref{psi_in} on should be understood as the cavity output in a particular time interval, in connection with this input/output picture. 
The fast cavity decay rate into the line (bad cavity assumption) is important in that it allows for the cavity to be removed from the discussion (e.g.~via adiabatic elimination); in this regime, we recover the picture discussed in the main text (and previous note), of the qubit directly coupled to the output line, and which allows us to write down Markovian dynamics for the qubit evolution conditioned on measurements of the line at different timesteps. 
} 

In order to develop a simple model of this type of situation, suppose we initially have a qubit in some arbitrary pure state, coupled to an empty output line that routes its spontaneous emission to a detector, characterized by the joint state $\ket{A_i} = (\zeta\ket{e} + \phi\ket{g})\otimes \ket{0}$.
After some short time evolution, $\ket{A_i}$ transforms to 
\be 
\ket{A_f} = \sqrt{e^{-\epsilon}} \:\zeta \ket{e,0} + \phi \ket{g,0} + \sqrt{1-e^{-\epsilon}} \:\zeta \ket{g,1}\label{psi_in}
\ee
where we have defined $\epsilon = \gamma \: dt$. 
{We have written a \emph{pure} state that explicitly respects the decay statistics that are detailed above, and expected on average. 
This state update could be equivalently written{, for small $\epsilon$,} as a operation}
\be \label{1QF-stateup}
\ket{A_f} = \left( \begin{array}{c} \sqrt{1-\epsilon}\, \zeta \\ \phi + \sqrt{\epsilon}\, \zeta\, \hat{a}^\dag \end{array} \right) \otimes \ket{0} = \left( \begin{array}{cc} 
\sqrt{1 - \epsilon} & 0 \\ \sqrt{\epsilon} \,\hat{a}^\dag & 1
\end{array} \right) \left( \begin{array}{c} \zeta \\ \phi \end{array} \right) \otimes \ket{0},
\ee
in the $\lbrace \ket{e,0},\ket{g,0} \rbrace$ qubit--field basis.\footnote{The notation here is set up such that the operators already appear similar to how they will once field states are chosen, leaving operators that act on the qubit basis $\lbrace \ket{e},\ket{g}\rbrace$ only. See e.g.~Refs.~\cite{FlorTeach2019, PL-Dissertation} for further comments.} 
The operator $\hat{a}^\dag$ creates a photon (i.e.~$\hat{a}^\dag \ket{0} = \ket{1}$) {in the transmission line, leading to a detection apparatus}.
The utility of this formulation is that we may choose a final state of the optical degree of freedom, and thereby obtain an operator that updates the qubit state \emph{conditioned} on the optical measurement outcome. 

{For example, if a photon number measurement is made after a short time ($\epsilon \ll 1$), then we may extract two Kraus operators \cite{FlorTeach2019,PL-Dissertation}
\be 
\hat{\mathcal{M}}_0 = \bra{0} \left( \begin{array}{cc} 
\sqrt{1 - \epsilon} & 0 \\ \sqrt{\epsilon} \,\hat{a}^\dag & 1
\end{array} \right) \ket{0} \quad\text{and}\quad \hat{\mathcal{M}}_1 = \bra{1} \left( \begin{array}{cc} 
\sqrt{1 - \epsilon} & 0 \\ \sqrt{\epsilon} \,\hat{a}^\dag & 1
\end{array} \right)\ket{0}
\ee
from the expression \eqref{1QF-stateup}. 
Conditional evolution of the qubit state is then implemented by 
\be 
\rho(t+dt) = \frac{\hat{\mathcal{M}}_0\,\rho(t)\,\hat{\mathcal{M}}_0^\dag}{\mathrm{tr}\left( \hat{\mathcal{M}}_0\,\rho(t)\,\hat{\mathcal{M}}_0^\dag\right)} \quad\text{or}\quad \rho(t+dt) = \frac{\hat{\mathcal{M}}_1\,\rho(t)\,\hat{\mathcal{M}}_1^\dag}{\mathrm{tr}\left( \hat{\mathcal{M}}_1\,\rho(t)\,\hat{\mathcal{M}}_1^\dag\right)},
\ee
where the first expression is used in the event that the photon counter does not click (no photon is emitted), and the second expression is used in the event that the detector does receive an emitted photon.}
{One may understand the process described from \eqref{psi_in} up to this point as one of Bayesian inference, in which the qubit state is updated conditioned on acquiring new information from its environment (and given a model of the qubit--field decay interaction, such that the meaning of the detector readout, relative to the qubit state, is clear).  
A similar approach may be applied in cases with different qubit--field interactions (e.g.~dispersive qubit--cavity coupling) as well \cite{Korotkov2011}.}
{This formal structure also lets us describe the unmonitored (unconditioned) evolution: If we trace out all possible outcomes that could have occurred, i.e.
\be 
\rho(t+dt) = \hat{\mathcal{M}}_0\,\rho(t)\,\hat{\mathcal{M}}_0^\dag + \hat{\mathcal{M}}_1\,\rho(t)\,\hat{\mathcal{M}}_1^\dag,
\ee 
and as in \eqref{dynmap},
and then expand to $O(dt)$, the unmonitored evolution \eqref{Lind1QF} is immediately recovered. 
This reduction of the Kraus operators to a single output channel is consistent with the reduction of the associated master equation to a single channel $\hat{L}_- = \sqrt{\gamma}\,\hat{\sigma}_-$, and is furthermore typical of operator--sum representations of a relaxation channel, as presented from a quantum information perspective \cite{BookNielsen}. 
The conditional (i.e.~continuously monitored) evolution of a single qubit has been studied in detail using this model \cite{Jordan2015flor, FlorTeach2019, PL-Dissertation}, leading to good agreement with experiments \cite{PCI-2013, Campagne-Ibarcq2016, PCI-2016-2, Naghiloo2016flor, Mahdi2016, Tan2017, Ficheux2018, Cottet_Thesis, FicheuxThesis, MahdiTeachThesis, Mahdi2017Qtherm}. }


\subsection{Conditional Two Qubit Dynamics} 

\begin{figure}[t]
\centering
\includegraphics[width = .49\columnwidth]{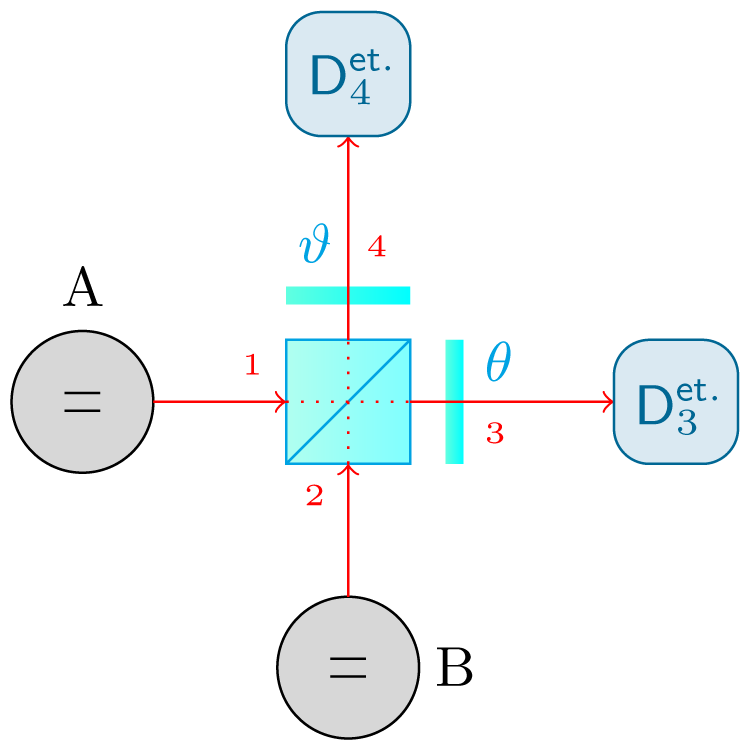}
\includegraphics[width = .49\columnwidth]{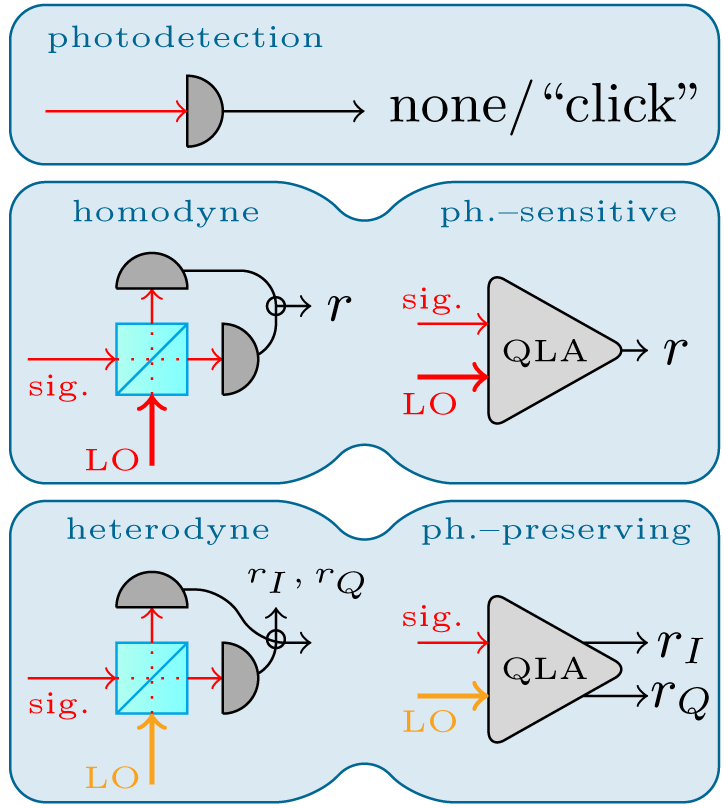}
\caption{(Left sketch) The kind of setup we envisage. Qubits in cavities A and B emit spontaneously into transmission lines 1 or 2, respectively. Each cavity and transmission line can be engineered to capture the fluorescence with high efficiency. These single--photon signals are mixed on a 50/50 beamsplitter and any phase in the two paths (relative to an external reference) is characterized with a pair of phase plates. The combined effect of these unitary transformations on modes 1 and 2 before they reach the detector at outputs 3 and 4 are summarized by the relations \eqref{bs}. \\ (Right sketch, in three parts)
We consider continuous monitoring ($dt \ll T_1$) at the outputs $\mathsf{D}^\mathsf{et.}_3$ and $\mathsf{D}^\mathsf{et.}_4$, with three different measurement options. Direct photodetection of the emitted signal leads to a number of clicks in each timestep (zero, one, or two photons may arrive at a detector). Homodyne and heterodyne detection involve measuring one or both quadratures of the field, respectively; both rely on mixing the signal with a strong coherent state local oscillator (LO).
The relative phase $\theta$ or $\vartheta$ between each signal and LO determines the particular quadrature(s) which are monitored. 
In the language of quantum--limited amplifiers (QLAs), pertinent to existing single--qubit circuit QED experiments,
homodyne detection corresponds to a ``phase--sensitive'' amplification, and heterodyne detection corresponds to ``phase--preserving'' amplification. 
See Sec.~\ref{sec-model} and/or \cite{FlorTeach2019,PL-Dissertation}, and references therein, for details.
}
\label{fig-splitterphase}
\end{figure}

The device geometry we focus on, shown in Fig.~\ref{fig-splitterphase}, has been used in conjunction with photodetection, and outside of the context of quantum trajectories, to entangle many types of solid state quantum systems that interact with their optical environment \cite{Moehring2007, Maunz2009, Hofmann2012, Hanson2013Heralded} {(see Sec.~\ref{sec-conclusions} for additional references and discussion)}. 
{One of the key features of the device geometry in question is that it allows for the creation of entanglement between distant qubits, without the need for them to interact directly; coherently correlated states of the emitters are established purely from the inferences that can be drawn from the optical measurements.} 
{Similar} methods have been leveraged to perform loophole--free Bell tests \cite{Brunner_RMP2014, HansonLoopholeFree}, and promising extensions to the above experiments, within the framework of quantum trajectories, have been proposed \cite{Santos2012}.

\par
We extend the single qubit treatment above to describe the system illustrated in Fig.~\ref{fig-splitterphase}. 
For simplicity, we only consider the case where the qubits and cavities, and therefore the decay rates $\gamma$ of qubit--cavity systems $A$ and $B$ are identical. 
{A beamsplitter illustrated in Fig.~\ref{fig-splitterphase}} implements a unitary mixing operation on the optical modes coming from either cavity, and some measurement devices can then be placed at the output ports 3 and 4. A two--qubit two--cavity system may be expressed by a pair of operators {as in \eqref{1QF-stateup}}, 
\be 
\mathcal{A} = \left( \begin{array}{cc} 
\sqrt{1 - \epsilon}~ & 0 \\ \sqrt{\epsilon}\, \hat{a}_1^\dag~ & 1
\end{array} \right), \quad \mathcal{B} = \left( \begin{array}{cc} 
\sqrt{1 - \epsilon}~ & 0 \\ \sqrt{\epsilon}\, \hat{a}_2^\dag~ & 1
\end{array} \right),
\ee
emitting into different transmission lines (where $\hat{a}_1^\dag$ and $\hat{a}_2^\dag$ create photons in paths 1 and 2, respectively).
Then the two--qubit two--mode state update goes as 
\be \label{combine-qubit-spaces}
\ket{\psi_{dt}}  = (\mathcal{A} \otimes \mathcal{B}) (\ket{A_i} \otimes \ket{B_i}),
\ee
which gives a short--time state update, now in the two--qubit two--mode basis $\lbrace \ket{ee,0_1 0_2},$ $\ket{eg,0_1 0_2},$ $\ket{ge,0_1 0_2},$ $\ket{gg,0_1 0_2} \rbrace$ (assuming some state $\ket{B_i} = (\xi \ket{e} + \varphi \ket{g})\otimes{0}$ which transforms like $\ket{A_i}$). This is equivalently notated  
\be \label{2Q_stateUp1}
 \ket{\psi_{dt}} =\underbrace{\left( \begin{array}{cccc}
1- \epsilon & 0 & 0 & 0 \\
\sqrt{\epsilon(1-\epsilon)} \hat{a}_2^\dag & \sqrt{1-\epsilon} & 0 & 0 \\
\sqrt{\epsilon(1-\epsilon)} \hat{a}_1^\dag & 0 & \sqrt{1-\epsilon} & 0 \\
\epsilon \hat{a}_1^\dag \hat{a}_2^\dag & \sqrt{\epsilon} \hat{a}_1^\dag & \sqrt{\epsilon}\hat{a}_2^\dag & 1
\end{array} \right)}_{\mathcal{M}} \underbrace{\left( \begin{array}{c} \zeta \xi \\ \zeta \varphi \\ \phi \xi \\ \phi \varphi \end{array} \right)\otimes \ket{0_1 0_2}}_{\ket{\psi_0}},
\ee
where the matrix $\mathcal{M}$ will become the object of primary interest in deriving Kraus operators that act on the two--qubit state {above}.

\par The effect of the beamsplitter and phase plates can be characterized by the unitary transformations
\begin{subequations} \label{bs} \be 
\hat{a}_1^\dag = \tfrac{1}{\sqrt{2}}\left( \hat{a}_3^\dag e^{i\theta} + \hat{a}_4^\dag e^{i\vartheta} \right), \quad \hat{a}_2^\dag = \tfrac{1}{\sqrt{2}}\left( \hat{a}_3^\dag e^{i\theta} - \hat{a}_4^\dag e^{i\vartheta} \right), 
\ee or, conversely \be 
\hat{a}_3^\dag = \tfrac{1}{\sqrt{2}} e^{-i\theta}\left( \hat{a}_1^\dag + \hat{a}_2^\dag \right), \quad \hat{a}_4^\dag = \tfrac{1}{\sqrt{2}} e^{-i\vartheta}\left( \hat{a}_1^\dag - \hat{a}_2^\dag \right).
\ee \end{subequations}
The state--update matrix $\mathcal{M}$ can be modified accordingly, to represent these optical transformations, leading to the outputs $\mathsf{D}^\mathsf{et.}_3$ and $\mathsf{D}^\mathsf{et.}_4$, and then reads
\be \label{idealMAT}
\mathcal{M} \rightarrow 
\left( \begin{array}{cccc}
1- \epsilon & 0 & 0 & 0 \\
\sqrt{\frac{\epsilon(1-\epsilon)}{2}} (\hat{a}_3^\dag e^{i\theta}-\hat{a}_4^\dag e^{i\vartheta}) & \sqrt{1-\epsilon} & 0 & 0 \\
\sqrt{\frac{\epsilon(1-\epsilon)}{2}} (\hat{a}_3^\dag e^{i\theta}+\hat{a}_4^\dag e^{i\vartheta}) & 0 & \sqrt{1-\epsilon} & 0 \\
\frac{\epsilon}{2}(\hat{a}_3^\dag \hat{a}_3^\dag e^{2i\theta} - \hat{a}_4^\dag \hat{a}_4^\dag e^{2i\vartheta}) & \sqrt{\frac{\epsilon}{2}} (\hat{a}_3^\dag e^{i\theta}+\hat{a}_4^\dag e^{i\vartheta}) & \sqrt{\frac{\epsilon}{2}} (\hat{a}_3^\dag e^{i\theta}-\hat{a}_4^\dag e^{i\vartheta}) & 1
\end{array} \right).
\ee 
The operator that updates the two--qubit state under particular measurement outcomes is then obtained by projecting out final optical states $\ket{\psi_{f:3,4}}$ consistent with {a particular detection process}, i.e.~$\hat{\mathcal{M}}_f  = \bra{\psi_{f:3,4}} \mathcal{M} \ket{0_3 0_4}$ acts purely on the two--qubit state, {updating it} (conditioned on optical measurement outcomes) via
\be \label{general-stateup}
\rho(t+dt) = \frac{\hat{\mathcal{M}}_f \rho(t) \hat{\mathcal{M}}_f^\dag}{\text{tr}\left( \hat{\mathcal{M}}_f \rho(t) \hat{\mathcal{M}}_f^\dag\right)}.
\ee
Such an approach will ultimately form the basis of all of our derivations and numerical modeling below.
Note that an update of this type is the state update an observer can make at $t+dt$, given previous knowledge of the state $\rho(t)$, and access to the measurement record at both outputs 3 and 4, under the assumption that no information is lost to unmonitored environmental degrees of freedom at any stage between the qubits and detectors. 
The joint qubit--field system remains perfectly isolated except for ideal, perfectly--efficient, measurements made at ports 3 and 4. 
{We will momentarily clarify below how the use of $\ket{\psi_{f:3,4}}$ in this presentation corresponds a type of measurement, with well--defined outcomes that are amplified to point where they correspond to the classical output of a measurement device.}
For generalizations to the case of inefficient measurements, see Sec.~\ref{sec-inefficient} and/or Appendix~\ref{app-inefficient}.
In assuming that the observer's state update applies in real time, we implicitly assume that the photon travel times between the qubits and detectors are negligible (as is the case, e.g., in any circuit--QED experiment in a single dilution refrigerator).\footnote{In the event that the photon travel times from qubits to detectors are not negligible, then one should understand the update above to apply at the time emission must have occurred in order to create the relevant detection event.}

\par Fluorescence moves the qubits from their excited states to their ground states. The clearest way to generate entanglement then involves starting with $\ket{ee}$, counting photons, and recognizing a Bell state $\ket{\Psi^\pm} = \tfrac{1}{\sqrt{2}}\ket{eg}\pm\tfrac{1}{\sqrt{2}}\ket{ge}$ when a detector clicks. Consider (with $\theta = 0 = \vartheta$)
\be \label{pd-ost}
\exval{1_3 0_4}{\mathcal{M}}{0_3 0_4} \ket{ee}  = \left( \begin{array}{cccc}
0 & 0 & 0 & 0 \\
{\color{Plum} \sqrt{\frac{\epsilon(1-\epsilon)}{2}} } & 0 & 0 & 0 \\
{\color{Plum} \sqrt{\frac{\epsilon(1-\epsilon)}{2}} } & 0 & 0 & 0 \\
0 & \sqrt{\frac{\epsilon}{2}} & \sqrt{\frac{\epsilon}{2}} & 0
\end{array} \right) \ket{ee},
\ee
which describes the update of the two--qubit state by the jump operator, which occurs conditioned on the detector at output 3 registering the arrival of a single photon in the requisite timestep. 
After {normalization}, this state corresponds to $\ket{\Psi^+}$. If a click occurs at output 4 instead, we take $\ket{ee}$ to $\ket{\Psi^-}$ (up to a sign) via the operation $\exval{0_3 1_4}{\mathcal{M}}{0_3 0_4} \ket{ee}$. 
The key point to take away from this simplest case, is that depending on which channel registers an event, we get a different Bell state, and the matrix elements highlighted in {\color{Plum} purple} are primarily responsible for generating entanglement, by correlating or anti--correlating $\ket{eg}$ and $\ket{ge}$ as amplitude decays out of $\ket{ee}$. When a second click is registered, we know that both qubits have emitted, and the state is updated to $\ket{gg}$ with the entanglement destroyed. 

\par Two--photon events (events in which both qubits emit ``simultaneously'') exhibit interference, similar to the type exhibited in the classic Hong--Ou--Mandel experiment \cite{HOM}. 
Our model coarse--grains the notion of simultaneity to mean merely that both emissions occur within the same detector integration interval $dt$. 
The probability of emission within the same interval $dt$ is sufficiently small (to $O(\epsilon^2)$ at worst) that the effects of this simplification to our model should be negligible. 
Related points are discussed in Appendix \ref{sec-onesteptest}. 
{The two-photon interference visibility is determined by the indistinguishability \cite{Mandel1991OL} (or identical-ness) of the two photons. 
Thus, the induced coherence with continuous fluorescence measurements can be viewed as a process of distilling the two-photon indistinguishability. 
Our use of these ideas is, in this sense, similar to that of induced coherence without induced emission first studied by Zou, Wang, and Mandel \cite{Zou1991PRL, Wang1991, Wiseman2000, Lahiri2019}.} 
{Fabricating solid--state qubits which emit genuinely indistinguishable photons is not necessarily easy, but the problem has been studied in the context of the device geometry of interest, and entanglement generation \cite{Legero2003, Legero2004, Metz2008, Hurst2019}.}

We do ultimately wish to proceed to considering homodyne or heterodyne measurements {in addition to} photodetections. Heterodyne monitoring can be modeled by projecting onto coherent state outcomes instead of Fock states \cite{Wiseman1996Review}, 
i.e., we use a Kraus operator $\hat{\mathcal{M}}_{\alpha\beta} = \bra{\alpha \beta} \mathcal{M} \ket{0 0} =  $
\be \label{M-het}
e^{-|\alpha|^2/2-|\beta|^2/2} \left( \begin{array}{cccc}
1-\epsilon & 0 & 0 & 0 \\
{ \sqrt{\frac{\epsilon(1-\epsilon)}{2}} (\alpha^\ast e^{i\theta} -  \beta^\ast e^{i\vartheta})} & \sqrt{1-\epsilon} & 0 & 0 \\
{\sqrt{\frac{\epsilon(1-\epsilon)}{2}} (\alpha^\ast e^{i\theta} + \beta^\ast e^{i\vartheta})} & 0 & \sqrt{1-\epsilon} & 0 \\
\tfrac{\epsilon}{2}\left( {\alpha^\ast}^2 e^{2i\theta} -{\beta^\ast}^2 e^{2i\vartheta} \right) & \sqrt{\frac{\epsilon}{2}} (\alpha^\ast e^{i\theta} + \beta^\ast e^{i\vartheta}) & \sqrt{\frac{\epsilon}{2}} (\alpha^\ast e^{i\theta} - \beta^\ast e^{i\vartheta}) & 1
\end{array} \right).
\ee 
Physically, this is achieved by mixing the signal beams with a strong coherent state local oscillator (LO), or equivalently doing phase preserving quantum--limited amplification (see Fig.~\ref{fig-splitterphase}). As in the one qubit case \cite{Jordan2015flor, FlorTeach2019,PL-Dissertation}, {the subsequent} readouts are related to the coherent state eigenvalues by
\be \label{hetd_readouts} 
\alpha = \sqrt{\frac{dt}{2}}\left( r_I + i r_Q \right), \quad \beta = \sqrt{\frac{dt}{2}} \left( r_X + i r_Y \right).
\ee
{Heterodyne detection (or phase--preserving amplification) involves the measurement of two non-commuting observables (measuring both quadratures of the field is like measuring the position and momentum of a quantum harmonic oscillator). This process is consequently limited by the Heisenberg uncertainty principle. 
Coherent states satisfy and saturate the Heisenberg uncertainty principle, and use of coherent state outcomes corresponds to the best possible balanced measurement of the two quadratures allowed by quantum mechanics. 
For additional details, see e.g.~\cite{Shapiro2009, Caves2012} and references therein. 
}

\par Homodyne detection is similar to heterodyne detection\footnote{The most substantial difference between optical homodyne and heterodyne detection is addressed directly in their names; homodyne detection involves a signal and LO at the same frequency, whereas in heterodyne detection the signal and LO have different frequencies.}, but information is only collected about one quadrature instead of both (the unmeasured one is effectively squeezed out) \cite{Tyc2004, Shapiro2009}. 
We model this by choosing our final optical states to be eigenstates of a particular quadrature, i.e.~we take the eigenstates of the $\hat{X} = (\hat{a}^\dag + \hat{a})/\sqrt{2}$ quadrature at both outputs \cite{Wiseman1996Review} (without loss of generality, since $\theta$ and $\vartheta$ are completely tunable), such that we have $\hat{\mathcal{M}}_{34} = \bra{X_3 X_4}\mathcal{M} \ket{00} =$
\be \label{M-hom} \begin{split}
& \frac{1}{\sqrt{\pi}} \exp\left[-\tfrac{1}{2}(X_3^2+X_4^2)\right] \times \\&\left( \begin{array}{cccc}
1-\epsilon & 0 & 0 & 0 \\
{ \color{MidnightBlue} \sqrt{\epsilon(1-\epsilon)} (e^{i\theta} X_3 - e^{i\vartheta} X_4)} & \sqrt{1-\epsilon} & 0 & 0 \\
{ \color{MidnightBlue} \sqrt{\epsilon(1-\epsilon)} (e^{i\theta} X_3 + e^{i\vartheta} X_4)} & 0 & \sqrt{1-\epsilon} & 0 \\
{\color{amaranth} \epsilon e^{2i\theta}(X_3^2-\tfrac{1}{2}) - \epsilon e^{2i\vartheta}(X_4^2-\tfrac{1}{2})} & 
\sqrt{\epsilon}(e^{i\theta} X_3 + e^{i\vartheta} X_4) & \sqrt{\epsilon}(e^{i\theta} X_3 - e^{i\vartheta} X_4) & 1
\end{array} \right).
\end{split} \ee 
We have used the standard Hermite polynomial solutions in the appropriate matrix elements,
\begin{subequations}\be 
\ip{X}{0} = \pi^{-\tfrac{1}{4}} e^{-X^2/2},
\ee \be 
\ip{X}{1} = \pi^{-\tfrac{1}{4}} e^{-X^2/2} X \sqrt{2},
\ee \be \label{hermite2}
\ip{X}{2} =\pi^{-\tfrac{1}{4}} e^{-X^2/2} \left( \frac{2X^2-1}{\sqrt{2}} \right),
\ee \end{subequations}
for the $X$--representation of the harmonic oscillator wavefunction (for dimensionless $X$), which we now use in each field mode, for all the matrix elements of $\mathcal{M}$. The readouts are related to the real numbers $X$ by
\be \label{hom-readouts}
X_3 = \sqrt{\frac{dt}{2}} r_3, \quad X_4 = \sqrt{\frac{dt}{2}} r_4,
\ee
using the same logic underpinning \eqref{hetd_readouts}. The operator describes a valid measurement and complete set of possible outcomes, i.e. 
\be 
\iint_{-\infty}^\infty dr_3 \: dr_4 \: \hat{\mathcal{M}}^\dag_{34} \hat{\mathcal{M}}_{34} \propto \hat{\mathbb{I}}
\ee
indicating that it forms a positive operator valued measure (POVM) \cite{BookNielsen} {(the same is true of the photodetection and heterodyne operators we have discussed above)}. 
For derivations and an overview of the analogous objects in the single qubit case, see~{Refs.}~\cite{FlorTeach2019,PL-Dissertation}, and for a very detailed treatment of the single qubit heterodyne case, see {Ref.}~\cite{Jordan2015flor}.

{A few remarks about the process above, and equations \eqref{hetd_readouts} and \eqref{hom-readouts} in particular, are warranted before continuing.} 
{In transitioning to the readout notations $r$, we imply that an amplification step has taken place, such that the readouts $r$ are e.g.~the current or voltage observed on a laboratory device at a macroscopic or classical scale.}
{When we use an update like \eqref{general-stateup}, we are imagining that an observer, in possession of the information encoded in any relevant $r$, is inferring the evolution of the two--qubit state conditioned on information gained by measuring the qubits' optical environment. 
Such an inference is drawn given a picture of the device that we have encoded in $\mathcal{M}$ \eqref{idealMAT}.}
{While the records $r$ take on sharp values in any individual measurement timestep, they are intrinsically noisy (stochastic) due to the limits quantum mechanics imposes on measurement and amplification.}
{For additional comments regarding quantum--optical measurement and/or amplification, as applied above and in connection with contemporary experiments, consult e.g.~\cite{Haus2000, Tyc2004, Shapiro2009, WallsMilburn, Caves2012, RoyDevoret2016, Flurin_Thesis, Bergeal2010, Macklin2015}.}

\par Note that it is very common to {express continuous quantum measurement in the language of stochastic differential equations}, by using the Stochastic Master Equation (SME) \cite{BookWiseman, Jacobs2006, Rouchon2015}. The typical It\^{o} SME for diffusive quantum trajectories reads
\begin{subequations} \label{SME}
\be 
d\rho = i[\rho,\hat{H}_S] dt + \sum_c \left( \hat{\mathcal{L}}[\rho,\hat{L}_c] dt + \sqrt{\eta_c} \hat{\mathcal{K}}[\rho,\hat{L}_c] dW_c \right).
\ee
The super--operators are the Lindblad dissipation term
\be 
\hat{\mathcal{L}}[\rho,\hat{L}_c] \equiv \hat{L}_c \rho \hat{L}_c^\dag - \tfrac{1}{2}\left(\hat{L}_c^\dag \hat{L}_c \rho + \rho \hat{L}_c^\dag \hat{L}_c \right),
\ee
and the measurement backaction term
\be 
\hat{\mathcal{K}}[\rho,\hat{L}_c] \equiv \hat{L}_c \rho + \rho \hat{L}_c^\dag - \rho \: \text{tr}\left(\hat{L}_c \rho + \rho \hat{L}_c^\dag \right).
\ee \end{subequations}
{The introduction of $\hat{\mathcal{K}}$ effectively supplements \eqref{LindME} with a stochastic term, implementing the conditional evolution given measurement outcomes.}
Each of the operators $\hat{L}_c$ describes a particular measurement channel, which is monitored with efficiency $\eta_c \in [0,1]$, where $\eta_c=1$ denotes a channel from which all possible information is collected, and $\eta_c = 0$ indicates that the channel is an opening to the environment but \emph{none} of the information leaking out is collected. 
Any unitary part of the dynamics can be applied using the Hamiltonian $\hat{H}_S$. 
The $dW_c$ denote a Wiener process associated with each measurement; this delta--correlated Gaussian white noise models the random nature of the measurement backaction, {and is responsible for the stochasticity of \eqref{SME}, as well as the noise in the measurement records 
\be 
r_c = \mathrm{tr}\left[\rho(\hat{L}_c + \hat{L}_c^\dag) \right] + \frac{dW_c}{dt}.
\ee}
{When the conditional state update \eqref{general-stateup} is expanded to $O(dt)$, using the operators \eqref{M-het} or \eqref{M-hom}, one may recover the SME, and thus ascertain that these two approaches agree (i.e.~the SME may be recovered as an approximation of the formalism we emphasize, for all cases of diffusive measurement dynamics considered in this manuscript). See Appendix~\ref{sec-KrausVSME} for details. }

The SME has been fruitfully applied in past work on systems similar to those we consider here \cite{Carvalho2007, Viviescas2010, Mascarenhas2010, Mascarenhas2010-1, Mascarenhas2011, Carvalho2011, Santos2012, Zhang2019Heralded}. 
While the SME is a powerful tool for the purposes of calculations, developing the corresponding Kraus operator treatment as we have done above has some advantages; specifically, our Kraus operators 1) {offer a conceptually transparent view of the measurements we consider here, and the inferences an observer may draw from their outcomes}, and 2) offer a good alternative to direct integration of the SME in numerical modeling, {as fewer approximations are necessarily made (see e.g.~Refs.~\cite{Rouchon2015, Cortez2017, Guevara2019} for closely--related comments)}.

\section{Which--Path Information and Interference \label{sec-erasure}}

We now consider the ``which--path information'' available in the measurement records obtained by photon counting, homodyne, and heterodyne detection.
In other words, we consider under what circumstances an observer using these protocols is able to ascertain which qubit makes particular contributions to the measurement record. 
Measurements that erase the which--path information (i.e.~do not carry information that disambiguates the origin of the emitted signal) are suitable for entanglement generation. 
Conversely, those that allow emission from qubit A or B to become distinguishable will reduce or spoil the possibility of entanglement generation.

\subsection{Photodetection and Interference \label{sec-phot-interference}} 

An ideal photodetector at port 3 measures the photon number $\hat{N}_3 = \hat{a}_3^\dag \hat{a}_3$ at each timestep, and the photodetector at port 4 measures $\hat{N}_4 = \hat{a}_4^\dag \hat{a}_4$. Notice that $\hat{N}_3$ and $\hat{N}_4$ are totally independent of the phases $\theta$ and $\vartheta$ (see Fig.~\ref{fig-splitterphase}), so these do not impact the measurement at all in this case.
If a photon is inserted with certainty at either port 1 or port 2, the probability that the ensuing click is registered at 3 or 4 is the same either way; this overlap in the probabilities associated with the measurement outcomes between our two paths indicates that the which--path information is erased by this measurement. 
{Moreover, the path erasure occurs unitarily, without any loss of coherence. In this sense, the beamsplitter functions as a \emph{quantum eraser} \cite{Scully1991, Kim2000, Walborn2002, Ma2012, Ma2013}}.
This means that, given an initial two--qubit state for which the photon origin is ambiguous (such as $\ket{ee}$), our photon counting measurement will be unable to disambiguate which qubit the photon came from, and can therefore generate entanglement {(coherent correlations) between emitters}. 

\par Our model predicts that certain qubit states lead to complete destructive interference at either output ports 3 or 4. This is because entangled states of the qubits / emitters map directly onto entangled photon states.
Consider the two--qubit Bell state 
$\ket{\Psi^\pm} = (\ket{eg} \pm \ket{ge})/\sqrt{2}$.
The resulting photon emission is given by
\be 
\tfrac{1}{\sqrt{2}}\left(\hat{a}_1^\dag \pm \hat{a}_2^\dag \right) \ket{0_1 0_2},
\ee
and the beamsplitter relations \eqref{bs} show that these then become either 
\be 
e^{i\theta} \hat{a}_3^\dag \ket{0_3 0_4} ~(+), \text{ or } e^{i\vartheta} \hat{a}_4^\dag \ket{0_3 0_4} ~(-).
\ee
This means that when the qubits are in state $\ket{\Psi^+}$ port 4 is completely dark, and $\ket{\Psi^-}$ leaves port 3 dark. 
A direct consequence is that in the photodetection case, the second photon measured must be seen at the same detector as the first because the first click creates one of the two Bell states $\ket{\Psi^\pm}$, which in turn creates an interference effect for the next photon. 
The interference occurs independently of the type of measurements performed after the beamsplitter, and some interesting consequences of this are developed in appendix~\ref{sec-het-simul}.

\subsection{Quadrature Measurements and Which--Path Information}

\par What happens when the observer makes some measurement along one (homodyne) or both (heterodyne) quadratures at ports 3 and 4 instead? Consider measuring {combinations of}
\be \begin{split} \label{quad-def}
\text{at }3:& \quad \hat{X}_3 = \tfrac{1}{\sqrt{2}}(\hat{a}_3^\dag+\hat{a}_3), \: \hat{P}_3 = \tfrac{i}{\sqrt{2}}(\hat{a}_3^\dag - \hat{a}_3), \\ \text{at }4:& \quad \hat{X}_4 = \tfrac{1}{\sqrt{2}}(\hat{a}_4^\dag+\hat{a}_4), \: \hat{P}_4 = \tfrac{i}{\sqrt{2}}(\hat{a}_4^\dag - \hat{a}_4).
\end{split} \ee
From the beamsplitter relations \eqref{bs}, it is apparent that (e.g.~for $\theta = 0 = \vartheta$), we can have situations where a photon originating from port 1 leads to an in--phase measurement event, as experienced between 3 and 4, whereas a photon originating from port 2 leads to an effect which is $180^\circ$ out of phase between ports 3 and 4. 
We need to be careful then: Depending on which quadrature(s) we measure at each output, we may be able to determine whether light was reflected or transmitted at a beamsplitter. 
We will confirm that in situations where we can thereby make inferences about the which--qubit origin of information in the measurement signals, the possibility to create entanglement between the qubits with that measurement is destroyed. 

\begin{figure}[t]
    \centering
    \includegraphics[width=.75\columnwidth]{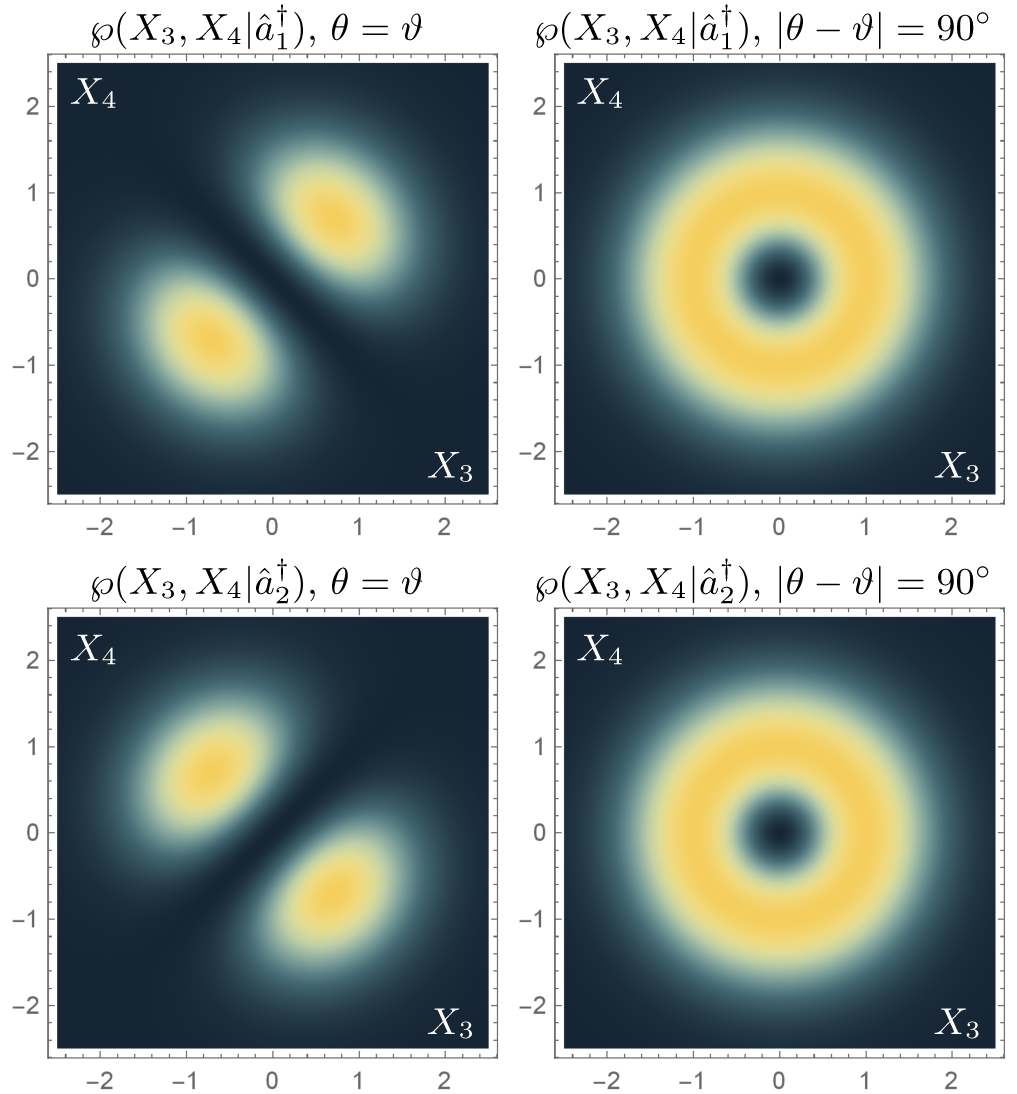}
    \caption{We plot the probability density \eqref{tQ-probdist}, corresponding to homodyne measurements at both system outputs, as a function of $X_3$ ($x$--axis) and $X_4$ ($y$--axis). A photon is allowed to enter at one port or the other; overlap of the subsequent probability density distributions for the measurement outcomes indicate that this which--path information is erased, while different distributions between the two cases indicate that the measurement can distinguish the photon source. In the left column we show the probability distributions for the homodyne measurement settings $\theta = 0 = \vartheta$; since the distributions differ between the case $\hat{a}^\dag_1$ (top) and $\hat{a}^\dag_2$ (bottom), we conclude that the which path information is not erased under these settings, which prevents measurement--induced entanglement genesis between our qubits. In the right column, by contrast, we see that the choice $\theta = 0$ and $\vartheta = 90^\circ$ leads to the same distribution of measurement outcomes for either photon input; the which--path information is thereby erased for these settings, which will be used for most of the homodyne examples developed later in the text. Generically, any choice which satisfies $|\theta - \vartheta| = 90^\circ$ erases the which--path information, yielding overlap as shown in the right column. 
    } \label{fig-QHom}
\end{figure}

\par We proceed by looking more closely at measurements involving information from only one quadrature i.e.~homodyne detection (equivalently, measurements made via phase--sensitive amplification of each output). We can consider a probability density associated with the {\color{MidnightBlue} blue} terms in \eqref{M-hom}. It is useful to consider {an optical} state $\hat{a}^\dag_1 \ket{0_1 0_2}$ ($+$) or $\hat{a}_2^\dag \ket{0_1 0_2}$ ($-$) (a single photon enters from one input or the other), for which the output optical state at the detectors is given by
\be \label{pathstate}
\ket{\psi_{3,4}} = \frac{1}{\sqrt{2}} \left( \hat{a}_3^\dag e^{i\theta} \pm \hat{a}_4^\dag e^{i\vartheta}\right) \ket{0_3 0_4}.
\ee
These states, by definition, carry perfect which--path information from the start, and the two--qubit states $\ket{ge}$ or $\ket{eg}$ which map onto them are therefore inappropriate initial states from which to begin a process leading to entanglement production. 
Our point here is to see which measurements preserve or erase that information, which we are presently inserting into the system in the most definite way we can.
The optical state \eqref{pathstate} leads to the probability density
\be \label{tQ-probdist} \begin{split}
\wp&(X_3,X_4|\hat{a}_1^\dag \:(+)\text{ or }\hat{a}_2^\dag\:(-) ) = |\ip{X_3 X_4}{\psi_{3,4}}|^2  \propto e^{-X_3^2-X_4^2} \left( X_3^2 + X_4^2 \pm 2 X_3 X_4 \cos(\theta-\vartheta) \right).
\end{split} \ee
It is obvious that the distributions \eqref{tQ-probdist} will be different between the $\pm$ cases (and therefore those cases are at least partially distinguishable), \emph{except} for a choice of $\theta$ and $\vartheta$ such that $\cos(\theta-\vartheta) = 0$. In other words, we can erase the which--path information by choosing, e.g., $\theta = 0$ and $\vartheta = 90^\circ$, which is effectively equivalent to measuring the $\hat{X}$ quadrature of mode 3, and the $\hat{P}$ quadrature of mode 4. See Fig.~\ref{fig-QHom}. The function $\wp$ is a proper probability density, because the states $\lbrace\ket{X}\rbrace$ form a complete set.

We can make similar comments about the heterodyne case by looking at the joint (two--mode) Husimi--$Q$ function at the outputs 3 and 4 \cite{Caves2012}. If $\ket{\psi_{3,4}}$ describes the output photon state, the $Q$ function is given by
\be 
Q(\alpha,\beta) = \tfrac{1}{\pi^2}| \ip{\alpha \beta}{\psi_{3,4}}|^2
\ee
where we are using a coherent state $\hat{a}_3 \ket{\alpha} = \alpha \ket{\alpha}$ at mode 3, and a coherent state $\hat{a}_4 \ket{\beta} = \beta \ket{\beta}$ at mode 4. We will decompose the complex coherent state eigenvalues according to $\alpha = X_3 + i P_3$ and $\beta = X_4 + i P_4$. 
Then we may write 
\be \label{Q-het} \begin{split}
Q &= \frac{1}{2\pi^2} e^{-|\alpha|^2-|\beta|^2}\left| \alpha^\ast e^{i\theta} \pm \beta^\ast e^{i\vartheta} \right|^2 \\
& = \frac{e^{-X_3^2-X_4^2-P_3^2-P_4^2}}{2\pi^2} \bigg[ X_3^2 + X_4^2 + P_3^2 + P_4^2 \pm 2(X_3 X_4 + P_3 P_4)\cos(\theta - \vartheta) \\& \quad\quad\quad\quad\quad\quad\quad\quad\quad\quad\quad\quad\quad\quad\quad~\,\pm 2(X_4 P_3 - X_3 P_4)\sin(\theta-\vartheta) \bigg],
\end{split} \ee
where the $+$ corresponds to the case where a photon started in port 1, and the $-$ corresponds to the case where a photon started in port 2.
This derivation works in a similar spirit to the one used in the homodyne case, with the notable difference that the $Q$--function is a quasiprobability distribution (because the states $\lbrace \ket{\alpha} \rbrace$, unlike the states $\lbrace \ket{X} \rbrace$, form an overcomplete basis). 
We can immediately see that functional form \eqref{Q-het} would allow for the $+$ and $-$ case to be distinguished; this suggests that heterodyne monitoring is always able to keep our sources distinguishable, and such measurements are consequently expected to be much less interesting to us from an entanglement genesis standpoint (which we confirm below and in Appendix~\ref{sec-het-simul}). 

Above we have provided simple arguments 1) for how joint homodyne detection can lead to erasure of the which--path information in our system (and therefore lead to some entanglement generation), and 2) that this information erasure is impossible for the cases of interest using {joint} heterodyne detection. 
{We briefly highlight how path distinguishability leads to separable states, before shifting our focus to the more interesting entangling cases.}

The heterodyne measurement, as discussed above, generates outcomes associated with coherent states $|\alpha\rangle$ and $|\beta\rangle$ respectively on the ports 3 and 4, satisfying the relations
{$
    \hat{a}_{3}\ket{\alpha_3 \beta_4}=\alpha\ket{\alpha_3\beta_4}$ and $\hat{a}_{4}\ket{\alpha_3 \beta_4}=\beta\ket{\alpha_3 \beta_4}
$}.
Inverting the beamsplitter relations \eqref{bs}, we can establish that the action of $\hat{a}_1$ and $\hat{a}_2$ on such a state follows
\begin{eqnarray}
        \hat{a}_{1}\ket{\alpha_3 \beta_4}&=&\tfrac{1}{\sqrt{2}}(e^{-i\theta}\alpha+e^{-i\vartheta}\beta)\ket{\alpha_3 \beta_4},\nonumber\\
       \hat{a}_{2}\ket{\alpha_3 \beta_4}&=&\tfrac{1}{\sqrt{2}}(e^{-i\theta}\alpha-e^{-i\vartheta}\beta)\ket{\alpha_3 \beta_4}.
\end{eqnarray}
This implies that $\ket{\alpha_3 \beta_4}$ can be written as a product of coherent states in modes 1 and 2 as well, i.e.
\be \label{het-sep}
\ket{\alpha_3 \beta_4} = \big| \underbrace{\tfrac{1}{\sqrt{2}}(e^{-i\theta}\alpha+e^{-i\vartheta}\beta)}_{\text{mode }1} ~ \underbrace{\tfrac{1}{\sqrt{2}}(e^{-i\theta}\alpha-e^{-i\vartheta}\beta)}_{\text{mode }2} \big\rangle.
\ee 
This proves that joint heterodyne measurement is effectively preparing separable states {of} modes one and two, leading to no entanglement generation between qubits. 

The same kind of separability argument can be made for the homodyne measurement in the case $\theta = \vartheta$ which maximizes which--path distinguishability (minimizes which--path information erasure). {To see this, we consider measurement of observables $\hat{X}_{3}$ and $\hat{X}_{4}$ in ports 3 and 4 respectively, yielding outcomes $X_3$ and $X_4$, such that,
$
    \hat{X}_{3}\ket{X_3\, X_4}=X_3 \ket{X_3\, X_4}$ and $\hat{X}_{4}\ket{X_3\, X_4}=X_4\ket{X_3\, X_4}
$.
For the choice $\theta = \vartheta$, we may write $\hat{X}_{3}=[\hat{X}_{1}(\theta)+\hat{X}_{2}(\theta)]/\sqrt{2}$ and $\hat{X}_{4}=[\hat{X}_{1}(\theta)-\hat{X}_{2}(\theta)]/\sqrt{2}$,
for $\hat{X}_{j}(\theta)=(e^{-i\theta}\hat{a}_{j}^{\dagger}+e^{i\theta}\hat{a}_{j})/\sqrt{2}$ is a local observable at port $j=1,2$. We therefore have
\begin{subequations}
\be
\big[\hat{X}_{1}(\theta)+\hat{X}_{2}(\theta)\big]\ket{X_3\,X_4}=\sqrt{2}\,X_3\ket{X_3\, X_4} \quad\text{\&}\quad \big[\hat{X}_{1}(\theta)-\hat{X}_{2}(\theta)\big]\ket{X_3\,X_4}=\sqrt{2}\,X_4\ket{X_3\, X_4},
\ee
which may be rearranged to read 
\be \hat{X}_{1}(\theta)\ket{X_3\,X_4}=\tfrac{1}{\sqrt{2}}(X_3+X_4)\ket{X_3\, X_4}\quad\text{\&}\quad \hat{X}_{2}(\theta)\ket{X_3\,X_4}=\tfrac{1}{\sqrt{2}}(X_3-X_4)\ket{X_3\, X_4}.
\ee \end{subequations}
As in the heterodyne case, we see that $\ket{X_3\, X_4}$ is a separable state also for modes 1 and 2, i.e.
\be \label{homo-sep}
\ket{X_3\, X_4} \leftrightarrow \big| \underbrace{\tfrac{1}{\sqrt{2}}(X_3+X_4)}_{\text{mode }1} ~ \underbrace{\tfrac{1}{\sqrt{2}}(X_3-X_4)}_{\text{mode }2} \big\rangle.
\ee 
We consequently see that the distributions in the left-hand column of Fig.~\ref{fig-QHom}, with the most path distinguishability, correspond to a measurement that projects the optical degrees of freedom in a separable state, rather than an entangled one. 
Related calculations, including the entangling case, appear in Appendix \ref{sec-onesteptest}. 
}

\subsection{Connecting Which--Path Information Erasure to Entanglement Swapping}
Here we argue that the generation of entanglement in our optimal homodyne scheme, as described above, can also be understood as entanglement swapping, using a continuous variable Einstein--Podolsky--Rosen (EPR) basis measurement \cite{EPR1935}. In entanglement swapping, one has two pairs of initially--entangled parties (four parties in total). 
To swap the entanglement, one generically performs a measurement in an entangled basis, on two parties---one from each initially entangled pair. This, in turn, entangles the remaining parties, effectively swapping the quantum entanglement between them. 
In our context, the fluorescence process naturally generates some time--dependent entanglement between each qubit and its cavity output mode (qubit A is entangled to mode 1, qubit B to mode 2, as per \eqref{psi_in}); by jointly measuring the fields after they are mixed on the beamsplitter (modes 3 and 4), we can swap the entanglement around so that the two qubits share correlations instead. 

This can be realized as follows: The observables $\hat{X}_{+}=\hat{X}_1+\hat{X}_2$ and $\hat{P}_-=\hat{P}_1-\hat{P}_2$ commute, and completely characterize the two--mode quantum state. Jointly measuring them can generate EPR correlations between the modes 1 and 2. A measurement of $\hat{X}_+$ with readout $r_{+}$ and a measurement of $\hat{P}_-$ yielding readout $r_{-}$, prepare a continuous variable EPR state with $X_1 + X_2=r_{+}$ and $P_1 - P_2 =r_{-}$. We first focus on the perfectly correlated scenario $X_1+X_2=0$ and $P_1-P_2=0$: the two mode wavefunction is then \cite{lvovsky2015squeezed}
\begin{subequations}
\begin{equation}\label{antiC}
\ip{X_1 X_2}{\psi_{1,2}} \propto\delta(X_1+X_2)
\end{equation}
in the position basis, and
\begin{equation}
\ip{P_1 P_2}{\psi_{1,2}} \propto\delta(P_1-P_2)
\end{equation} \end{subequations}
in the momentum basis.
The particular state \eqref{antiC} (which has the same symmetry of a photon pair produced from  vacuum) can also be written in the Wigner form as~\cite{pirandola2006quantum},
\begin{equation}
    \mathcal{W}(X_1,X_2,P_1,P_2) = \delta(X_1+X_2)\delta(P_1-P_2).
\end{equation}
An arbitrary set of readouts $\{r_{-},r_{+}\}$ preparing maximally--entangled field modes are related to the state \eqref{antiC} by a local unitary operation in either of the modes, which is a generic single mode displacement operation that preserves the entanglement. The state \eqref{antiC} is also the limit of maximal squeezing in a two mode squeezed vacuum state, 
\begin{equation}
  \ket{\psi^s_{1,2}} =   \sum_{n=0}^{\infty}\frac{1}{\text{cosh}(s)}\tanh(s)^{n}\ket{n_1}\ket{n_2}.
\end{equation}
The wavefunction $\ip{X_1,X_2}{\psi^s_{1,2}}$ is identical to \eqref{antiC} in the limit $s\rightarrow\infty$~\cite{lvovsky2015squeezed}, highlighting the role of the squeezing operation implicit in {quantum--limited} detection. 
{Implementation of such an operation via a single two--input quantum--limited amplifier operating in the large--gain limit is described by e.g.~Refs.~\cite{Reid1989, Flurin2012,Silveri2016}.} 

In our setting, the modes $\hat{X}_{+}$ and $\hat{P}_{-}$ are realized from modes $1$ and $2$ by passing them through the beamsplitter. 
Then local measurements of quadratures on the outgoing ports realize the measurements of the sum  $\hat{X}_{+}$ and difference $\hat{P}_{-}$ quadratures of the input modes. 
We can formalize this statement 
by noting that $(\hat{X}_1+\hat{X}_2)/\sqrt{2} \rightarrow \hat{X}_3$ and $(\hat{P}_1-\hat{P}_2)/\sqrt{2} \rightarrow \hat{P}_4$ under the beamsplitter relations \eqref{bs}, with $\theta = 0 = \vartheta$. Equivalently, we have $(\hat{P}_1-\hat{P}_2)\sqrt{2} \rightarrow \hat{X}_4$ with $\theta = 0$ and $\vartheta = 90^\circ$ (for the generalized $\hat{X}_4 = (\hat{a}_4^\dag e^{i\vartheta} + \hat{a}_4 e^{-i\vartheta})/\sqrt{2}$); thus, the measurement we have claimed erases which--path information, with $|\theta- \vartheta| = 90^\circ$ is exactly of the EPR form just {introduced}.

We note that there is some precedent for the double homodyne detection device we emphasize. 
Quite similar devices have been used to verify the properties of continuous--variable (optical) EPR states \cite{EPR1935, Reid1989, Ou1992}, as well as in related experiments concerned with the steerability of such states \cite{Reid2009, Handchen2012}. 
The entanglement swapping interpretation we give above has also been used in explaining the effect of such homodyne measurements {elsewhere} \cite{Pirandola2015, Takeda2015}, {and applies to the photodetection case as well \cite{Santos2012}}. 
Implementations of these concepts directly using microwave amplification hardware, which is critical in realizing quantum trajectory experiments with superconducting qubits, have been proposed and realized. 
Specifically, the sorts of EPR measurements of interest have been performed with microwave quantum optics (without qubits at the source) \cite{Flurin2012}. Furthermore, an implementation of the measurement of interest has been proposed in the context of dispersive qubit measurement, based on typical amplifiers used in superconducting circuit experiments \cite{Silveri2016}. 
The homodyne measurement of fluorescence we derive above as optimal for erasing which path information has previously been shown \emph{also} to be the optimal diffusive unraveling of the two--qubit master equation based on decay channels for the purposes of entanglement preservation \cite{Viviescas2010}. 
As we explore numerical results for this case in Secs.~\ref{sec-homodyne} and \ref{sec-inefficient}, we will be able to further confirm that result, and elaborate substantially on it.

\par This concludes our general overview of the measurements we wish to consider. 
{We have formalized several continuous measurement schemes above, are may go forward with an understanding of how they build on an established conceptual and experimental foundation.}
The remainder of the paper is dedicated to detailing the dynamics of the two--qubit states for specific measurement cases, using numerical simulation. 
We review the case of photodetection and jump trajectories in Sec.~\ref{sec-photodetect}. 
We then develop the homodyne detection case in Sec.~\ref{sec-homodyne}, drawing comparisons with the photodetection case. 
{Characterization and comparison of these dynamics allows us to illustrate in detail how different measurements create different types of two--qubit correlations, putting the principles we have discussed above into practice.}
{With the ideal case established we then also consider the more realistic case of} inefficient measurements in Sec.~\ref{sec-inefficient}. 

\section{Jump Trajectories from Continuous Photodetection \label{sec-photodetect}}

We turn our attention to photodetection and jump trajectories. Three types of events that can occur within a single measurement timestep are of primary interest; either no photons are detected, as described by $\hat{\mathcal{M}}_{00} = \bra{0_3 0_4} \mathcal{M} \ket{0_3 0_4}$, a photon is measured in output 3 as described by $\hat{\mathcal{M}}_{10} = \bra{1_3 0_4} \mathcal{M} \ket{0_3 0_4}$, or a photon is measured in output 4 as described by $\hat{\mathcal{M}}_{01} = \bra{0_3 1_4} \mathcal{M} \ket{0_3 0_4}$. There is also the more remote possibility that both cavities emit at once (within the same detector integration interval $dt$), described by $\hat{\mathcal{M}}_{20}$ or $\hat{\mathcal{M}}_{02}$. These Kraus operators are given by
\begin{subequations}
\be \label{photodect_mat_first}
\hat{\mathcal{M}}_{00} = \left( \begin{array}{cccc}
1-\epsilon & 0 & 0 & 0\\
0 & \sqrt{1-\epsilon} & 0 & 0 \\
0 & 0 & \sqrt{1-\epsilon} & 0 \\
0 & 0 & 0 & 1
\end{array} \right),
\ee \be \label{PD1}
\hat{\mathcal{M}}_{10} = \left( \begin{array}{cccc}
0 & 0 & 0 & 0 \\
{\color{Plum} \sqrt{\frac{\epsilon(1-\epsilon)}{2}} } & 0 & 0 & 0 \\
{\color{Plum} \sqrt{\frac{\epsilon(1-\epsilon)}{2}} } & 0 & 0 & 0 \\
0 & \sqrt{\frac{\epsilon}{2}} & \sqrt{\frac{\epsilon}{2}} & 0
\end{array} \right),
\quad 
\hat{\mathcal{M}}_{01} =\left( \begin{array}{cccc}
0 & 0 & 0 & 0 \\
{\color{Plum} -\sqrt{\frac{\epsilon(1-\epsilon)}{2}} } & 0 & 0 & 0 \\
{\color{Plum} \sqrt{\frac{\epsilon(1-\epsilon)}{2}} } & 0 & 0 & 0 \\
0 & \sqrt{\frac{\epsilon}{2}} & -\sqrt{\frac{\epsilon}{2}} & 0
\end{array} \right),
\ee \be 
\hat{\mathcal{M}}_{20} = \left( \begin{array}{cccc}
0 & 0 & 0 & 0 \\
0 & 0 & 0 & 0 \\
0 & 0 & 0 & 0 \\
\epsilon/\sqrt{2} & 0 & 0 & 0 
\end{array} \right), \quad \label{photodect_mat_last}
\hat{\mathcal{M}}_{02} = \left( \begin{array}{cccc}
0 & 0 & 0 & 0 \\
0 & 0 & 0 & 0 \\
0 & 0 & 0 & 0 \\
-\epsilon/\sqrt{2} & 0 & 0 & 0 
\end{array} \right).
\ee \end{subequations}
These form a complete set of outcomes, such that $\sum \hat{\mathcal{M}}_{ij}^\dag \hat{\mathcal{M}}_{ij} = \hat{\mathbb{I}}$, where the sum is over all five matrices above.
Simulations of this situation simply involve applying the appropriate $\hat{\mathcal{M}}_{ij}$ to $\rho$ according to 
\be
\rho(t+dt) = \frac{\hat{\mathcal{M}}_{ij} \rho(t) \hat{\mathcal{M}}_{ij}^\dag}{\text{tr} \left( \hat{\mathcal{M}}_{ij} \rho(t) \hat{\mathcal{M}}_{ij}^\dag \right)},
\ee
where the outcomes of different combinations of detector clicks are generated randomly over time, according to the correct statistics. To do this, we derive the probabilities
\be 
w_{ij} = \text{tr} \left( \hat{\mathcal{M}}_{ij} \rho \hat{\mathcal{M}}_{ij}^\dag \right),
\ee
assigned to each of the outcomes above, which are normalized such that $\sum w_{ij} = 1$. Then we may draw a number from a multinomial distribution at every timestep, each possibility of which corresponds to a given detector outcome. The weight factors are
\begin{subequations}
\be \begin{split}
w_{00} = 1 &- dt \: \gamma \: \Xi
+ dt^2 \gamma^2 \Theta,
\end{split} \ee 
\be \begin{split}
w_{10} = & \: \gamma \: dt \left(\frac{\Xi}{2} - \frac{q_{4}}{\sqrt{2}} \right) - \gamma^2 dt^2 \Theta,
\quad 
w_{01} = \: \gamma \: dt \left(\frac{\Xi}{2} + \frac{q_{4}}{\sqrt{2}} \right)
 - \gamma^2 dt^2 \Theta,
\end{split} \ee \be 
w_{02} = \frac{\gamma^2 dt^2}{2}\Theta = w_{20},
\ee
for 
\be \label{XiTheta} \begin{split}
\Xi &\equiv 1 + \frac{q_1}{\sqrt{2}} + \frac{q_2}{\sqrt{6}} + \frac{2 q_3}{\sqrt{3}},
\quad 
\Theta \equiv \frac{1}{4} + \frac{q_1}{\sqrt{2}} + \frac{q_2}{\sqrt{6}} + \frac{q_3}{2\sqrt{3}}.
\end{split} \ee \end{subequations}
We have introduced a set of two--qubit generalized Bloch coordinates $q_j$, with $1\leq j \leq 15$ (see Appendix \ref{sec-bloch-coord} for details).

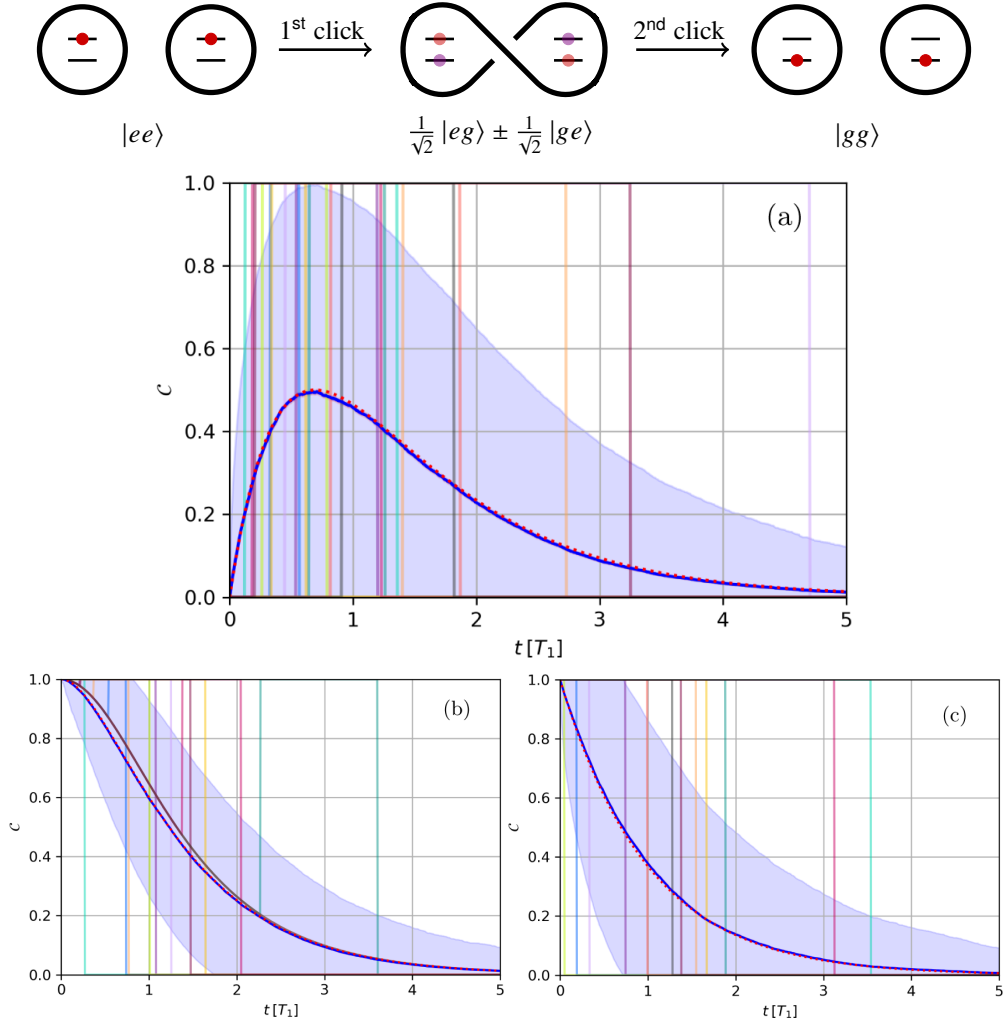
\begin{figure}
    \centering
    \vspace{-10pt}
    \begin{tikzpicture}[scale = 0.95]
\draw[line width = 2pt] (0.75,0.9) circle (0.6 cm);
\draw[line width = 2pt] (2.55,0.9) circle (0.6 cm);
\draw[line width = 1pt] (0.55,1.05) -- (0.95,1.05);
\draw[line width = 1pt] (0.55,0.75) -- (0.95,0.75);
\draw[line width = 1pt] (2.35,1.05) -- (2.75,1.05);
\draw[line width = 1pt] (2.35,0.75) -- (2.75,0.75);
\filldraw[draw = black!20!red, fill = black!20!red] (0.75,1.05) circle (0.08 cm);
\filldraw[draw = black!20!red, fill = black!20!red] (2.55,1.05) circle (0.08 cm);
\node[] at (1.6,-0.3) {$\ket{ee}$};
\node[] at (4.1,1.2) {1$^{\mathsf{st}}$ click};
\draw[->, line width = 1pt] (3.5,0.9) -- (4.8,0.9);
\draw[line width = 2pt, rounded corners = 0.6 cm] (6.5,0.75) -- (5.75,0) -- (5.0,0.9) -- (5.75,1.8) -- (6.65,0.9) -- (7.55,0.0) -- (8.3,0.9) -- (7.55,1.8) -- (6.8,1.05);
\draw[line width = 1pt] (5.55,1.05) -- (5.95,1.05);
\draw[line width = 1pt] (5.55,0.75) -- (5.95,0.75);
\draw[line width = 1pt] (7.35,1.05) -- (7.75,1.05);
\draw[line width = 1pt] (7.35,0.75) -- (7.75,0.75);
\filldraw[draw = black!20!red, fill = black!20!red,opacity = 0.5] (5.75,1.05) circle (0.08 cm);
\filldraw[draw = black!20!red, fill = black!20!red,opacity = 0.5] (7.55,0.75) circle (0.08 cm);
\filldraw[draw = patriarch, fill = patriarch,opacity = 0.5] (5.75,0.75) circle (0.08 cm);
\filldraw[draw = patriarch, fill = patriarch,opacity = 0.5] (7.55,1.05) circle (0.08 cm);
\node[] at (6.6,-0.3) {$\tfrac{1}{\sqrt{2}} \ket{eg} \pm \tfrac{1}{\sqrt{2}}\ket{ge}$};
\node[] at (9.1,1.2) {2$^{\mathsf{nd}}$ click};
\draw[->, line width = 1pt] (8.5,0.9) -- (9.8,0.9);
\draw[line width = 2pt] (10.75,0.9) circle (0.6 cm);
\draw[line width = 2pt] (12.55,0.9) circle (0.6 cm);
\draw[line width = 1pt] (10.55,1.05) -- (10.95,1.05);
\draw[line width = 1pt] (10.55,0.75) -- (10.95,0.75);
\draw[line width = 1pt] (12.35,1.05) -- (12.75,1.05);
\draw[line width = 1pt] (12.35,0.75) -- (12.75,0.75);
\filldraw[draw = black!20!red, fill = black!20!red] (10.75,0.75) circle (0.08 cm);
\filldraw[draw = black!20!red, fill = black!20!red] (12.55,0.75) circle (0.08 cm);
\node[] at (11.6,-0.3) {$\ket{gg}$};
\end{tikzpicture}
    \includegraphics[width=.7\textwidth]{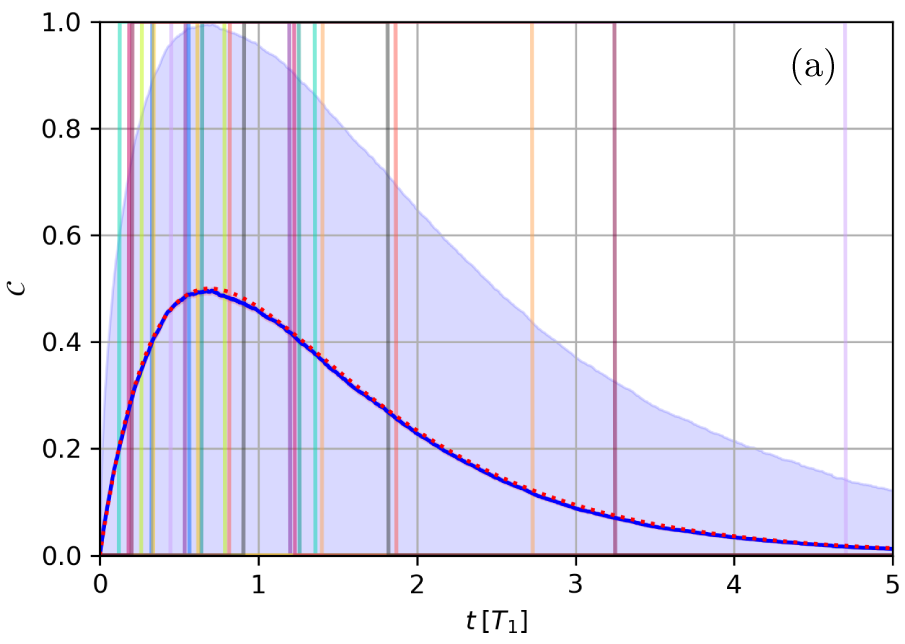} \\
    \includegraphics[width = \textwidth]{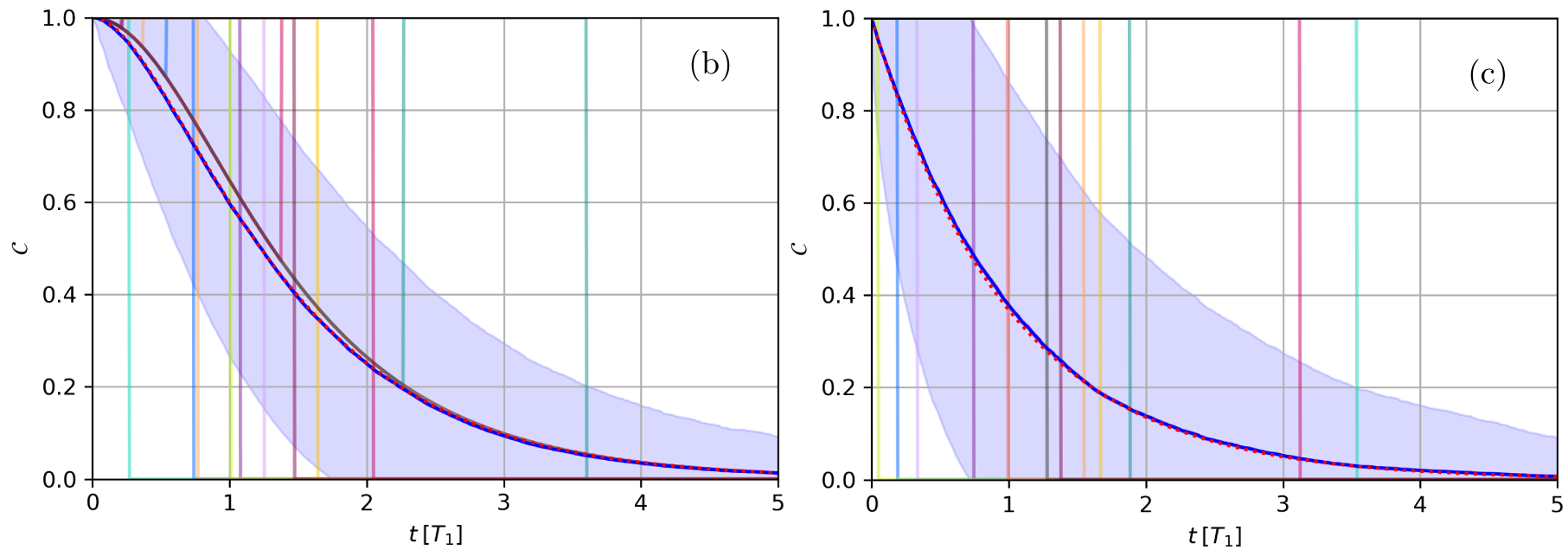}
    \caption{
We illustrate the entanglement generation process, as experienced by an observer with access to the measurement record obtained by perfect photodetectors, and simulate ensembles of the corresponding quantum trajectories for various initial states. The concurrence $\mathcal{C}$ of a dozen individual jump trajectories (low opacity, in multiple colors unique to each trajectory), and the average concurrence over an ensemble of 10,000 quantum trajectories (blue) are plotted above. The surrounding pale blue envelope denotes the standard deviation of the concurrence of the underlying ensemble. 
All figures are generated with simulations described in Sec.~\ref{sec-photodetect} and assume that ideal photodetectors are placed at ports 3 and 4, as illustrated in Fig.~\ref{fig-splitterphase}. 
We use $\gamma = 1~\mathrm{MHz} = 1~(\mathrm{\mu s})^{-1}$, and $dt \leq 5~\mathrm{ns}$ for numerical purposes, with a total duration $T = 5 T_1$. Both qubits are assumed to have the same decay rate $\gamma = T_1^{-1}$. The qubits are initialized in the state $\ket{ee}$ (a), and as illustrated above, $\ket{\Phi^+}$ (b), and $\ket{\Psi^+}$ (c), where $\ket{\Phi^\pm} = (\ket{ee}\pm\ket{gg})/\sqrt{2}$ and $\ket{\Psi^\pm} = (\ket{eg}\pm\ket{ge})/\sqrt{2}$ are Bell states. 
In (a) we see the rise and fall of entanglement generated by the measurement given the initial state $\ket{ee}$, which follows $\bar{\mathcal{C}} = 2 e^{-\gamma t}(1-e^{-\gamma t})$ (dotted red; see \eqref{barC-jhe}). In (b) and (c) we see that the measurement gradually erodes the initial two--qubit entanglement, which asymptotically approaches $\mathcal{C} = 0$ for $t \gg T_1$. The averages from simulation (solid blue) are in good agreement with the expressions $\bar{\mathcal{C}} = e^{-\gamma t}(2-e^{-\gamma t})$ and $\bar{\mathcal{C}} =e^{-\gamma t} $ in (b) and (c), respectively (dotted red).
    }
    \label{fig-photodect-conc}
\end{figure}

\par Some simulations of this scenario are shown in Fig.~\ref{fig-photodect-conc}, and we find that we can create substantial entanglement between the two qubits, as expected. Specifically, if both qubits are prepared in the excited state, the overwhelming majority of trajectories involve two photons coming out within a few $T_1 = \gamma^{-1}$ of the start of the experiment. 
A Bell state is prepared when the first photon comes out, and then the qubits must be in $\ket{gg}$ when the second exits. {The concurrence of jump trajectories simulated according to the scheme above, as well as their ensemble average, are plotted in} Fig.~\ref{fig-photodect-conc}. {Concurrence is a measure of two qubits' entanglement \cite{wooters1998}, which is defined formally in Appendix~\ref{sec-concurrence}.} 
{The concurrence takes on values $\mathcal{C}\in[0,1]$, with $\mathcal{C} = 1$ denoting maximal entanglement (as in a Bell state), and $\mathcal{C} = 0$ denotes a separable state.}

{The average concurrence yield over an ensemble of trajectories may be inferred from statistical arguments.} 
{Given the initial state $\ket{ee}$, it} is the projection of the optical field into either $\ket{0_3 1_4}$ or $\ket{1_3 0_4}$ (i.e.~into the $\lbrace \ket{1_1 0_2}, \ket{0_1 1_2} \rbrace$ subspace) that generates {one of the entangled states $\ket{\Psi}$ under photodetection.} {A single qubit, initially in $\ket{e}$, has a probability $\wp_e(t)  = e^{-\gamma\,t}$ to remain in $\ket{e}$, and a probability $\wp_g(t) = 1 - \wp_e(t)$ to have decayed to ground. 
It follows that in the two qubit case, with the initial state $\ket{ee}$, the probability for either one (but not both!) of the qubits A and B to have decayed goes as
} 
\be \label{barC-jhe}
\wp_e^{(A)}(t) \,\wp_g^{(B)}(t) + \wp_g^{(A)}(t)\, \wp_e^{(B)}(t) = 2 e^{-\gamma t}(1-e^{-\gamma t}) = \bar{\mathcal{C}}(t).
\ee
We denote this $\bar{\mathcal{C}}$, because the probability to get a state with concurrence $1$ is precisely the average concurrence generated over an ensemble of jump trajectories. 
This may equivalently be derived from the two qubit generalization of \eqref{psi_in}; the average concurrence given other initial states may be inferred via the same strategy, consistent with our earlier comments about our model effectively implementing a statistical inference procedure.\footnote{We further clarify that the argument of \eqref{barC-jhe} is a purely statistical one. Our use of \eqref{psi_in} to derive Kraus operators is, in general, restricted to short times (in parallel with collision models of quantum optics). The photodetection case, however, depends simply on how many photons have emerged up to a given time, and thus statistical arguments about the emission times can be used over longer intervals to obtain correct results. We see later, in \eqref{viviescas_cbar}, that the result for $\bar{\mathcal{C}}$ in \eqref{barC-jhe} can be derived in the homodyne case as well, as a solution to a differential equation \cite{Viviescas2010} (which necessarily arises from having gone into a time--continuum limit with individual measurements of an infinitesimal duration). } 
See Fig.~\ref{fig-photodect-conc}(a) for a comparison between the analytic expression $\bar{\mathcal{C}}$ and average from the simulated measurement process. The protocol described above can also be understood as an entanglement swap, and has been interpreted in this way elsewhere \cite{Santos2012}. 

\par We comment briefly on the decay of concurrence from different Bell states under this measurement protocol, as illustrated in Fig.~\ref{fig-photodect-conc}(b--c). 
Initializing our qubits in $\ket{\Phi^\pm} = (\ket{ee}\pm\ket{gg})/\sqrt{2}$ leads to the longest--lived average concurrence under photodetection; the slope of the average concurrence $\bar{\mathcal{C}} = e^{-\gamma t}(2-e^{-\gamma t})$ is zero at $t = 0$, indicating a prolonged entanglement lifetime before the exponential decay sets in. 
This is best understood in comparison with Fig.~\ref{fig-photodect-conc}(c); there the average concurrence from $\ket{\Psi^\pm} = (\ket{eg}\pm\ket{ge})/\sqrt{2}$ decays simply as $\bar{\mathcal{C}} = e^{-\gamma t}$. 
This is because an initial state $\ket{\Psi^\pm}$ generates only one jump in any realization, and that jump then drops the concurrence from $\mathcal{C} = 1$ to $\mathcal{C}=0$. 
Then the characteristic decay dynamics of the individual qubits maps directly onto the decay of the concurrence. 
By contrast, the states $\ket{\Phi^\pm}$ lead to either two jumps or no jumps in any given realization.  The concurrence stays high at the beginning of the no--jump case, before we can infer with high confidence that we aren't going to get a photon, \emph{and} the first jump in the two--jump case does not kill the concurrence, but rather creates the other Bell state $\ket{\Psi^\pm}$, which decays to $\mathcal{C} = 0$ only after the second jump. 
Thus the states $\ket{\Phi^\pm}$ exhibit a longer entanglement lifetime on average, by effectively delaying their decay to the separable state $\ket{gg}$. 
This brings to mind other works, which have shown that changing the encoding of an entangled state can make it more or less susceptible to disentanglement via certain environmental interactions \cite{Aolita_2015, YuEberly_2004, Frowis2011, Chaves2012}; we will be able to elaborate further on additional connections present in the continuous measurement case \cite{Mascarenhas2011, Viviescas2010} in the following section.

We conclude this section with some remarks about entanglement genesis and entanglement sudden death. 
Certain states are prone to finite--time disentanglement in the case where the system we have discussed is unmonitored (i.e.~when the detectors are turned off) \cite{YuEberly_2004}. 
Measurement both alters the average concurrence dynamics of the system, and opens the possibility for many distinct trajectories about that average, conditioned on the stochastic measurement outcomes. 
Under measurement, generically, a reverse process of sudden entanglement genesis after a finite waiting time becomes possible \cite{Williams2008}. 
The individual trajectories described above may be regarded as further examples of such lines of thinking; the jumps discussed above can herald the sudden genesis of entanglement from {a separable state}, or its complete destruction.


\section{Entanglement by Joint Homodyne Detection of Fluorescence \label{sec-homodyne}}

We now investigate the case where both of the outputs are homodyned, i.e.~we consider dynamics generated by a measurement of the type \eqref{M-hom}. 
Recall from the discussion above that we can optimally erase the which--path information by choosing $|\theta-\vartheta|=90^\circ$ in this scenario, and therefore expect these settings to correspondingly be optimal for generating two--qubit entanglement. 

\subsection{Concurrence Yield}

We develop additional expressions from \eqref{M-hom} that we need to perform simulations, and then work towards understanding the two--qubit state dynamics generated by this measurement.
The denominator of the state update equation \eqref{general-stateup} describes the probability density from which the readouts are drawn at each time step. As in the single qubit cases \cite{Jordan2015flor, FlorTeach2019,PL-Dissertation}, it is useful to expand the logarithm of that probability density to $O(dt)$, in much the same way we have when doing optimal path analysis \cite{Chantasri2013, Weber2014, Chantasri2015, Jordan2015flor, Areeya_Thesis, Lewalle2016, Mahdi2016, Lewalle2018, FlorTeach2019, PL-Dissertation}.
Expanding in this way gives us an expression $\mathcal{G}$ such that $\text{tr}(\hat{\mathcal{M}}_{34} \rho \hat{\mathcal{M}}_{34}^\dag) = e^{C + \mathcal{G} dt + O(dt^2)}$; the term $\mathcal{G}$ typically leaves expressions which are quadratic (Gaussian) in the readout, and this case is no exception. The readout statistics obey 
\begin{subequations} \be \label{G34} \begin{split}
\mathcal{G}_{34} =& -\tfrac{1}{2}\left(r_3 - \sqrt{\gamma} \chi_3 \right)^2 - \tfrac{1}{2}\left(r_4 - \sqrt{\gamma} \chi_4 \right)^2 \\    
& + \tfrac{\gamma}{2}\left( \chi_3^2 + \chi_4^2 \right) - \gamma \: \Xi  + \tfrac{\gamma}{\sqrt{2}} \big[ q_{13}(\sin(2\vartheta)-\sin(2\theta))  +q_7 (\cos(2\vartheta)-\cos(2\theta)) \big],
\end{split} \ee
\be \label{chi34} \begin{split}
\text{with} \quad \chi_3 =& (q_{11} + q_{12} + q_{14} + q_{15})\sin\theta  + (q_5+q_6 + q_8 + q_9)\cos\theta, \quad\text{and}
\end{split} \ee \be \begin{split}
\chi_4 =& -(q_5-q_6-q_8+q_9)\cos\vartheta - (q_{11} - q_{12} - q_{14} + q_{15})\sin\vartheta.
\end{split} \ee
\end{subequations}
We consequently see that our readouts $r_3$ and $r_4$ are drawn from Gaussians of variance $1/dt$, with means $\sqrt{\gamma}\chi_3(\theta)$ and $\sqrt{\gamma}\chi_4(\vartheta)$, respectively. 
The coordinates $\mathbf{q}$ parameterize arbitrary two--qubit states; they and the associated generalized Gell--Mann matrices $\boldsymbol{\hat{\Gamma}}$ are defined in Appendix \ref{sec-bloch-coord}.
Simulations are implemented by iteratively updating the density matrix over small timesteps, using readouts generated stochastically from the Gaussians just described. 
In the language of the SME, measurement records are $r_3 = \langle \hat{L}_3 + \hat{L}_3^\dag \rangle + \xi_3$, and $r_4 = \langle \hat{L}_4 + \hat{L}_4^\dag \rangle + \xi_4$, where the $\xi_j \sim dW_j/dt$, for $j=3,4$ are the noise terms. The $dW_j$ are Wiener increments, i.e.~$\xi_j$ are Gaussian variables of zero mean and variance $1/dt$ \cite{Jacobs2006, BookGardiner2}. 
The Gaussian form of $\exp\left(\mathcal{G}_{34}\right)$ is key in demonstrating that the form of the SME \eqref{SME} written in terms of Wiener increments $dW$, is in fact suitable for describing the scenario of interest.
We infer that the appropriate operators for the SME, which reproduce the correct signal means for $\theta = 0$ and $\vartheta = 90^\circ$, are $\hat{L}_3 + \hat{L}_3^\dag = \sqrt{\gamma}(\hat{\Gamma}_5 + \hat{\Gamma}_6 + \hat{\Gamma}_8 + \hat{\Gamma}_9)$ and $\hat{L}_4 + \hat{L}_4^\dag = \sqrt{\gamma}(-\hat{\Gamma}_{11} + \hat{\Gamma}_{12} + \hat{\Gamma}_{14} - \hat{\Gamma}_{15})$, or equivalently
\be \label{homodyne-SME-ops} 
\hat{L}_3 = \sqrt{\gamma/2}\left( \hat{\mathbb{I}}_A \otimes \hat{\sigma}_-^B + \hat{\sigma}_-^A \otimes \hat{\mathbb{I}}_B \right),
\quad 
\hat{L}_4 = i\sqrt{\gamma/2} \left(\hat{\sigma}_-^A \otimes \hat{\mathbb{I}}_B -  \hat{\mathbb{I}}_A \otimes \hat{\sigma}_-^B\right).
\ee
The factor $i$ on $\hat{L}_4$ relative to $\hat{L}_3$ is the $90^\circ$ phase difference which ensures the erasure of which--path information. For further details about the connection between the SME and {our} Kraus operator {approach, see Appendix}~\ref{sec-KrausVSME}. 

\begin{figure}
    \centering
    \includegraphics[width=\textwidth]{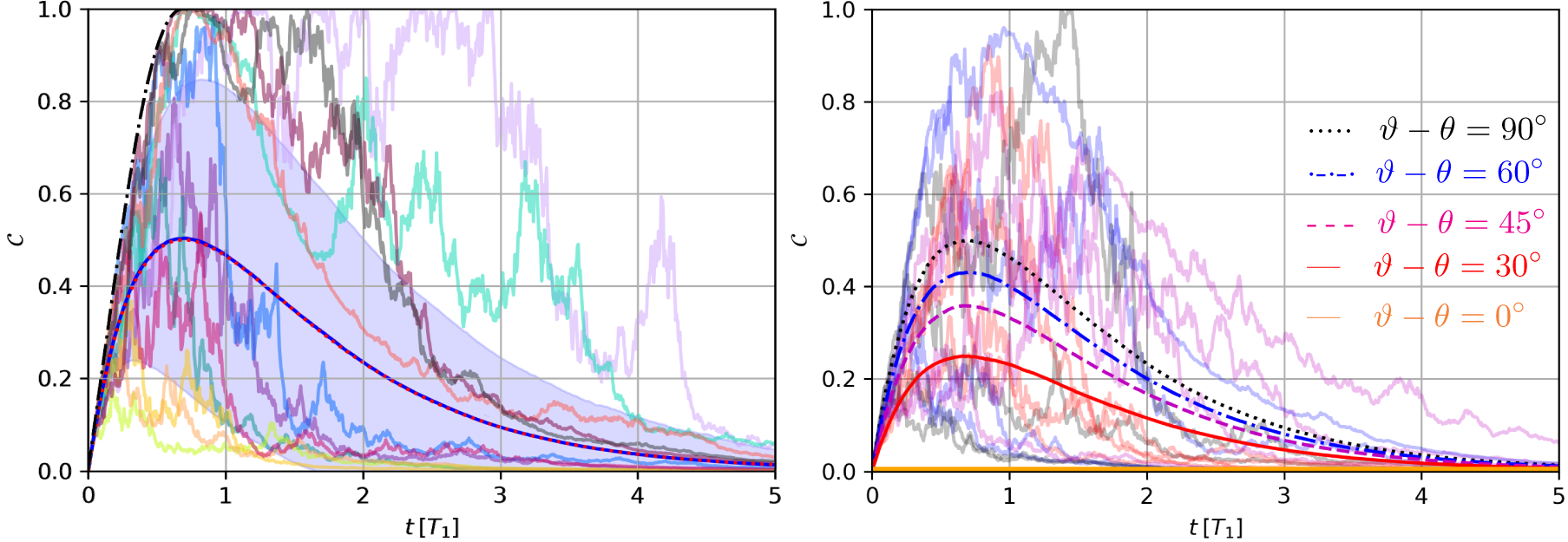}
    \caption{We show the average concurrence, and that of some trajectories, obtained from two homodyne detectors monitoring the output ports 3 and 4 of the device illustrated in Fig.~\ref{fig-splitterphase}. The initial state is $\ket{ee}$ for all trajectories. We plot the average two--qubit concurrence, and that of individual trajectories, in the left panel, using relative measurement phases $\theta = 0^\circ$ and $\vartheta = 90^\circ$ which are ideal for generating two--qubit entanglement. We see that many trajectories do much better than the average, reaching maximal concurrence $\mathcal{C} = 1$. These best trajectories are bounded by \eqref{pure_Cmax} (shown in dash--dotted black), as discussed in the main text. By comparing again with the expression \eqref{barC-jhe} in dotted red, we see that the average concurrence from these diffusive trajectories is in good agreement with the average concurrence in the photodetection case (see Fig.~\ref{fig-photodect-conc}(a)). Note that the colors of individual trajectories in the left panel match those same trajectories as they appear in Appendix \ref{sec-extraplots} (shown here with lower opacity). In the right panel we show how entanglement genesis is hurt by changing the relative phases of the two homodyne measurements; the optimal choice (dotted black, or the top panel) eliminates competition between the two measurements and allows for deterministic entanglement genesis, while the least--optimal choice (orange) destroys any possibility of entanglement genesis entirely. A few trajectories for each case are plotted, matching the colors assigned to the averages.}
    \label{fig-homodyne-eeGen}
\end{figure}

We run simulations initialized from $\ket{ee}$, and show some plots in Fig.~\ref{fig-homodyne-eeGen} highlighting the most basic features of the entanglement dynamics. Comparing the homodyne case in Fig.~\ref{fig-homodyne-eeGen}(a) to the photodetection case of Fig.~\ref{fig-photodect-conc}(a), we immediately see that there are, of course, stark differences in character between individual trajectories under photodetection, as compared with quadrature measurements. 
The diffusive trajectories we obtain from homodyning do not even allow us to say that the photon was emitted at any particular time, as in the one--qubit case \cite{FlorTeach2019, PL-Dissertation}; the system diffuses from $\ket{ee}$ to $\ket{gg}$ without any single well--defined emission event. 
Despite these differences, however, the average concurrence over the duration of the simulation is identical to the expression \eqref{barC-jhe} we derived in the photodetection case.
It has already been shown by Viviescas et al.~\cite{Viviescas2010} (using different arguments) that a measurement satisfying $|\theta-\vartheta| = 90^\circ$ is the optimal one among those utilizing decay channels to generate two--qubit diffusive trajectories for the purposes of \emph{preserving} two--qubit entanglement. 
One of their key results is the derivation of a differential equation describing the evolution of the average concurrence: They use the stochastic Schr\"{o}dinger equation (SSE)\footnote{The SSE is a stochastic differential equation modeling ideal measurements and pure state evolution, which generalizes to the SME \eqref{SME} when $\eta \neq 1$ or mixed states are otherwise necessary.} to derive
\be \label{viviescas_cbar}
\dot{\bar{\mathcal{C}}} = -\gamma \bar{\mathcal{C}} + 2 \gamma \rho_{ee} e^{-2\gamma t},
\ee
where $\bar{\mathcal{C}}$ again denotes the average concurrence, dots denote time derivatives, and $\rho_{ee}$ is the initial excited state population. 
We again consider the case $\rho_{ee} = 1$ (which implies the initial value $\bar{\mathcal{C}}(t=0) = 0$), and find that the solution to \eqref{viviescas_cbar} is precisely the function \eqref{barC-jhe} we derived from the photodetection case. 
This suggests that the average entanglement yield between these two measurements are not just similar, but formally equivalent when $\theta$ and $\vartheta$ are chosen optimally; such an equivalence has been noted before \cite{Mascarenhas2011}. 
While it may be surprising that the two protocols we have discussed lead to identical concurrence yield on average, given the differences between jump and diffusive trajectories, it follows naturally from the entanglement swapping ideas \cite{Takeda2015,Santos2012} we have discussed. 
From that viewpoint, the fluorescence process generates a certain amount of entanglement in the system (between each qubit and its output mode) as a function of time, irrespective of the subsequent measurements; both photodetection and the EPR/homodyne strategy are able to perform an optimal swap in this case, rearranging that entanglement. 
They do this in very different ways, but the two measurements are ultimately manipulating the same resources in the system, leading to the same average concurrence yield.

\par Less--than--ideal measurements {(which fail to completely erase the which--path information)} could be understood as wasting some of that potential entanglement;
for example, in Fig.~\ref{fig-homodyne-eeGen}(b) we see that changing the relationship between $\theta$ and $\vartheta$ retains the shape of the curve from Fig.~\ref{fig-homodyne-eeGen}(a), but modulates it down by an overall factor $\sim |\sin(\theta-\vartheta)|$ as the degree of which--path distinguishability is changed (flatlining to zero two--qubit concurrence for all time, in the case of total distinguishability $\theta = 0 = \vartheta$). 
We finally note that under the ideal homodyne measurements with $|\theta-\vartheta| = 90^\circ$, some trajectories do reach the maximum $\mathcal{C} = 1$ within a few $T_1$ of the start of the simulation. 

{We continue by extending the quantum trajectory analysis of the homodyne scheme beyond previous discussions \cite{Viviescas2010}, first by characterizing the dynamics of trajectories leading to strong entanglement, and then by deriving a bound on the fastest possible rise to a Bell state due to joint homodyne fluorescence detection.}

\subsection{Maximally--Entangled States}

\begin{figure*}[t]
\includegraphics[width = \textwidth]{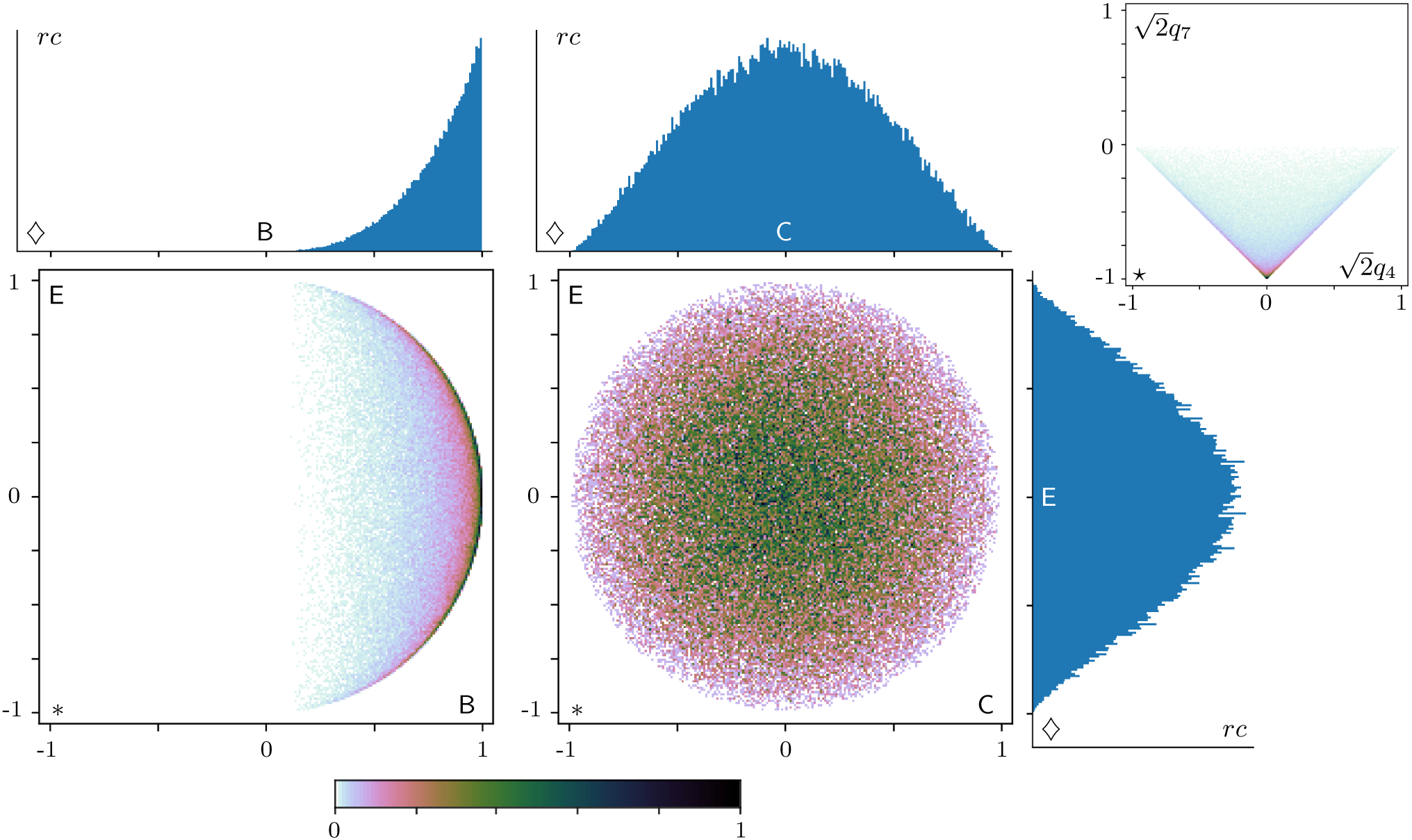}
\caption{We plot one-- and two--dimensional histograms describing the statistics with which different combinations of the Bell--basis amplitudes $\mathsf{B}$, $\mathsf{C}$, and $\mathsf{E}$ defined in \eqref{maxCstate} appear. Simulations are of the double homodyne measurement with $\theta = 0$ and $\vartheta  = 90^\circ$. We use an ensemble of 100,000 states, each obtained from the first timestep in which a quantum trajectory initialized at $\ket{ee}$ reaches $\mathcal{C} \geq 0.999$. 
The colorbar denotes count density per bin in the two--dimensional histograms ($\ast$~and $\star$), while relative counts ($rc$) per bin are plotted in the one--dimensional (marginal) histograms ($\diamondsuit$). 
In the figures marked $\ast$ and $\diamondsuit$ we plot using our Bell basis amplitudes $\mathsf{B}$, $\mathsf{C}$, and $\mathsf{E}$; one--dimensional histograms $\diamondsuit$~are aligned with the two--dimensional histograms $\ast$~such that summing out a row or column of bins in the two--dimensional plots would give the accompanying one--dimensional plot. 
We see that the distributions in $\mathsf{C}$ and $\mathsf{E}$ are symmetric, and centered about 0 with their peak there. Normalization then demands that the single most--likely state about which the distribution is peaked occurs at $\mathsf{B} = 1$, i.e.~the state which is most--likely to occur when the concurrence is maximized, under the given measurement settings, is $\ket{\Phi^-} = (\ket{ee}- \ket{gg})/\sqrt{2}$. However, the maximally concurrent states which we obtain are generically superpositions of the Bell states \eqref{maxCstate}, and while states with $\mathsf{B} = 0$ are the least--likely, the system does explore the full space of states for $\mathsf{C},\mathsf{E} \in [-1,1]$, and $\mathsf{B} \in [0,1]$, which satisfy the normalization condition $\mathsf{B}^2 + \mathsf{C}^2 + \mathsf{E}^2 = 1$. 
The additional density plot $\star$~in which we histogram $q_4$ against $q_7$ is significant in that it shows that $q_7$ is never positive for the maximally concurrent states in the simulated sample; this indicates that $\mathsf{A} = 0$, as discussed in the main text, justifying its exclusion from the other plots. }
\label{fig-MaxCStats}
\end{figure*}

The states we reach at times of maximal concurrence $\mathcal{C} = 1$ are superpositions of three of the four Bell states. Any pure two--qubit state may be expressed in the Bell basis according to 
\be \label{bell-state-basis}
\ket{\psi} = \mathsf{A} \ket{\Phi^+} + \mathsf{B} \ket{\Phi^-} + \mathsf{C} \ket{\Psi^+} + \mathsf{D} \ket{\Psi^-}.
\ee
The concurrence of the state is given, in this representation, by $\mathcal{C}  = |\mathsf{A}^2+\mathsf{D}^2-\mathsf{B}^2-\mathsf{C}^2|$. 
For trajectories initialized at $\ket{ee}$, the {\color{MidnightBlue} blue} terms in \eqref{M-hom} guarantee that we generate correlations with a real $\mathsf{C}$ (generated by $r_3$) and imaginary $\mathsf{D}$ (generated by $i r_4$) in the odd--parity subspace. 
The amplitude $\ket{ee}$ at any given time is given by $\tfrac{1}{\sqrt{2}}(\mathsf{A}+\mathsf{B})$, which remains real and nonnegative along the duration of any trajectory initialized at $\ket{ee}$. 
Re--parameterizing $\mathsf{D} = i \mathsf{E}$, and assuming $\mathsf{A}$, $\mathsf{B}$, and $\mathsf{C}$ are real (consistent with all our simulations initialized at $\ket{ee}$ for $\theta = 0 $ and $\vartheta = 90^\circ$), we then have 
\be \label{conc-pure-bell}
\mathcal{C} = |\mathsf{A}^2 - \mathsf{B}^2 - \mathsf{C}^2 - \mathsf{E}^2|.
\ee
Under these conditions, we find that $\mathcal{C}$ is maximized when $\mathsf{A} = 0$, i.e.~we maximize $\mathcal{C}$ when the measurement pushes the two--qubit state into a form
\be \label{maxCstate}
\ket{\psi} = \mathsf{B} \ket{\Phi^-} + \mathsf{C} \ket{\Psi^+} + i \mathsf{E} \ket{\Psi^-},
\ee
for real $\mathsf{B}$, $\mathsf{C}$, and $\mathsf{E}$.
The particular values of these normalized amplitudes depend on the measurement record in a given realization. 
Upon obtaining such a maximally--entangled state, one can shift it to a single Bell state through a local unitary operation on a single qubit. For example, consider applying a pulse to qubit B to {implement} the rotation
\be 
\mathcal{U}_B\ket{\psi} = \hat{\mathbb{I}}_A \otimes \left(\begin{array}{cc}
\mathsf{B} & \mathsf{C} - i\mathsf{E} \\
-(\mathsf{C}+i\mathsf{E}) & \mathsf{B}
\end{array} \right)\ket{\psi} = \ket{\Phi^-},
\ee
where $\ket{\psi}$ is of the form in \eqref{maxCstate}. {Thus, correlations may be rearranged into a particular form or encoding, if desired, using local operations and knowledge of the state acquired through continuous monitoring.} 
The operation shown above, for example, allows an observer located at qubit B to make a local rotation which guarantees that the outcomes of subsequent local Pauli-$z$ measurements on qubits A and B will be correlated in individual runs of the experiment. 
Such an operation requires that $\ket{\psi}$ of the form above is created by continuous joint homodyne monitoring, and that the measurement record (or two--qubit state $\ket{\psi}$) be communicated classically to the observer at qubit B, so that they may perform a local operation to transform $\ket{\psi}\,\rightarrow\,\ket{\Phi^-}$.

\par We may consider the statistics of the maximally--concurrent states created by our homodyne measurement. 
Specifically, we look at an ensemble of states that successful trajectories initialized at $\ket{ee}$ generate, in the first timestep at which they attain $\mathcal{C} \geq 0.999$ (still with $\theta = 0$ and $\vartheta = 90^\circ$); we then confirm that $\mathsf{A} = 0$ for all such states, and look at the distribution of the non--zero Bell basis amplitudes $\mathsf{B}$, $\mathsf{C}$, and $\mathsf{E}$ in the ensemble. 
The analysis in question is shown in Fig.~\ref{fig-MaxCStats}, and 
highlights a substantial difference between the homodyne and photodetection cases. 
Recall that for the photodetection case, the concurrence appears in the form $\ket{\Psi^+}$ or $\ket{\Psi^-}$ only. 
The homodyne measurement not only adds the possibility of having $\ket{\Phi^-}$ appear, but in fact, $\ket{\Phi^-}$ is the single most--likely state for a trajectory to reach as it attains $\mathcal{C}=1$; in Fig.~\ref{fig-MaxCStats}, we clearly see that the histogrammed distributions (probability density) of state amplitudes are peaked about $\mathsf{B} = 1$, and $\mathsf{C} = 0 = \mathsf{E}$. 
However, a generic realization is not restricted to any one Bell state, or even a small subset of them; the distribution covers the entire space of normalized states for $\mathsf{C},\mathsf{E} \in [-1,1]$, and $\mathsf{B} \geq 0$. 
The probability density within those possibilities has positive amplitude $\mathsf{B}$, and is symmetric in $\mathsf{C}$ and $\mathsf{E}$, but no other constraints appear on the range of possible random $\mathcal{C} = 1$ states which arise from this measurement.
For further details about the dynamics leading to this distribution of states, see Appendix~\ref{sec-extraplots}. 

\subsection{Upper Bound on Concurrence Generation}

{We next demonstrate that there is a finite time before which a given degree of entanglement may be generated by joint homodyne fluorescence detection, given the initial state $\ket{ee}$.}
{We may derive a bound on the fastest rise in entanglement by} considering the case of perfectly correlated outcomes for joint quadrature measurements on the emitted photons (recall \eqref{antiC}), creating EPR--like correlations~\cite{pirandola2006quantum}. In other words, we consider the readouts $r_3 = 0 = r_4$, equivalent to the perfectly--correlated and maximally--symmetric outcomes $X_1+X_2=0$ and $P_1-P_2=0$ (still for $\theta = 0$ and $\vartheta = 90^\circ$). 
In this case the state dynamics will be 
restricted purely to the subspace $\mathsf{a}\ket{ee} - \sqrt{1-\mathsf{a}^2}\ket{gg}$ for $\mathsf{a} \in [0,1]$, and we are effectively deriving the ideal trajectory that travels to $\ket{\Phi^-}$ as quickly as possible.
By choosing these particular values of the measurement record (which are smooth), the dynamics of the system can be reduced to an ordinary differential equation, rather than a stochastic one. 
We find that while concurrence is increasing (i.e.~while $\mathsf{a} > \sqrt{1-\mathsf{a}^2}$), the relevant differential equation is 
\be \label{dotCbound}
\dot{\mathcal{C}} = \gamma (\mathcal{C}+1) \left[1-\mathcal{C}+\sqrt{1-\mathcal{C}^2} \right],
\ee
which, for $\mathcal{C}_0 = 0$, admits the solution
\be \label{pure_Cmax}
\mathcal{C}(t) = \frac{2e^{\gamma t}-2}{2- e^{\gamma t}(2- e^{\gamma t})}.
\ee
This curve sets the bound on the speed at which concurrence is created from the initial state $\ket{ee}$ via our double--homodyne measurement, until it hits its maximum at $t = T_1 \text{ln}(2)$. 
After this point, the concurrence may fluctuate stochastically for an arbitrarily long time, taking on any value $\mathcal{C} \in [0,1]$, as illustrated in Fig.~\ref{fig-homodyne-eeGen}(a).
While we note that this particular best--case measurement record $r_3 = 0 = r_4$, which bounds the fastest possible path to maximum concurrence, requires blind luck to acquire (we have no control over what measurement outcomes we get, in practice), similar records, or records exhibiting qualitatively the same concurrence dynamics, do occur naturally in this system with relatively high probability; indeed, many trajectories which stay close to \eqref{pure_Cmax} and achieve a Bell state within $t < T_1$ are visible in Fig.~\ref{fig-homodyne-eeGen}(a). 
{Recent works have shown that feedback may be used to enforce a condition $r_3 = 0 = r_4$ throughout the two--qubit evolution \emph{on average}, which optimizes entanglement creation \cite{Leigh2019, Lewalle2020feed, PL-Dissertation}.}

\begin{figure}[t]
    \centering
    \includegraphics[width=\textwidth]{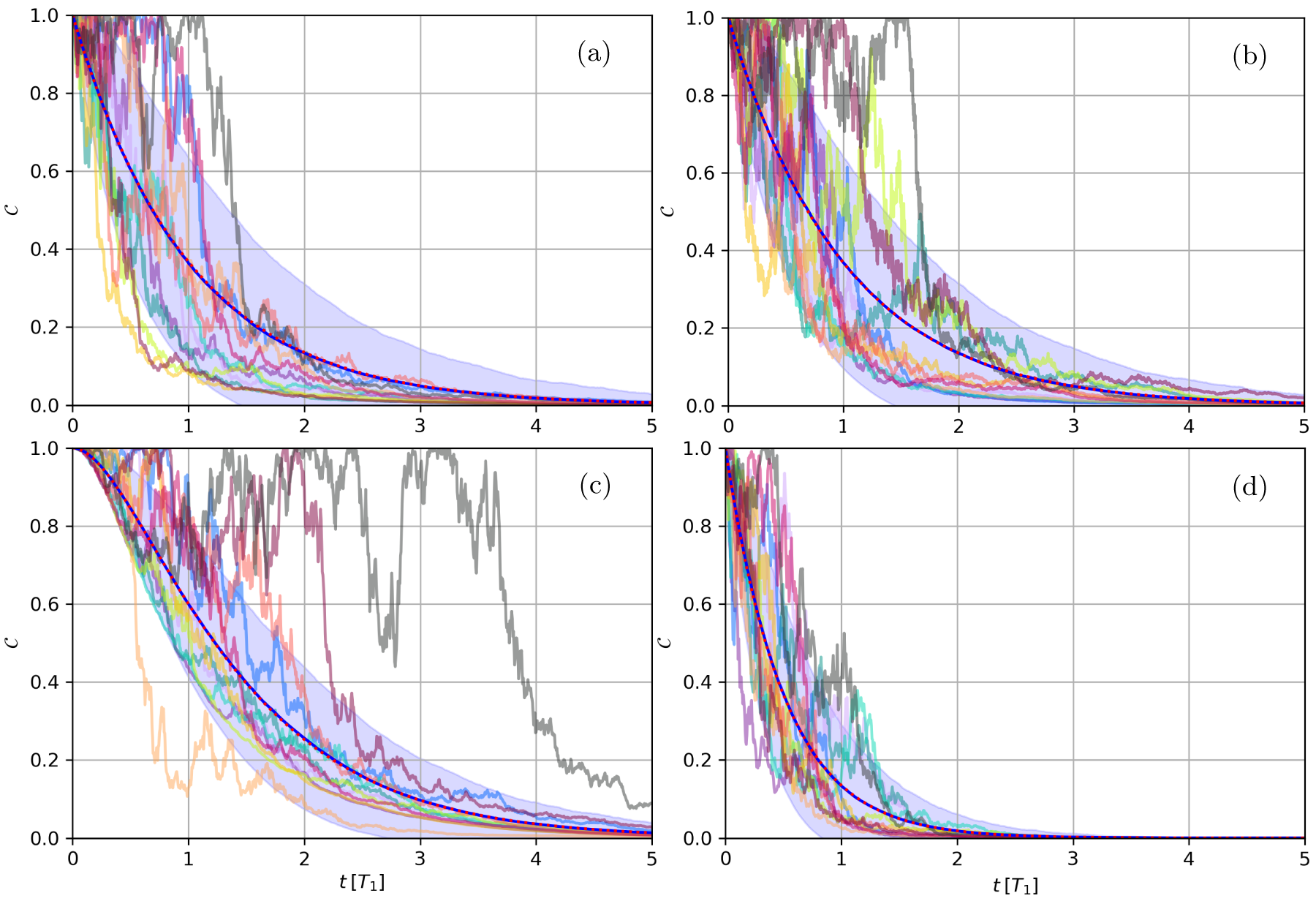}
    \caption{We plot the concurrence decay from different Bell states, under our double homodyne measurement dynamics, using the optimal settings $\theta = 0$ and $\vartheta = 90^\circ$. In (a) the initial state is $\ket{\Psi^+}$; its pair $\ket{\Psi^-}$ exhibits qualitatively the same concurrence dynamics, and is shown in (b). Note that the concurrence decay from these states is the same as in the photodetection case on average (compare with Fig.~\ref{fig-photodect-conc}(c), and the dotted red line $\bar{\mathcal{C}} = e^{-\gamma t}$). In (c) the initial state is $\ket{\Phi^-}$; as in the photodetection case (Fig.~\ref{fig-photodect-conc}(b)), we see that the initial slope of the average concurrence decay is zero; this extends the concurrence lifetime somewhat on average, with the decay from simulation (solid blue) matching $\bar{\mathcal{C}} = e^{-\gamma t}(2-e^{-\gamma t})$ (dotted red). The final simulation, plotted in (d), is initialized at the one Bell state $\ket{\Phi^+}$ that doesn't play a helpful role creating concurrence from our double homodyne measurement for $\theta = 0$ and $\vartheta = 90^\circ$. Its built--in correlations are anathema to the type created by the measurement, and lead to exponential decay as $\bar{\mathcal{C}} = e^{-2\gamma t}$ (dotted red). 10,000 trajectories were simulated to compute the averages, and the envelope of $\pm$ one standard deviation around it. For further comments, see Sec.~\ref{sec-homBelldecay}. }
    \label{fig-Homx2-BellDecay}
\end{figure}

\subsection{Entanglement Preservation \label{sec-homBelldecay}}

\par A final example, shown in Fig.~\ref{fig-Homx2-BellDecay}, serves to illustrate the behavior of different types of correlations in response to our measurement; we choose each of the Bell states as initial two--qubit states, and look at the evolution of the concurrence under homodyne $\theta = 0$ and $\vartheta = 90^\circ$ measurement. We find that they exhibit different average lifetimes, such that the Bell state $\ket{\Phi^+}$ whose correlation type runs against entangling dynamics of the measurement $\theta = 0$ and $\vartheta = 90^\circ$ decays on average at least twice as fast as any of the others.
This is in contrast with the photodetection case, where this faster decay does not appear.
Dynamics originating from the other three Bell states, under our homodyne scheme, have clear counterparts in the photodetection case, however, which are apparent from comparing Figs.~\ref{fig-photodect-conc} and \ref{fig-Homx2-BellDecay}. 
This again reinforces the notion that our {joint homodyne} measurement exhibits some preferred type of correlations {(as well as reinforcing the comments we have made about entanglement swapping connecting different types of measurements)}.
Further discussion can be found in appendix \ref{sec-onesteptest}. 
The solutions obtained analytically for the photodetection case in Fig.~\ref{fig-photodect-conc}, and in the homodyne case Figs.~\ref{fig-homodyne-eeGen} and \ref{fig-Homx2-BellDecay}(a,b,c), are all solutions to the equation \eqref{viviescas_cbar} derived by \cite{Viviescas2010}. 
Note however, that the solution in Fig.~\ref{fig-Homx2-BellDecay}(c) is not a solution to \eqref{viviescas_cbar}, because the choice $\theta = 0$ and $\vartheta = 90^\circ$ is the \emph{least optimal} choice of measurement settings given $\ket{\Phi^+}$, and \eqref{viviescas_cbar} assumes the optimal unraveling. 
The problem can be rotated in several ways; if we want to preserve e.g.~$\ket{\Phi^+}$ instead of $\ket{\Phi^-}$, we may change our measurement settings to $\theta = -90^\circ$ and $\vartheta = 0$, and thereby swap the behavior seen in Figs.~\ref{fig-Homx2-BellDecay}(b) and \ref{fig-Homx2-BellDecay}(c). 
While our discussion above has focused on $\theta = 0$ and $\vartheta = 90^\circ$ for clarity, all the results we discuss there are conceptually correct for any choice satisfying $\vartheta = \theta + 90^\circ$, up to a corresponding rotation of all the amplitudes. 
Given a state
\be 
\ket{\psi} = \mathsf{a}\ket{ee} + \mathsf{d}\, e^{i\delta}\ket{gg} + e^{i\theta}\left(\mathsf{X} \ket{\Psi^+} + i\mathsf{Y} \ket{\Psi^-}\right),
\ee
moderately more general than \eqref{maxCstate}, assuming $\vartheta = \theta + 90^\circ$, and with $\mathsf{a}$, $\mathsf{d}$, $\mathsf{X}$, $\mathsf{Y}$ all real, the optimal choice of measurement parameters is given by $\theta = (\delta - \pi)/2$.
Such considerations are formally derived by \cite{Viviescas2010} using the SSE.

\subsection{Summary of Pure State Results}

\par In summary, we find that the entangling double homodyne measurement \eqref{M-hom} creates a set of dynamics in which the overall decay from $\ket{ee} \rightarrow \ket{gg}$ is achieved by a process that exhibits correlation between different two--qubit basis states, at every step of the decay. 
For the measurement settings $\theta = 0$ and $\vartheta = 90^\circ$, with an initial amplitude $1$ on $\ket{ee}$, every Bell state \emph{except} $\ket{\Phi^+}$ contributes constructively to the two--qubit entanglement, as per \eqref{conc-pure-bell}. 
In this sense, the measurement exhibits some asymmetry, admitting only truncated manifestations of $\ket{\Phi^+}$, but allowing for completely coherent manifestations of the other Bell states. 
While the average entanglement yields \eqref{barC-jhe} are identical between photodetection and optimal homodyne detection, the measurements differ in virtually every other qualitative sense. 
Notably, jumps can occur at any time, such that some realizations of the photodetection process initialized at $\ket{ee}$ lead to immediate entanglement; our diffusive trajectories, however, are bounded by \eqref{pure_Cmax}, such that no measurement record allows these trajectories to reach a Bell state before $t = T_1 \text{ln}(2) \approx 0.69 \,T_1$. 
{Taken together, the discussion leading to this point, from concepts to numerics, constitutes a complete study of two--qubit entanglement generated via \emph{ideal} measurements of spontaneously--emitted photons.}


\section{Impact of Measurement Inefficiency \label{sec-inefficient}}

{While it is conceptually useful to study the behavior of an idealized measurement process, completely lossless measurement devices do not exist in any laboratory of which we are yet aware.}
{Consequently, it is of great practical importance that we consider how \emph{inefficient} measurements affect the ideas presented above.}
{We now extend our model to this case, again deriving some useful upper bounds on concurrence generation in the process.}
{It will become apparent as we go that the case of inefficient measurement has a great deal to offer pedagogically, as well as practically.}

\begin{figure}
    \centering 
    \includegraphics[width=.75\columnwidth]{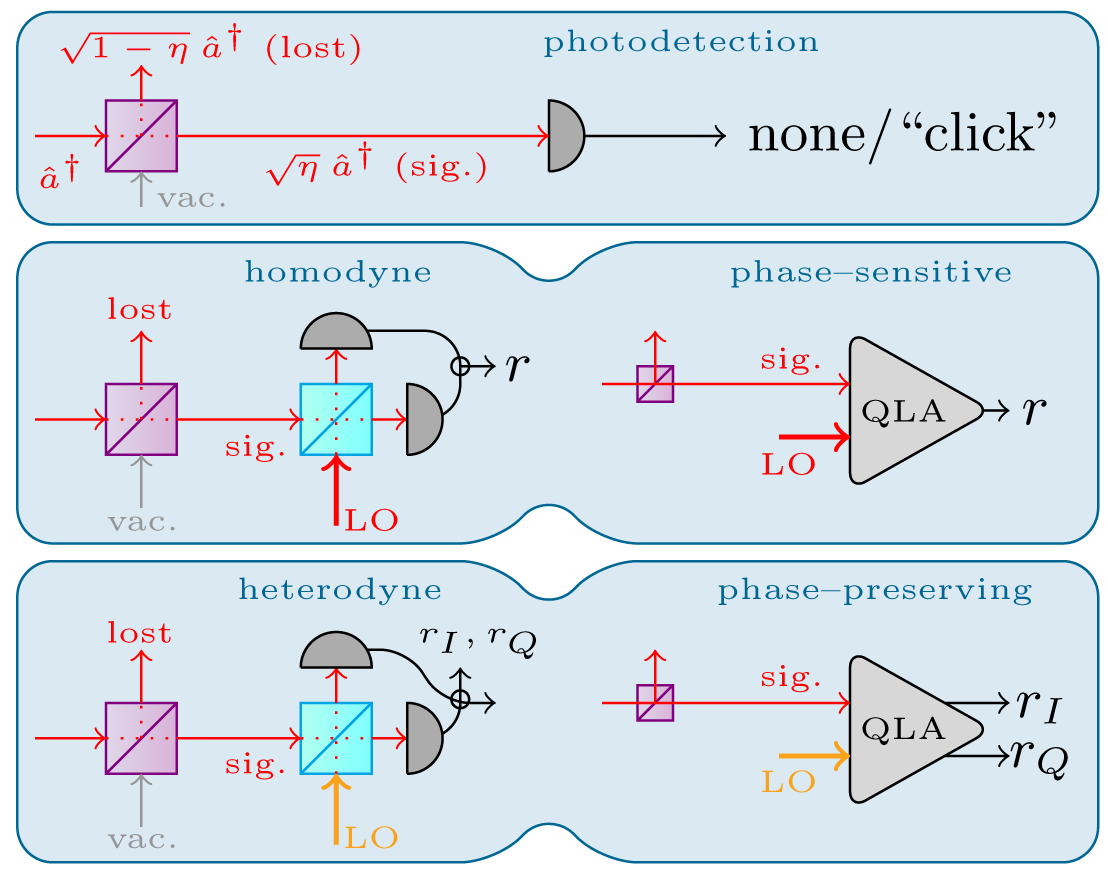}
    \caption{We sketch a modification of Fig.~\ref{fig-splitterphase}, which includes unbalanced beamsplitters (purple) in the monitored channels, used to model measurement inefficiency. These unbalanced beamsplitters allow signal to reach an (otherwise ideal) detector with probability $\eta$, but the signal may be lost with probability $1-\eta$. See \eqref{adageta}. The value of $\eta \in [0,1]$ then denotes the measurement efficiency, where $\eta = 1$ recovers the ideal case of lossless measurement.}
    \label{fig-inefficient}
\end{figure}

A simple model for measurement inefficiency is shown in Fig.~\ref{fig-inefficient}, which elaborates on Fig.~\ref{fig-splitterphase}; after modes 3 \& 4 are mixed on the main beamsplitter, {we then imagine that they are} split into a ``signal'' portion and ``lost'' portion (i.e.~3 \& 4 are mixed with vacuum modes on unbalanced beamsplitters). 
Algebraically, we express this according to 
\be \label{adageta} \begin{split}
\hat{a}^\dag_3 &\rightarrow \sqrt{\eta_3}\, \hat{a}_{3s}^\dag + \sqrt{1-\eta_3}\, \hat{a}_{3\ell}^\dag, \quad\text{and}\quad
\hat{a}^\dag_4 \rightarrow \sqrt{\eta_4}\, \hat{a}_{4s}^\dag + \sqrt{1-\eta_4}\, \hat{a}_{4\ell}^\dag,
\end{split} \ee
where $\eta$ are the efficiency of each measurement (a signal photon may be transmitted to the detector with probability $\eta$, or lost with probability $1-\eta$).
Finally, the surviving signal passes through phase plates which effectively set the field quadrature that each homodyne device monitors.
The overall transformation of the optical modes, from emission up to the detector, can then be summarized by the unitary transformations
\begin{subequations} \label{aeta} \be 
\hat{a}_1^\dag \rightarrow
 e^{i\theta} \sqrt{\frac{\eta_3}{2}}\hat{a}_{3s}^\dag + \sqrt{\frac{1-\eta_3}{2}}\hat{a}_{3\ell}^\dag+e^{i\vartheta}\sqrt{\frac{\eta_4}{2}}\hat{a}_{4s}^\dag + \sqrt{\frac{1-\eta_4}{2}}\hat{a}_{4\ell}^\dag,
\ee \be 
\hat{a}_2^\dag \rightarrow
 e^{i\theta} \sqrt{\frac{\eta_3}{2}}\hat{a}_{3s}^\dag + \sqrt{\frac{1-\eta_3}{2}}\hat{a}_{3\ell}^\dag-e^{i\vartheta}\sqrt{\frac{\eta_4}{2}}\hat{a}_{4s}^\dag - \sqrt{\frac{1-\eta_4}{2}}\hat{a}_{4\ell}^\dag.
\ee \end{subequations}
The ideal case is recovered by the choice $\eta_3 = 1 = \eta_4$ (complete information goes to the signal mode), while inefficient measurements have $\eta_3 < 1$ and/or $\eta_4 < 1$. 
A matrix $\mathcal{M}_\eta$ can be obtained by applying the transformations \eqref{aeta} within the matrix $\mathcal{M}$, defined in Eq.~\eqref{2Q_stateUp1}.
A series of Kraus operators can then be derived from $\mathcal{M}_\eta$, for different types of measurements, much as they were in Sec.~\ref{sec-model}; these will necessarily now depend on outcomes not only in the signal channels $s$, but also in the lost channels $\ell$. The essential operational meaning of measurement inefficiency, and of the channels $\ell$ being ``lost'', is that all outcomes that \emph{could have occurred} in the lost channels must be traced out, leaving a modified state update equation that will tend to generate partially--mixed states. 
{In this sense, the inefficient measurements we are here considering form a case intermediate between the ideal conditional evolution \eqref{general-stateup}, and the completely dissipative evolution \eqref{LindME} arising when no measurement is made at all.}
We will leave the algebraic details of this to Appendix~\ref{app-inefficient}, and proceed here to discuss the qualitative impact of measurement inefficiency in greater detail, based on numerical simulations.
The comparable analysis for the one qubit case can be found in Refs.~\cite{FlorTeach2019, PL-Dissertation}.

\subsection{Inefficient Photodetection}

\begin{figure}[t]
    \centering
    \includegraphics[width=\textwidth]{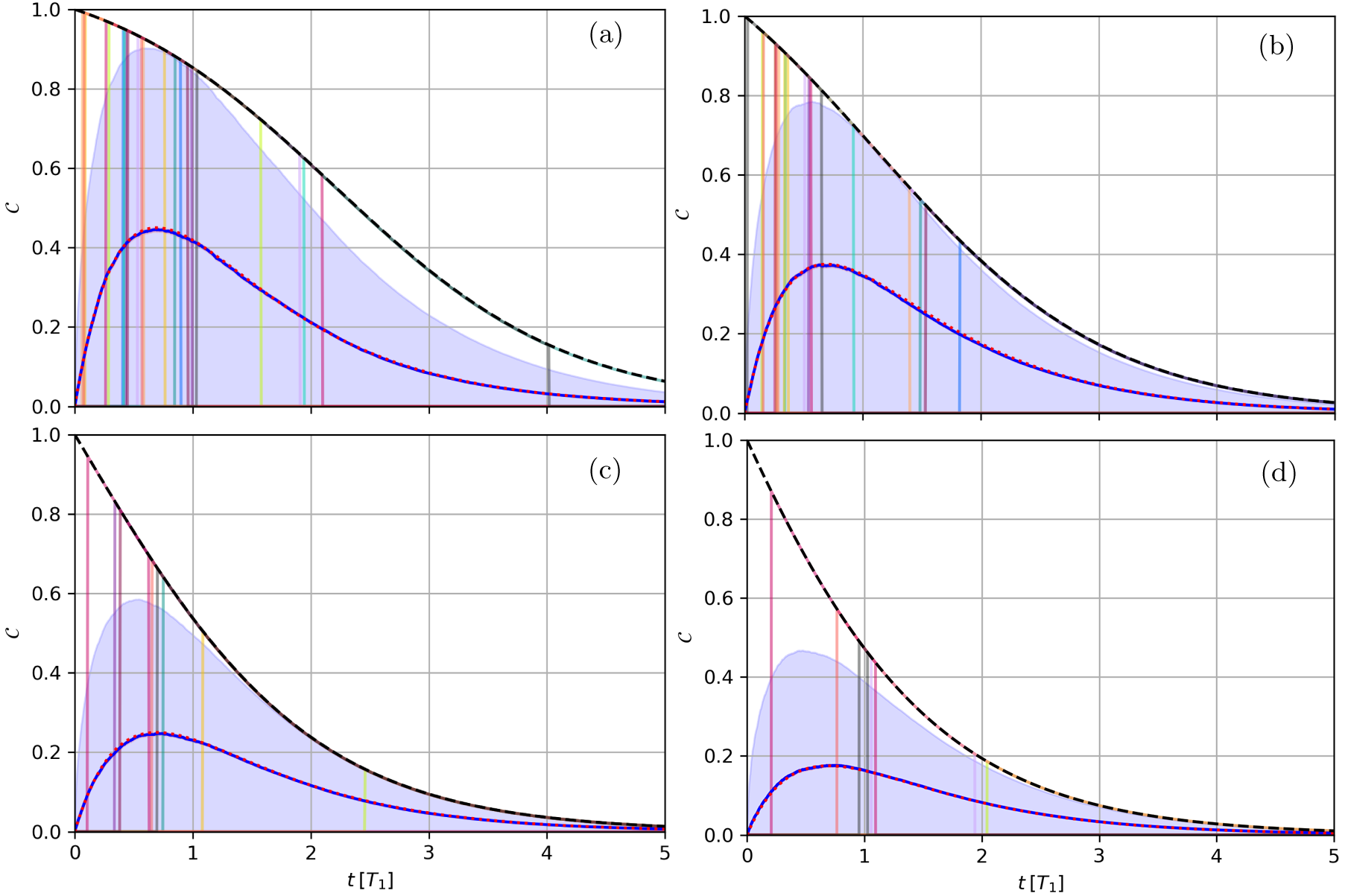}
    \caption{We plot the average concurrence (solid blue) and that of a dozen jump trajectories (multiple colors), obtained from performing inefficient photodetection on qubits initialized in $\ket{ee}$. We use $\eta_3 = 0.90 = \eta_4$ (a), $\eta_3 = 0.75 = \eta_4$ (b), $\eta_3 = 0.50 = \eta_4$ (c), and $\eta_3 = 0.35 = \eta_4$ (d). We notate $\eta_3  = \eta = \eta_4$ below. We observe that the dynamics of the average concurrence are similar to those in Fig.~\ref{fig-photodect-conc}(a) or \ref{fig-homodyne-eeGen}(a); although we have not formally derived an analytic expression for the average concurrence in the case $\eta<1$, we observe that that attenuating the ideal solution \eqref{barC-jhe} according to $\bar{\mathcal{C}}(t) = 2 \eta e^{-\gamma t}(1-e^{-\gamma t})$ (shown in dotted red), leads to good agreement with the average from simulation (solid blue, and computed from an ensemble of $10,000$ trajectories).
    We also see that the maximum concurrence attainable from a jump event decreases as a function of time, because the no--click dynamics of the system cause any entangled two--qubit states to lose purity when the measurement efficiency is imperfect. The analytic solution \eqref{CmaxPDeta} for this maximal concurrence is superposed atop the simulation curves in dashed black, and defines a tight upper bound on the maximum concurrence attainable at any time in this scenario. }
    \label{fig-PDeta}
\end{figure}

We simulate jump trajectories, from the initial state $\ket{ee}$, for efficiencies $\eta_3 = 0.9 = \eta_4$, $\eta_3 = 0.75 = \eta_4$, and $\eta_3 = 0.5 = \eta_4$, and plot the subsequent concurrence in Fig.~\ref{fig-PDeta}.
Generically, the accumulated loss of information over time, due to continuous inefficient measurement, causes the state of the system to lose purity.
We see this manifested in two distinct features of the plots in Fig.~\ref{fig-PDeta}.
First, the shape of the curve \eqref{barC-jhe} representing the average concurrence from the ideal case has effectively been preserved, and is simply shrunk by an overall factor $\eta$ (where $\eta_3 = \eta = \eta_4$). 
Second, the concurrence that is generated in \emph{individual} realizations no longer tops out at $\mathcal{C}=1$, but instead hits a ceiling $\mathcal{C}_{max}^\eta(t)$. 
We may derive the latter bound, {and thereby straightforwardly quantify the cumulative effect of successive inefficient measurements on the two--qubit entanglement generation.} 

The decay of the purity and maximum concurrence is determined by the no--click dynamics, initialized from $\ket{\Psi^\pm}$ (the state we infer is created from an immediate jump, before the losses due to accumulated inefficient measurements).
A density matrix of the form
\be \begin{split} \label{rhox}
\rho &= \left( \begin{array}{cccc}
0 & 0 & 0 & 0 \\
0 & x & \pm x & 0 \\
0 & \pm x & x & 0 \\
0 & 0 & 0 & 1- 2x
\end{array}\right) = 2 x \ket{\Psi^\pm}\bra{\Psi^\pm} + (1-2x)\ket{gg}\bra{gg},
\end{split} \ee
is then adequate derive the solutions $\mathcal{C}_{max}^\eta(t)$.
The concurrence of \eqref{rhox} is simply given by $\mathcal{C} = 2x$.
In the time--continuum limit, the dynamics inbetween the first click and a possible second, with $\eta_3  = \eta = \eta_4$, may be summarized by
\be \label{xdot_pd-eta}
\dot{\mathcal{C}} = \eta \: \gamma \: \mathcal{C}^2 - \gamma \: \mathcal{C}.
\ee
For the initial condition $x_0=\tfrac{1}{2}$ or $\mathcal{C}_0 = 1$, the solution is given by
\be \label{CmaxPDeta}
\mathcal{C}_{max}^\eta(t) = \frac{1}{(1-\eta) e^{\gamma t} + \eta}.
\ee
This expression is found to be in good agreement with the maximum concurrence curves visible in the numerics from Fig.~\ref{fig-PDeta}.
The impact of measurement inefficiency is easily quantified from such an upper bound, and is substantial after significant time evolution; for example, a jump at $t = 5T_1$, which would still generate a perfect Bell state for $\eta = 1$, now generates $\mathcal{C} < 0.1$ even for a reasonably good measurement characterized by $\eta = 0.9$.
Qualitatively, this indicates that after this relatively long wait for the first click, the likelihood that the click is genuinely from the first emitted photon is low; the likelier option is that the measured photon was the second, while the first was lost and never recorded. 
Without the possibility of certainty about which photon was caught, the observer's state is necessarily a statistical mixture of the two possible options, that is, after longer waiting times times $t \gg T_1$, heavily weighted towards the separable part $\ket{gg}\bra{gg}$, rather than the entangled part $\ket{\Psi^\pm}\bra{\Psi^\pm}$. 
{This is a direct illustration of how imperfect information collection degrades an observer's knowledge of the two--qubit correlations in an individual experimental run.}

\subsection{Inefficient Homodyne Detection \label{sec-hom-eta}}

\begin{figure}
    \centering
    \includegraphics[width = \textwidth]{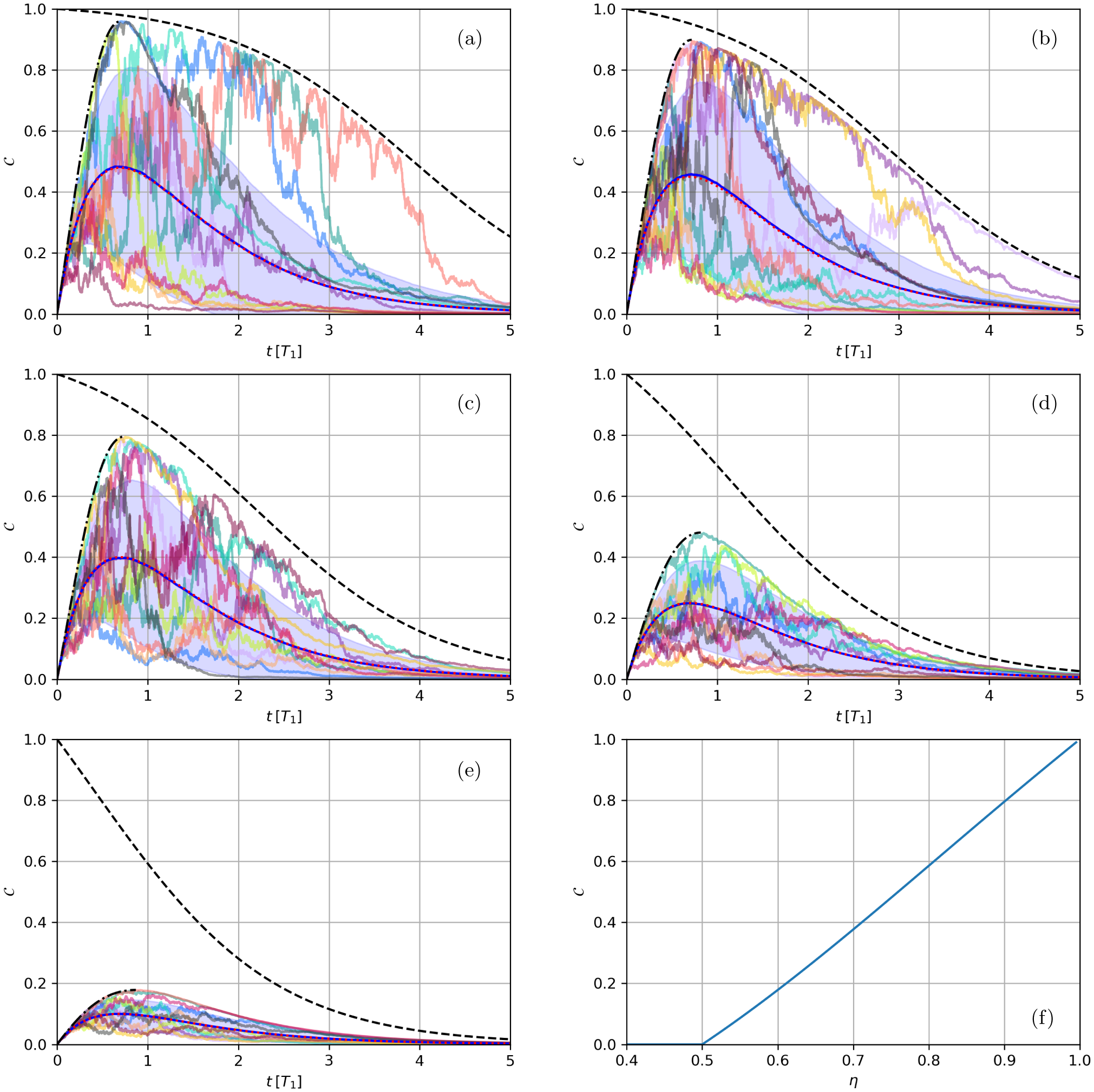}
    \caption{We plot the average concurrence (solid blue) and that of a dozen diffusive trajectories (various colors), obtained by simulating inefficient homodyne detection on qubits initialized in $\ket{ee}$. An ensemble of 10,000 trajectories are used to compute the average concurrence, and envelope of $\pm$ one standard deviation around it. We show simulations for $\eta_3 = 0.98 = \eta_4$ (a), $\eta_3 = 0.95 = \eta_4$ (b), $\eta_3 = 0.90 = \eta_4$ (c), $\eta_3 = 0.75=\eta_4$ (d), and $\eta_3= 0.60 = \eta_4$ (e). As in the ideal case, the trajectory defined by $r_3 = 0 = r_4$ (i.e.~the generalization of \eqref{pure_Cmax} to $\eta\leq 1$) sets a tight bound on the maximum attainable entanglement, and the fastest rise time to it (see the main text and Appendix~\ref{sec-hometa-Cmaxderiv} for details). This bound is plotted in dash--dotted black in (a) through (e). In (f), we plot the maximum concurrence, as determined from the peak of the $r_3 = 0 = r_4$ trajectory, as a function of $\eta$. We see that for homodyne detection, no concurrence at all is generated for $\eta \leq 50\%$, in contrast with the equivalent photodetection case. The maximum homodyne concurrence yield decreases approximately linearly from the ideal, to $\mathcal{C}_{max} = 0$ at $\eta = \tfrac{1}{2}$. We again plot the bound \eqref{CmaxPDeta} we derived from the photodetection case, in dashed black in (a) through (e); these are still correct as upper bounds, although they are no longer tight, further confirming that the homodyne measurement never yields more concurrence than photodection for $\eta < 1$. While we do not derive an analytic expression for the average concurrence, we find that attenuating the ideal solution \eqref{barC-jhe} according to $\bar{\mathcal{C}}(t) = 2(2\eta-1)e^{-\gamma t}(1-e^{-\gamma t})$ (shown in dotted red) leads to good agreement with the numerical average (shown in blue) in plots (a) through (e). }
    \label{fig-HOMeta}
\end{figure}

A similar analysis can be performed for the homodyne case.
We again leave a discussion of the generalized state update expressions to Appendix~\ref{app-inefficient}, and summarize two {main} points that emerge from that analysis: The mixed states generated by inefficient homodyne detection are of a more complex form than those in the photodetection case (e.g.~\eqref{rhox}), and the measurement signal in each channel is attenuated {relative to the noise} by a factor $\sqrt{\eta}$, but is otherwise unchanged. 
Numerical simulations easily account for these features, and the evolution of the concurrence they give is plotted in Fig.~\ref{fig-HOMeta}. Several important features of those dynamics are then readily apparent.
First, the upper bounds \eqref{CmaxPDeta} we just derived in the photodetection case still apply, but are no longer tight, meaning that the joint homodyne measurement here \emph{never} outperforms the corresponding photodetection scheme in terms of concurrence generation. 
Second, this measurement's ability to generate two--qubit concurrrence is eliminated entirely for $\eta \leq 50\%$, which is a substantial disadvantage compared with the photodetection case. 
Our homodyne measurement is, in this sense, substantially more sensitive to inefficiency than the corresponding photodetection protocol. 
{This result is consistent with related analyses performed in the context of feedback control \cite{Leigh2019,Lewalle2020feed}.}

We are able to derive an upper bound on the entanglement creation in the same way we did for the pure state case. 
Although we are not able to get a closed--form expression as we did with \eqref{pure_Cmax},
the condition $r_3 = 0 = r_4$ still has the same effect, and we are able to obtain curves from this concept using numerical methods in the case $\eta_3  = \eta = \eta_4$ (see Appendix \ref{sec-hometa-Cmaxderiv} for the computational details). 
For the initial density matrix $\ket{ee}\bra{ee}$, and $\theta = 0$ and $\vartheta = 90^\circ$, these bounds on the maximum entanglement and fastest rise to it are obtained by numerically integrating a system of three coupled differential equations. 
These numerical upper bounds apply until the concurrence hits its maximum, and are shown in Fig.~\ref{fig-HOMeta}.
{This constitutes a straightforward characterization of the impact of lossy measurements, allowing us to solve both the maximum possible concurrence arising from this measurement, as well as the time at which that maximum may be reached.}

\section{Discussion and Conclusions \label{sec-conclusions}}

We have presented a comprehensive and self--contained quantum trajectory analysis of entangling measurements based on continuous monitoring of two identical qubits via their mixed decay channel.
Entanglement is widely understood to be a quantum information resource. 
Spontaneous emission is, in many contexts, an error channel which impedes quantum information processing and/or destroys entanglement \cite{Aolita_2015}; by monitoring this $T_1$ channel, however, we see that this natural process can be used to our advantage instead.
The device geometry we have used to do this, shown in Fig.~\ref{fig-splitterphase}, resembles that typically associated with performing a Bell--state measurement. 
We have gone beyond the typical analysis of a Bell--state measurement by considering time--continuous measurements (i.e.~stochastic quantum trajectories). 
We have developed a consistent theoretical framework suitable for efficient numerical implementation in the device geometry of interest, allowing us to explore different measurement dynamics in detail.
Our Kraus operators are motivated by a physically intuitive picture, {highlighting the distinct roles of the qubit--field interaction (characterized by $\mathcal{M}$, e.g.~as in \eqref{idealMAT}) and eventual detection device (corresponding to $\ket{\psi_{f:3,4}}$) in making generalized joint measurements to infer coherent correlations between distant subsystems. Furthermore, our approach allows} us to simulate SQTs more efficiently than via direct Euler integration of the SME, in the same spirit as {in Refs.}~\cite{Rouchon2015, Cortez2017, Guevara2019}.

\par We have reviewed the major known results pertaining to the device shown in Figs.~\ref{fig-splitterphase} {and expanded on them with our numerical modeling of the measurement--induced dynamics, including analysis of inefficient measurements, and the derivation of useful bounds on the the maximum amount of entanglement generated by different measurements}. 
Specifically, we have shown that homodyne monitoring of quadratures $90^\circ$ apart is the optimal quadrature measurement for entangling our two qubits in our context, consistent with other works which have considered such homodyne measurements \cite{pirandola2006quantum, Reid1989, Reid2009, Viviescas2010, Handchen2012, Flurin2012, Takeda2015, Silveri2016, Pirandola2015, Ou1992}. 
That measurement scenario is often understood to be performing an entanglement swapping operation \cite{Pirandola2015}; in our context, this manifests as taking correlations between individual qubits and their field modes created in the spontaneous emission process, and swapping such that we entangle the qubits with each other instead.
The degree to which this double--homodyne measurement can entangle the emitters is tunable, and depends on the relative phase between the quadratures measured at each output. 
We are able to explain this tunability, and the success or failure of any of the fluorescence measurements we have considered to generate two--qubit entanglement, in terms of the erasure of information about which qubit originated any given signal.
Such considerations arise in the device geometry we have considered even when {a different qubit--field coupling is used as the basis of measurement} \cite{Silveri2016}. 
We find that the average entanglement yield of the optimal double homodyne measurement is equivalent to that of the photodetection case (from the initial state $\ket{ee}$), consistent with past works which have found the same equivalence from the standpoint of entanglement preservation \cite{Mascarenhas2011}. 
Thus we are able to clearly explain our numerical results by invoking a number of conceptual elements present in the literature. 
 Despite the equivalence of optimal photodetection and homodyne schemes on average, all details of the dynamics are markedly different. 
Photodetection behaves optimally regardless of measurement phase settings (as long as the mixing beamsplitter is balanced, the which--path information remains indistinguished), and creates Bell states when entanglement is generated. 
In contrast, even for optimal phase settings, homodyne detection generates both partially--entangled states in some realizations, and a wide variety of maximally--entangled states in its best realizations. 
We have considered the case of inefficient measurement, and leveraged some of the conceptual connections we have made to the wider literature to derive bounds on the maximum entanglement generated by our measurements. 
A particularly important result, practically--speaking, is that homodyne detection requires a minimum of $50\%$ efficiency in order to generate any two--qubit entanglement {(consistent with recent and closely--related results leveraging the measurements of interest for feedback control \cite{Leigh2019,Lewalle2020feed})}.

\par We see many immediate opportunities to test and expand on our analysis. 
It is grounded in methods that are experimentally feasible on a variety of systems: Works that consider or incorporate measurement devices with the geometry we have emphasized include e.g.~\cite{Furusawa1998, Bose1999, Beige2000, Opatrny2000, DLCZ2001, Zhang2002, BarrettKok, DuanKimble2003, Browne2003, Simon2003, Takei2005, Takei2005-2, Lim2005, Moehring2007, Metz2008, Maunz2009, Reid2009, Sun2010, Lee2011, Hofmann2012, Santos2012, Flurin2012, Takeda2013, Hanson2013Heralded, HansonLoopholeFree, Nourmandipour2016, Krutyanskiy2019, wein2020analyzing}. 
{The DLCZ (Duan, Lukin, Cirac, Zoller) \cite{DLCZ2001} protocol, which uses similar principles to generate entanglement between atomic ensembles over distance, is especially widespread, and has been developed in connection with quantum repeaters \cite{Sangouard2011, Zwerger_2017}. 
Quantum repeaters are a technology currently under development, which aim to use entanglement to counteract photon losses which inhibit long--distance communication over quantum networks.
The EPR-like (i.e.~homodyne) measurements we discuss have also been applied} to quantum repeaters \cite{Brask2010, Borregaard2012, Borregarrd2019} and discussed in the context of steering \cite{Handchen2012}, and are of interest both for applications to quantum computing and quantum communication.

A variety of related devices and schemes, which use different measurement channels or device geometries to generate entanglement, have also been proposed and/or realized \cite{Cabrillo1998, Plenio1999, Ruskov2003-2, Trauzettel2006, Williams2008, Slodicka2013, Roch2014, Motzoi2015, Silveri2016, Chantasri2016, Greplova2016, Ohm2017, Magazzu2018, Araneda2018, Zhang2019Heralded, Hurst2019, Bultink2020, Qian2020}. 
Those alternatives that implement a quantum trajectory approach often rely on dispersive measurements \cite{Blais2004, Clerk2010Review, Korotkov2016, Murch2015teach, blais2020circuit}, instead of continuous monitoring of fluorescence. 

We have emphasized the rich recent literature on stochastic quantum trajectories of a decay channel~\cite{Viviescas2010, Santos2011, PCI-2013, bolund2014stochastic, Jordan2015flor, FlorTeach2019, Campagne-Ibarcq2016, PCI-2016-2, Naghiloo2016flor, Mahdi2016, Tan2017, Ficheux2018, FicheuxThesis, MahdiTeachThesis,  Mahdi2017Qtherm, PL-Dissertation, Cottet_Thesis}; the scheme we have described could be viewed as scaling up these existing single--qubit experiments to two qubits. 
The majority of recent quantum trajectory experiments have been performed using superconducting qubits and/or microwave quantum optics \cite{Gu_2017_review, blais2020circuit}.
While efficient photodetection is relatively well--developed e.g.~for optical photons, effective photocounting in the microwave regime is still emerging, and substantially more difficult \cite{Kono2018, Lescanne2019, Essig2020, dassonneville2020numberresolved}, and it is consequently of practical interest to understand the behavior of alternative entangling measurements. 
The quadrature measurements we discuss offer such an alternative, and require tools that are well--established on these circuit--QED platforms (although the measurement efficiencies required for strong entanglement may still be challenging, in the near term). 
The types of measurements we have emphasized are possible on other platforms as well \cite{Pirandola2015, Lenzini2018, Li:19}.
Scaling our theoretical methods to larger numbers of qubits, as has been proposed elsewhere \cite{Santos2012} using SME--based methods, also appears feasible. 

We have described a method by which entanglement between individual quantum systems may be created by the measurement process that tracks its formation. 
One does not need to stop there however; given real--time measurement outcomes, an observer has the option to intervene in the system dynamics with further conditional operations in order to promote some desired behavior. 
In other words, advances in continuous quantum measurement are a pre-requisite for feedback--based quantum control strategies \cite{Zhang2017}, which continue to be an important avenue in contemporary research \cite{Cottet_PNAS, Minev2018}.
Feedback has been incorporated into quantum trajectory schemes, including from the measurement of a single qubit's decay channel \cite{PCI-2016-2, Cottet_Thesis, Mahdi2017Qtherm}. It has also been used to generate and/or preserve entanglement \cite{Ahn2003-2, Mancini2007,  Carvalho2007-2, Carvalho2008, Hill2008, Serafini2010, Riste2013, Rheda2014, martin2015remote, Rafiee2016,  Martin_2017, Andersen2019, Zhang2018}. 
Such strategies are of interest in that they typically allow for entanglement generation and/or entanglement lifetimes exceeding those from the measurement dynamics alone. 

As of recently, feedback schemes based precisely on the kind of device and measurements we have emphasized above have been proposed \cite{Leigh2019, Lewalle2020feed, PL-Dissertation}; these aim to increase the degree of two--qubit correlations on average, and extend the lifetime of the entangled state, using Local Operations and Classical Communication (LOCC). 
This means that these schemes function based on continually sending the measurement record to observers located at each qubit, such that operations may be applied to each qubit conditioned on the real--time measurement outcomes. 
In experiments involving qubits that are far apart, there will necessarily be a fundamental delay time in applying feedback operations due to the time required to communicate classical measurement outcomes to the locations of the individual emitters. 
Even if the qubits and their respective cavities are placed close together (e.g.~in the same dilution refridgerator or on the same lab bench), feedback delay due to signal processing times can never be completely eliminated in experiments (see Ref.~\cite{Leigh2019} for further discussion of these timescales in an experimental realization). 
In an ideal setup however, in which these delay times are negligible compared to the qubit decay timescale and the measurement efficiency is perfect, the feedback protocols detailed in Refs.~\cite{Leigh2019, Lewalle2020feed, PL-Dissertation} allow for preservation of concurrence near 1 on average for arbitrary durations, using the measurements described above. 
Specifically, ideal feedback based on the joint homodyne detection we have described allows for deterministic generation of a Bell state, which is possible due to the incremental nature of diffusive trajectories (in contrast with jump trajectories).
The diffusive entangling scheme we have discussed here exhibits more complicated behavior than the photodetection case. However, such advances in feedback schemes reinforce the motivation for considering other options, and further demonstrate that the complexity of the homodyne scheme we have discussed can be both manageable and useful.

\section*{Funding}
 PL, CE, SKM, and ANJ acknowledge funding from NSF grant no.~DMR-1809343, and US Army Research Office grant no.~W911NF-18-10178. PL acknowledges support from the US Department of Education grant No.~GR506598 as a GAANN fellow. XFQ and JHE acknowledge support from NSF grants PHY-1505189 and PHY-1539859.

\section*{Acknowledgements}
The work above has benefited from conversations with Benjamin Huard, Tathagata Karmakar, Alexander Korotkov, Spencer Rogers, and Irfan Siddiqi. 
We are grateful to Marcelo F.~Santos for pointing out a number of important references, including \cite{Viviescas2010, Mascarenhas2011}, after we posted v1 of our pre--print. PL thanks the Quantum Information Machines school at \'{E}cole de Physique des Houches for their hospitality during part of this manuscript's preparation. Numerical codes used to generate many of our figures have been written using Python 2.7.

\section*{Disclosures}
The authors declare no conflicts of interest.

\appendix

\section{Additional Simulations of Pure--State Diffusive Dynamics \label{sec-extraplots}}

\par
\begin{figure*}
\centering
\includegraphics[width = \textwidth]{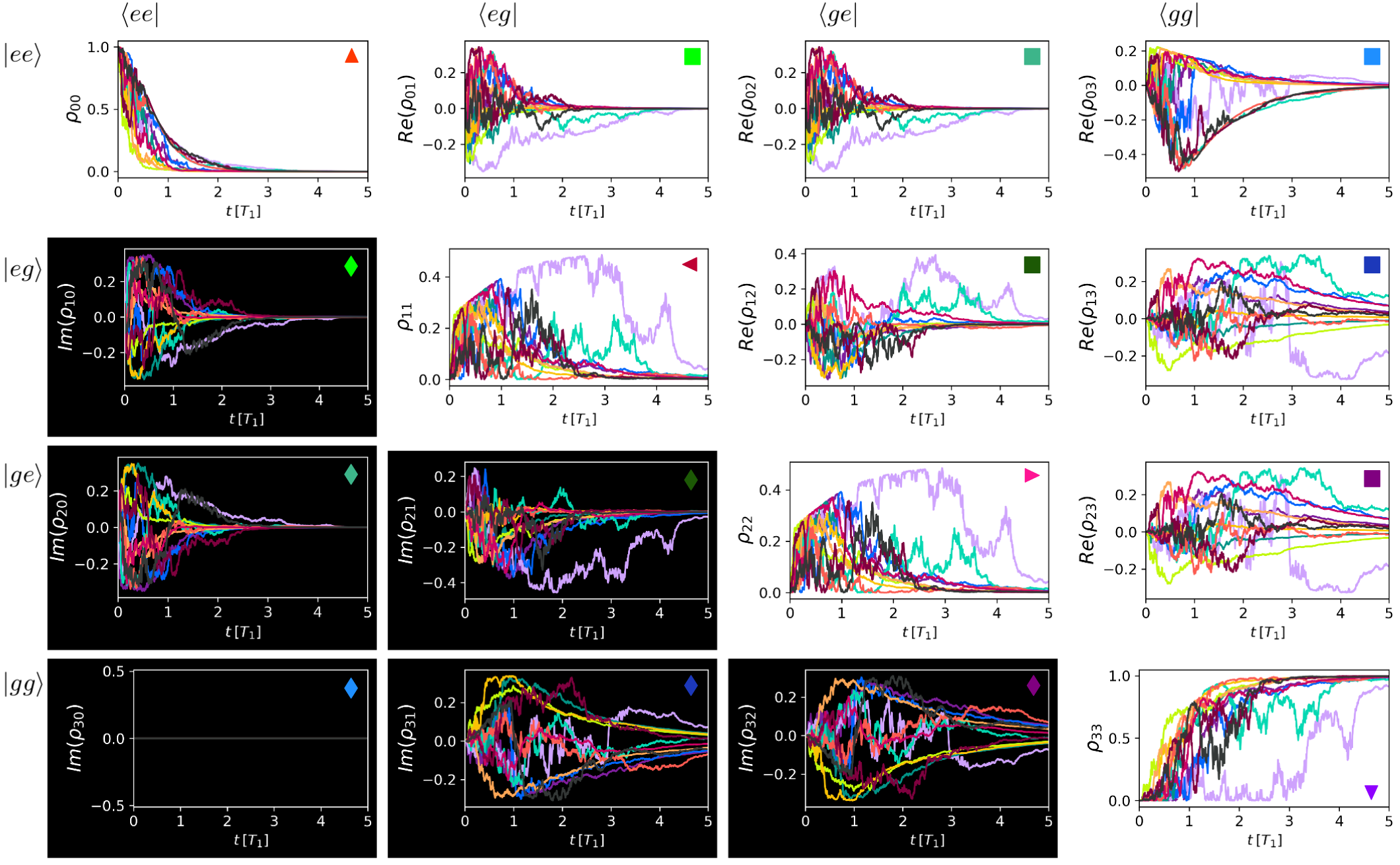}
\caption{We plot a dozen individual simulated SQTs initialized at $\ket{ee}$, monitored according to our double homodyne detection scheme with $\theta = 0$ and $\vartheta = 90^\circ$; as discussed in the main text, these parameters are ideal for erasing which--path information and generating two--qubit entanglement. The sampling of trajectories shown here are the same as those plotted in Fig.~\ref{fig-homodyne-eeGen}(a), with matched colors. They lead to the average concurrence in the ensemble peaking at $\mathcal{C} = \tfrac{1}{2}$, with the best realizations reaching $\mathcal{C} = 1$ at points in their evolution. The plots above are arranged similarly to the density matrix. The population is plotted down the diagonal, the real parts of the coherences are plotted in the upper triangular region, and the imaginary parts of the coherences are plotted in the lower triangular region (in inverse color). A key clarifying this layout and the colored plot markers is provided in \eqref{colorrho}. The correlations between different elements of $\rho$, in individual realizations, are visible. For instance, the populations in $\ket{eg}$ {$\color{crimsonglory} \blacktriangleleft$} \& $\ket{ge}$ {$\color{deeppink} \blacktriangleright$} are perfectly correlated in all realizations. Similarly, the real parts of the coherence / transition elements from $\ket{ee}$ towards $\ket{eg}$ and $\ket{ge}$ ({$\color{electricgreen} \blacksquare$} \& {$\color{mint} \blacksquare$}) are perfectly correlated, as are those transitioning from $\ket{eg}$ and $\ket{ge}$ toward $\ket{gg}$ ({$\color{persianblue} \blacksquare$} \& {$\color{patriarch} \blacksquare$}); the imaginary parts of these same elements are perfectly anti--correlated, i.e.~{$\color{electricgreen} \blacklozenge$} \& {$\color{mint} \blacklozenge$}, and {$\color{persianblue} \blacklozenge$} \& {$\color{patriarch} \blacklozenge$} form anti--correlated pairs. (Equivalently, $q_5$ \& $q_6$, and $q_8$ \& $q_9$, exhibit perfect correlations in all realizations, at all times. Likewise, $q_{11}$ \& $q_{12}$, and $q_{14}$ \& $q_{15}$, exhibit perfect anti--correlations in all realizations, at all times.)
This indicates that we have a correlated and coherent link between $\ket{ee}$ and $\ket{gg}$; every possible transition of amplitude from $\ket{ee}$ toward $\ket{gg}$ exhibits internal correlations. That the measurement at hand generates entanglement between our two emitters is a reflection of this. We note the assymmetry in the $\ket{ee}\bra{gg}$ {$\color{dodgerblue} \blacksquare$} element; the system clearly exhibits a preference for correlations of the type $\ket{\Phi^-} = \tfrac{1}{\sqrt{2}}(\ket{ee}-\ket{gg})$ over those of type $\ket{\Phi^+} = \tfrac{1}{\sqrt{2}}(\ket{ee}+\ket{gg})$.
\label{fig-SQT_table-HomEnt}}
\end{figure*}

\begin{figure*}[t]
\centering
\includegraphics[width = \textwidth]{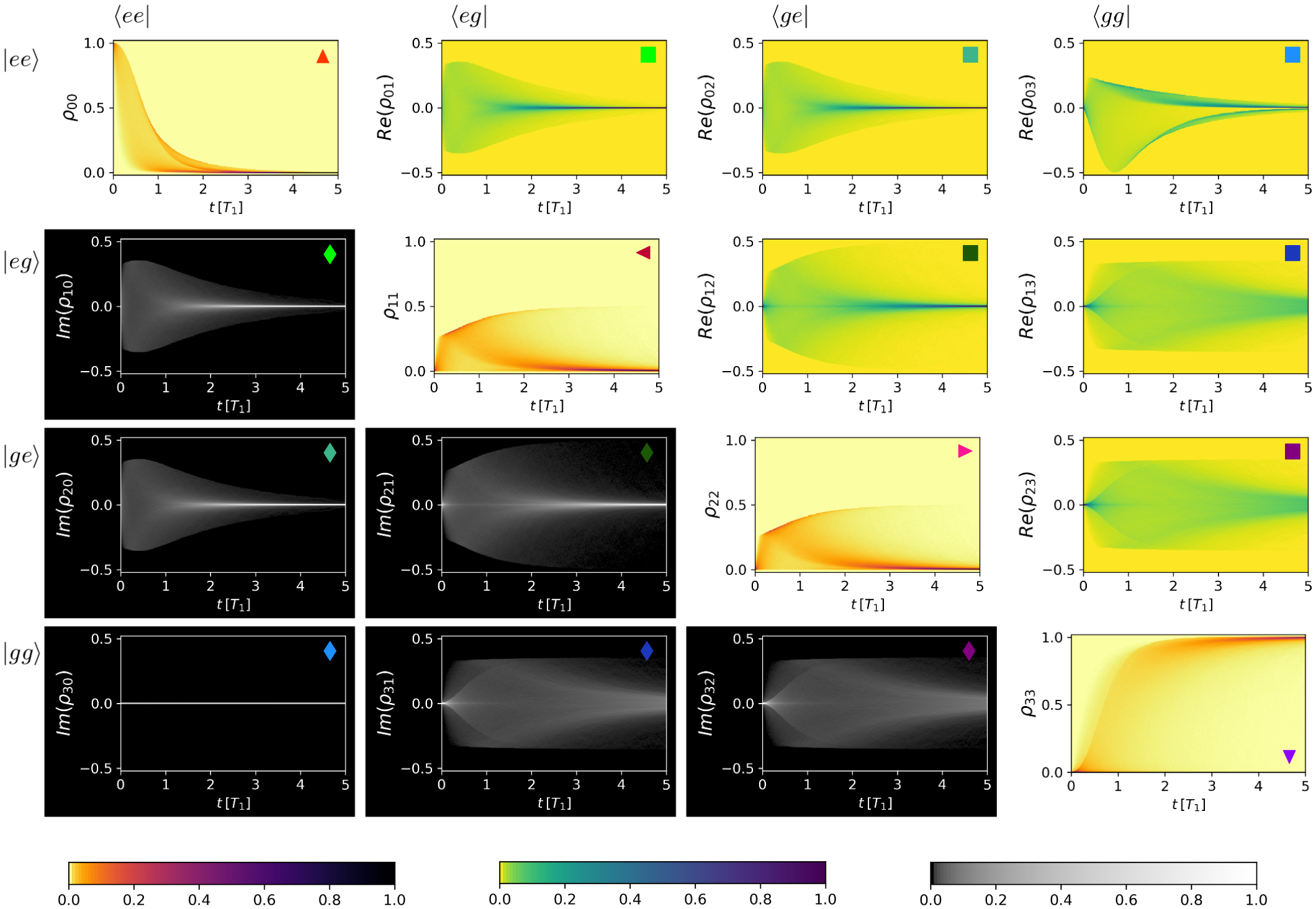}
\caption{We plot the density of SQTs as a function of time, with the ensemble of 10,000 initialized from $\ket{ee}$ with $\theta = 0$ and $\vartheta = 90^\circ$. This gives the density profile of each matrix element, corresponding to the individual realizations plotted in Fig.~\ref{fig-SQT_table-HomEnt}. The layout follows \eqref{colorrho}, with the populations down the diagonal (yellow and orange), the real parts of the coherence in the upper triangular region (yellow and green), and the imaginary parts of the coherences in the lower triangular region (inverse color). Correlations between the density matrix elements on aggregate are visible in this representation; the ability of the measurement to generate entanglement is however also captured by the fact these correlations exist \emph{not just on agreggate}, but also within \emph{individual} realizations of the continuous measurement process, as described in Fig.~\ref{fig-SQT_table-HomEnt}. This becomes clearer through comparison with the same plots for the non--entangling measurements, e.g.~as in Figs.~\ref{fig-SQT_table-HomFail} and \ref{fig-DENS_table-HomFail}.
\label{fig-DENS_table-HomEnt}}
\end{figure*}

 We here reconsider entanglement generation due to ideal homodyne detection, with $\theta = 0$ and $\vartheta = 90^\circ$ and the initial state $\ket{ee}$, in greater detail. We turn our attention to the stochastic trajectories, in order to understand how individual realizations of the measurement process generate the class of states \eqref{maxCstate}. In Figs.~\ref{fig-SQT_table-HomEnt} and \ref{fig-DENS_table-HomEnt} we show trajectories according to their density matrix components, and the ensemble density of SQTs.

Several additional insights emerge from these figures. First, we can see that at the level of individual trajectories there are perfect correlations between the real parts of the coherences involved with amplitudes moving in and out of the $\ket{eg}$ and $\ket{ge}$ subspace (coordinates $q_5$ \& $q_6$, and $q_8$ \& $q_9$, in the notation of Appendix~\ref{sec-bloch-coord}), and perfect anti--correlations in the imaginary parts of those coherences ($q_{11}$ \& $q_{12}$, and $q_{14}$ \& $q_{15}$). 
Note that the means of the signals represent combinations of precisely these terms.
Reaching maximal entanglement requires that the coherences within the $\ket{eg}$ and $\ket{ge}$ subspace be able to explore their full range $[-\tfrac{1}{2},\tfrac{1}{2}]$ (we refer to elements described by coordinates $q_4$ associated with $\mathsf{C}$, and $q_{10}$ associated with $\mathsf{E}$). The central elements of Figs.~\ref{fig-SQT_table-HomEnt} and \ref{fig-DENS_table-HomEnt} show that we are able to do this. Likewise, entanglement in the even--parity Bell subspace depends on the coherences between $\ket{ee}$ and $\ket{gg}$ being able to explore their whole range (coordinates $q_7$ and $q_{13}$). We see that the imaginary part of this coherence is never used by the measurement we consider now ($q_{13}$ is zero for all time), while the real part is able to explore its full \emph{negative} range $[-\tfrac{1}{2},0]$, but \emph{not} its full positive range. 
The range of the real part of the $\ket{ee}\bra{gg}$ element is only able to access $[0,\tfrac{1}{4}]$ while the range $[\tfrac{1}{4},\tfrac{1}{2}]$ associated with fully manifesting the state $\ket{\Phi^+}$ appears forbidden. 
We infer that our homodyne measurement with $\theta=0$ and $\vartheta = 90^\circ$ ``prefers'' generating correlations of the type $\ket{\Phi^-}$ as opposed to those of type $\ket{\Phi^+}$ (consistent with the arguments made in and around \eqref{conc-pure-bell}). 
One clear expression which contributes to this are the factors $-\tfrac{1}{2}$ in the {\color{amaranth} red} matrix element of \eqref{M-hom}, which moves population directly from $\ket{ee}$ to $\ket{gg}$. 
For futher comments in this vein, see appendix \ref{sec-onesteptest}. 
We can spot particular realizations in Figs.~\ref{fig-homodyne-eeGen} and \ref{fig-SQT_table-HomEnt} which conform especially well to the different options we see in the statistical discussion surrounding Fig.~\ref{fig-MaxCStats}; the burgundy path in Figs.~\ref{fig-homodyne-eeGen} and \ref{fig-SQT_table-HomEnt}, for instance, is a prototypical path which maximizes $\mathcal{C}$ by generating large $\mathsf{B}$ and low $|\mathsf{C}|$ and $|\mathsf{E}|$. 
It is a good example of a ``real'' trajectory with behavior very close to the idealized one underlying the best--case bound derived in \eqref{pure_Cmax}.
By contrast, the lavender--colored path exhibits the other extreme behavior, maintaining an unremarkable $\mathsf{B}$ and maximizing its concurrence by generating amplitude in $\mathsf{C}$ and $\mathsf{E}$ instead. 

\begin{figure*}
\centering
\includegraphics[width = \textwidth]{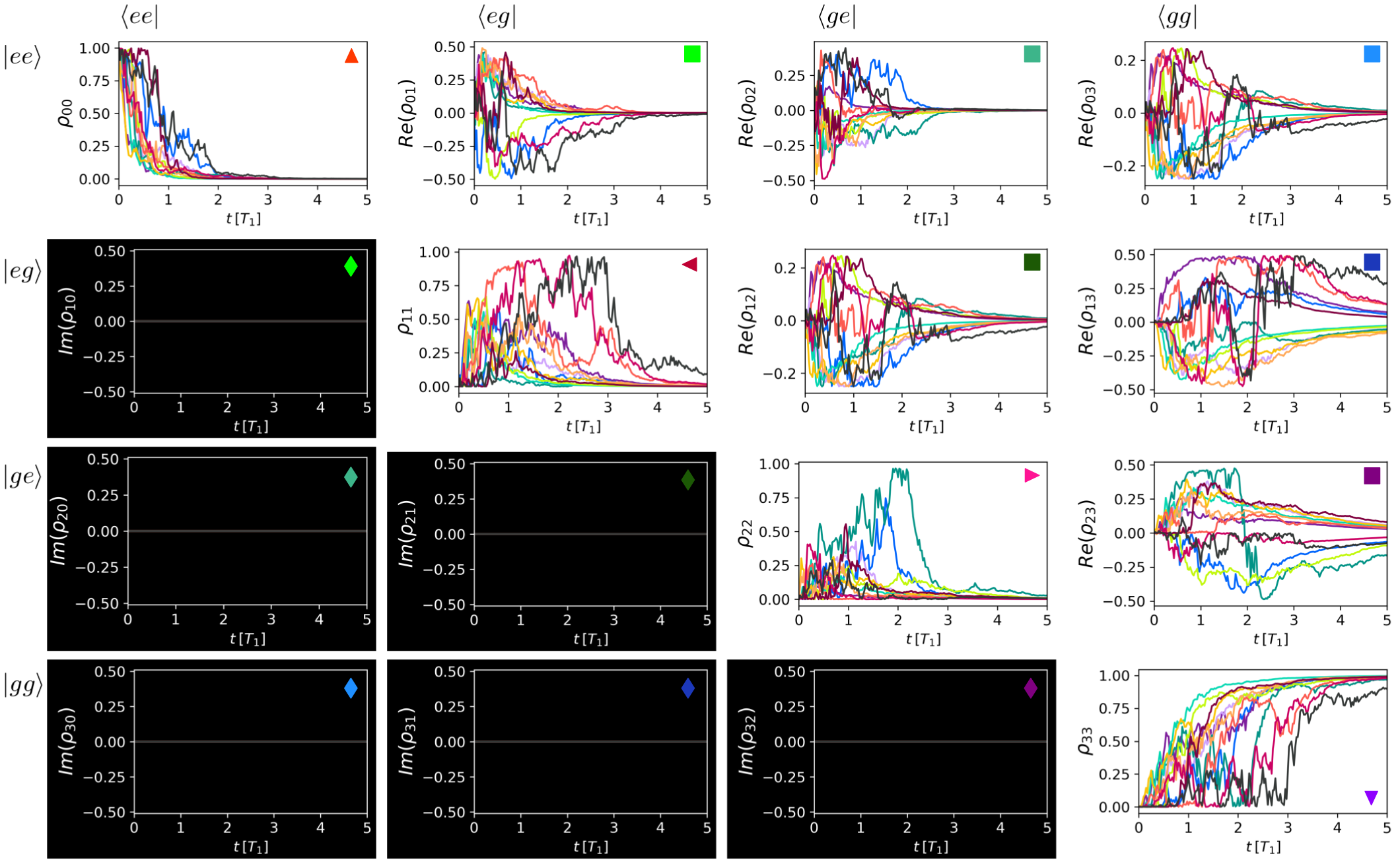}
\caption{We plot a dozen simulated SQTs, initialized from $\ket{ee}$ and computed with $\theta =  0 = \vartheta$. As noted throughout the main text, this measurement scenario does not generate entanglement. Note the lack of clear (anti--)correlations in individual realizations of the measurement process, in contrast with the entangling case (see Fig.~\ref{fig-SQT_table-HomEnt}). Instead of getting clear correlations among the real parts of coherences according the measurement record $r_3$, and anti--correlations in the imaginary parts of the coherences according to the measurement record $r_4$, both measurement records send their uncorrelated noise to the real parts of the density matrix; this effectively generates a competition between $\ket{\Psi^+}$--type correlations and those of the $\ket{\Psi^-}$--type, destroying entanglement. No trajectory plotted above exhibits any two--qubit concurrence at any point in its evolution. The layout follows \eqref{colorrho}, reflecting the two--qubit density matrix, with populations down the diagonal, the real parts of the coherences in the upper triangular region, and the imaginary parts of the coherences plotted in the lower triangular region (in inverse color). 
\label{fig-SQT_table-HomFail}}
\end{figure*}

\par
\begin{figure*}[t]
\centering
\includegraphics[width = \textwidth]{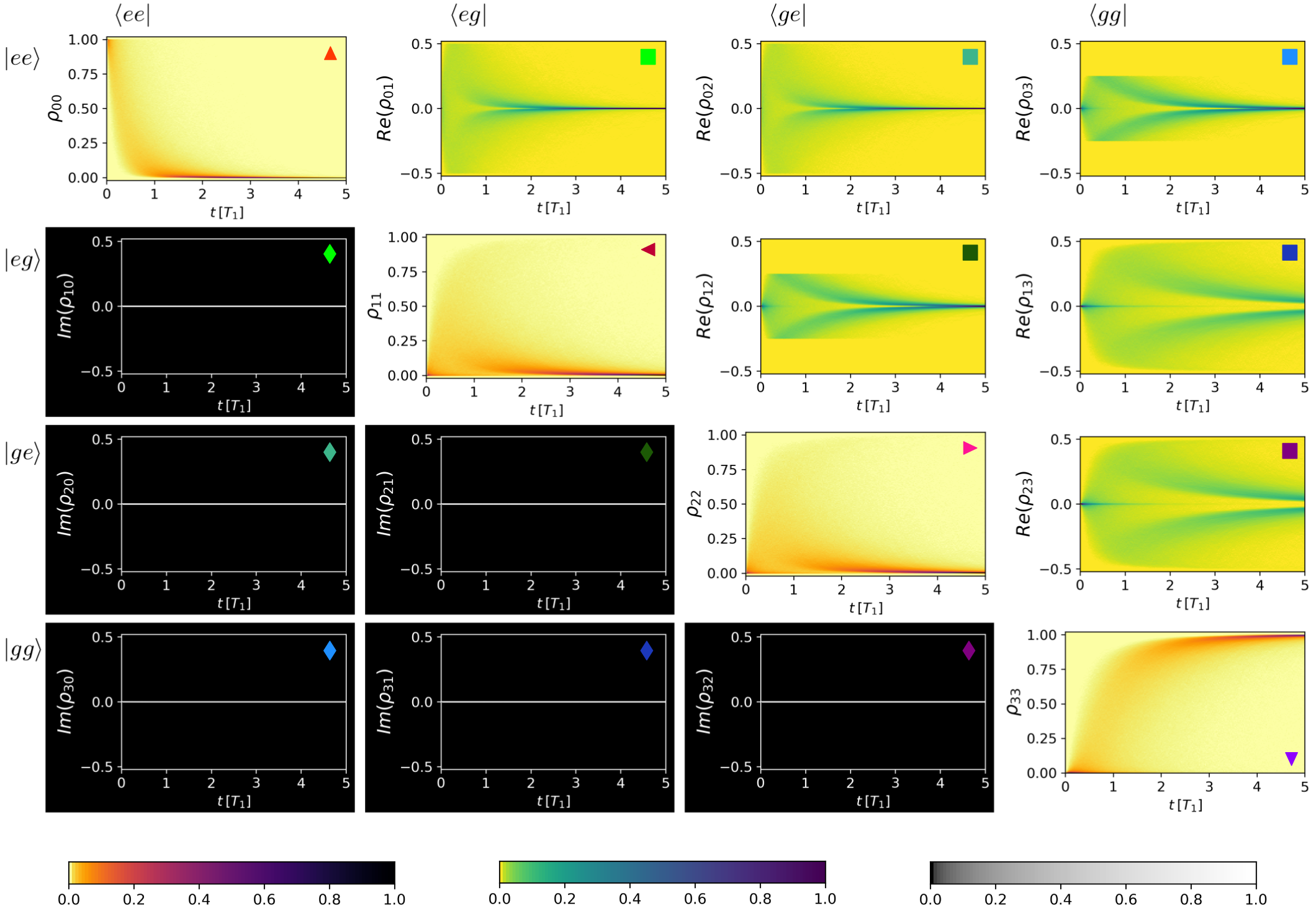}
\caption{We plot the density of simulated SQTs in the case of homodyne detection, with $\theta = 0 = \vartheta$. As discussed in the main text, this choice of quadrature measurements leads to the acquisition of which--path information, and does not generate entanglement at any point in time, in any of the underlying SQTs (see Fig.~\ref{fig-SQT_table-HomFail}). The plots are arranged to reflect the layout of the density matrix, with the populations down the diagonal (yellow and orange), the real parts of the coherences in the upper triangular region (corresponding to $q_4$ through $q_9$, in yellow and green), and the imaginary parts of the coherences in the lower triangular region (corresponding to $q_{10}$ through $q_{15}$, in inverse color). See appendix \ref{sec-bloch-coord}, and equation \eqref{colorrho} in particular, for details about this labeling scheme. Notice that many of the density matrix elements appear correlated on aggregate, but they aren't in individual realizations (see Fig.~\ref{fig-SQT_table-HomFail}), which spoils the possibility of entanglement. Furthermore, the coherences {$\color{lincolngreen} \blacksquare$} and {$\color{dodgerblue} \blacksquare$} are both truncated to half of their full range, indicating that Bell states are not represented in this ensemble with high fidelity.
\label{fig-DENS_table-HomFail}}
\end{figure*}

 Some of the points we make are clearer in contrast with a non--entangling case of the dynamics. We show trajectories and densities for the case $\theta = 0 = \vartheta$ in Figs.~\ref{fig-SQT_table-HomFail} and \ref{fig-DENS_table-HomFail}. Coherences associated with moving amplitude in and out of the $\lbrace \ket{eg}, \ket{ge} \rbrace$ subspace appear correlated on aggregate, but are \emph{not} at the level of individual trajectories.
Furthermore, the coherences within the $\lbrace \ket{ee}, \ket{gg} \rbrace$ ($q_7$) and $\lbrace \ket{eg},\ket{ge} \rbrace$ ($q_4$) subspaces which are key to generating Bell states are \emph{all} restricted to a truncated range. 
Both of these features are consistent with our observation that no trajectory achieves any concurrence at any point in its evolution for these measurement settings. Only the real parts of the density matrix are utilized for $\theta = 0  = \vartheta$. 
In the Bell state basis notation of \eqref{bell-state-basis}, this corresponds to having $\mathsf{D}$ be completely real instead of completely imaginary. 
We can consequently understand the entangling measurement $\theta = 0$ and $\vartheta = 90^\circ$ as allowing the readouts $r_3$ and $r_4$ to work cooperatively in generating concurrence. 
By contrast, the non--entangling measurement $\theta = 0 = \vartheta$ causes $\mathsf{C}$ and $\mathsf{D}$ to be forced into direct competition, destroying the possibility of generating concurrence (in contrast with the expression \eqref{conc-pure-bell}).

\section{One--Step Entanglement Tests \label{sec-onesteptest}}

We define a ``one--step test'' for entanglement genesis, starting from the excited state $\bra{ee} = (1,0,0,0)$. Effectively, we take one step from $\ket{ee}$, which is separable ($\mathcal{C} = 0$), with an idealized measurement (such that the state is still pure), and see how the concurrence behaves. We have already dealt with this problem both by several analytical arguments (which path information, separability of the optical states), and numerical methods (longer--time simulations) in the main body of the text. The single step test we consider now is simple enough to keep useful analytic expressions in play; while it is less general than the numerics already presented, some features of these simple arguments can help us understand what we see numerically, and add to the analytic arguments we have already presented. 

\par If we heterodyne both outputs, our state update after one step goes like
\be 
\hat{\mathcal{M}}_{\alpha\beta} \ket{ee} \propto \left( \begin{array}{c}
1-\epsilon  \\
\sqrt{\frac{\epsilon(1-\epsilon)}{2}} (\alpha^\ast e^{i\theta} -  \beta^\ast e^{i\vartheta}) \\
\sqrt{\frac{\epsilon(1-\epsilon)}{2}} (\alpha^\ast e^{i\theta} + \beta^\ast e^{i\vartheta})  \\
\tfrac{\epsilon}{2}\left( {\alpha^\ast}^2 e^{2i\theta} -{\beta^\ast}^2 e^{2i\vartheta} \right) 
\end{array} \right),
\ee
such that we find
\be \begin{split}
\mathcal{C} \propto \tfrac{\epsilon(1-\epsilon)}{2}\big|& (\alpha^\ast e^{i\theta})^2 - (\beta^\ast e^{i\vartheta})^2 - (\alpha^\ast e^{i\theta} - \beta^\ast e^{i\vartheta})(\alpha^\ast e^{i\theta} + \beta^\ast e^{i\vartheta}) \big| = 0.
\end{split} \ee
So we see again that in the heterodyne case, in which we always acquire information about the photon source, there is no possibility to get any entanglement from $\ket{ee}$, independent of the choices of LO phases $\theta$ and $\vartheta$. Simulations show that the situation does not improve as the system continues to evolve; see appendix~\ref{sec-het-simul}. 

\par Let us contrast this with the corresponding calculation in the homodyne case. We have
\be \label{hom-one-step}
\hat{\mathcal{M}}_{34}\ket{ee} \propto 
\left( \begin{array}{c}
1-\epsilon \\
\sqrt{\epsilon(1-\epsilon)} (e^{i\theta} X_3 - e^{i\vartheta} X_4)\\
\sqrt{\epsilon(1-\epsilon)} (e^{i\theta} X_3 + e^{i\vartheta} X_4) \\
\epsilon e^{2i\theta}(X_3^2-\tfrac{1}{2}) - \epsilon e^{2i\vartheta}(X_4^2-\tfrac{1}{2}),
\end{array} \right)
\ee
where we note that both $X_3$ and $X_4$ are real numbers (as opposed to $\alpha^\ast$ and $\beta^\ast$, which were complex; this allows for the cooperative behavior between different types of correlations, rather than competition, as described in and around \eqref{conc-pure-bell}). The concurrence after one measurement step goes like
\be \begin{split}
\mathcal{C} &\propto ~\epsilon(1-\epsilon) \big| e^{2i\theta}(X_3^2-\tfrac{1}{2}) - e^{2i\vartheta}(X_4^2 - \tfrac{1}{2}) - (e^{i\theta} X_3 - e^{i\vartheta} X_4)(e^{i\theta} X_3 + e^{i\vartheta} X_4) \big| \\& \quad = \tfrac{\epsilon(1-\epsilon)}{2}\left| e^{2i\vartheta} - e^{2i\theta} \right|.
\end{split} \ee
Notice that we now have $\mathcal{C} = 0$ for $\theta = 0 = \vartheta$ as above, but have $\mathcal{C} > 0$ (and with the greatest possible increase) for the cases which maximize the photon indistinguishability, e.g.~$\theta = 0$ and $\vartheta = 90^\circ$.

The constant term in the Hermite polynomials $X^2 + \tfrac{1}{2}$ [recall \eqref{hermite2}] is very important for avoiding complete cancellation of terms in the concurrence. Furthermore, this actually leads to a potentially desirable property: Entanglement genesis in the first step does not depend at all on the measurement records, and is therefore deterministic. It has recently been shown that this property can be retained in subsequent steps as well, using suitable feedback control \cite{Leigh2019}. 
The appearance of the second order Hermite polynomials is connected to our beamsplitter relations \eqref{bs}, and pertains to the matrix element which (at least in the photodetection case) is best ascribed to double/simultaneous emission events. 
Given the apparent importance of these terms in the concurrence generation, we infer that they are enforcing the source--indistinguishability requirement we have discussed at length ({i.e.~we see a clear connection between our assumption / requirement that our qubits emit indistinguishable photons, and subsequent possibility of entanglement genesis between the emitters by measurement}). 
{We reiterate that e.g.~Refs.\cite{Legero2003, Legero2004, Metz2008, Hurst2019} consider photon indistinguishability in more detail, in the context of setups pertinent to our own.}

\par It is possible to take a second step in the evolution, and still obtain expressions which help us to understand the dynamics apparent from simulation. Consider, for $\theta = 0$ and $\vartheta = 90^\circ$, a sequence of two measurements
\be \label{hom-two-step}
 \hat{\mathcal{M}}'_{34} \hat{\mathcal{M}}_{34}\ket{ee} \propto  \left(\begin{array}{c} 
1- 2\epsilon \\
\sqrt{\epsilon}(X_3+X_3'-iX_4-iX_4')\\
\sqrt{\epsilon}(X_3+X_3'+iX_4+iX_4')\\
-2\epsilon
\end{array} \right) + O(\epsilon^2).
\ee
We see the continued growth of correlations between $\ket{eg}$ and $\ket{ge}$; in the language of \eqref{bell-state-basis}, the sequence of outcomes $X_3$ promote the growth of $\mathsf{C}$ across sequential measurements, and the sequence of outcomes $X_4$ perform the same role for $\mathsf{D}$ or $\mathsf{E}$. What is more striking however, is the way amplitude appears in $\ket{gg}$; there are higher order (in $\epsilon$) corrections to this term $-2\epsilon$ which depend on the measurement outcomes, but what we essentially see to $O(\epsilon)$ is that there is quasi--deterministic growth of correlations of the type $\mathsf{B}$ over a sequential pair of measurements. This helps to underscore what we mean when we say that the system ``prefers'' correlations of type $\ket{\Phi^-}$ over $\ket{\Phi^+}$, and offers hints as to how the state $\ket{\Phi^-}$, which never plays a role in the photodetection scenario, actually ends up being the single most--likely maximally--concurrent state which can emerge from the homodyne scenario, under our chosen measurement settings. 
It is also apparent, from the expressions \eqref{hom-one-step} and \eqref{hom-two-step}, that selecting $X_3 = 0 = X_4$ provides an express route to the state $\ket{\Phi^-}$, as discussed in the arguments leading to \eqref{pure_Cmax}.

\section{Mixing Diffusive and Jump Dynamics: Role of Interference \label{sec-het-simul}}

\begin{figure}
\centering
    \includegraphics[width = .75\columnwidth]{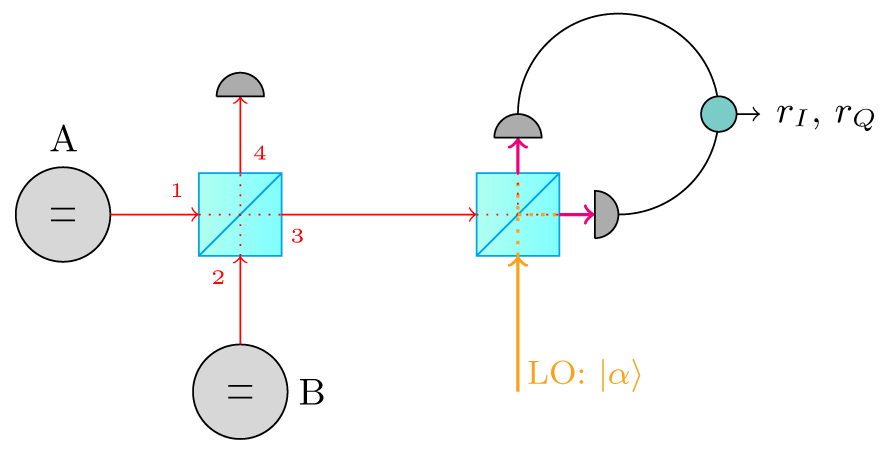}
    \caption{We sketch an apparatus which employs heterodyne detection at output 3, and photodetection at output 4. Trajectories behave purely diffusively conditioned on no photons exiting at output 4, but jumps may also occur unless interference effects prohibit a click event at port 4. Both beamsplitters are assumed to be 50/50 (such that the cavities' signals are mixed symmetrically, and the heterodyne detection is balanced).} 
    \label{fig-MixedMeas}
\end{figure}

{This appendix aims to fulfill two aims: First, we describe our simulation procedures in slightly more detail than in the main text, illustrating the flexibility of our method by applying it to a situation that uses heterodyne detection at one output, and photodetection at another (see Fig.~\ref{fig-MixedMeas}).
Secondly, and more interestingly, we show that (assuming our qubits emit indistinguishable photons) correlations in the qubit state which create interference at the mixing beamsplitter may select one detector or the other with complete certainty, thereby selecting a different type of measurement backaction.}
{This example highlights the consequences of the interference described in Sec.~\ref{sec-phot-interference}, and the ways it is reflected in our formalism and measurement statistics.}

The particular forms of the Kraus operators $\hat{\mathcal{M}}_{\alpha j}$ we are now interested in (using $\theta = 0 =\vartheta)$ are:
\be \label{tMa0}
\hat{\mathcal{M}}_{\alpha 0} = e^{-|\alpha|^2/2} \left( \begin{array}{cccc}
1- \epsilon & 0 & 0 & 0 \\
{\color{huntergreen} \alpha^\ast \sqrt{\frac{\epsilon(1-\epsilon)}{2}}} & \sqrt{1-\epsilon} & 0 & 0 \\
{\color{huntergreen} \alpha^\ast \sqrt{\frac{\epsilon(1-\epsilon)}{2}}} & 0 & \sqrt{1-\epsilon} & 0 \\
\frac{\epsilon}{2} (\alpha^\ast)^2 & \alpha^\ast \sqrt{\frac{\epsilon}{2}} & \alpha^\ast \sqrt{\frac{\epsilon}{2}} & 1
\end{array} \right)
\ee 
for $\bra{\alpha_3 0_4}\mathcal{M}\ket{0_3 0_4}$ (no click at output 4),
\be \label{tMa1}
\hat{\mathcal{M}}_{\alpha 1} = e^{-|\alpha|^2/2} \sqrt{\frac{\epsilon}{2}} \left( \begin{array}{cccc}
0 & 0 & 0 & 0 \\
-\sqrt{1-\epsilon} & 0 & 0 & 0 \\
 \sqrt{1-\epsilon} & 0 & 0 & 0 \\
0 & 1 & -1 & 0
\end{array} \right)
\ee
for $\bra{\alpha_3 1_4}\mathcal{M}\ket{0_3 0_4}$ (one click at output 4), and finally
\be \label{tMa2} 
\hat{\mathcal{M}}_{\alpha 2} = e^{-|\alpha|^2/2} \frac{ \epsilon}{2} \left( \begin{array}{cccc}
0 & 0 & 0 & 0 \\
0 & 0 & 0 & 0 \\
0 & 0 & 0 & 0 \\
-\sqrt{2} & 0 & 0 & 0
\end{array} \right) 
\ee
for $\bra{\alpha_3 2_4}\mathcal{M}\ket{0_3 0_4}$ (two clicks at output 4). These also form a proper POVM (verifiable by summing over $j  = 0,1,2$, and integrating out $d^2\alpha$). In the event that we turn on the photodetector, we update our state by
\be \label{stateup_MAJ}
\rho(t+dt) = \frac{\hat{\mathcal{M}}_{\alpha j} \rho(t) \hat{\mathcal{M}}_{\alpha j}^\dag}{\text{tr}\left( \hat{\mathcal{M}}_{\alpha j} \rho(t) \hat{\mathcal{M}}_{\alpha j}^\dag\right)},
\ee 
whereas if the output of channel 4 is irretrievably lost {(i.e.~if the photodetector is turned off)}, the state update is given by
\be \label{stateup_MAdisc}
\rho(t+dt) = \frac{\sum_{j=0,1,2} \hat{\mathcal{M}}_{\alpha j} \rho(t) \hat{\mathcal{M}}_{\alpha j}^\dag}{\text{tr}\left( \sum_{j=0,1,2} \hat{\mathcal{M}}_{\alpha j} \rho(t) \hat{\mathcal{M}}_{\alpha j}^\dag\right)}
\ee
instead.

\begin{figure}
    \centering
    \includegraphics[width = .75\columnwidth]{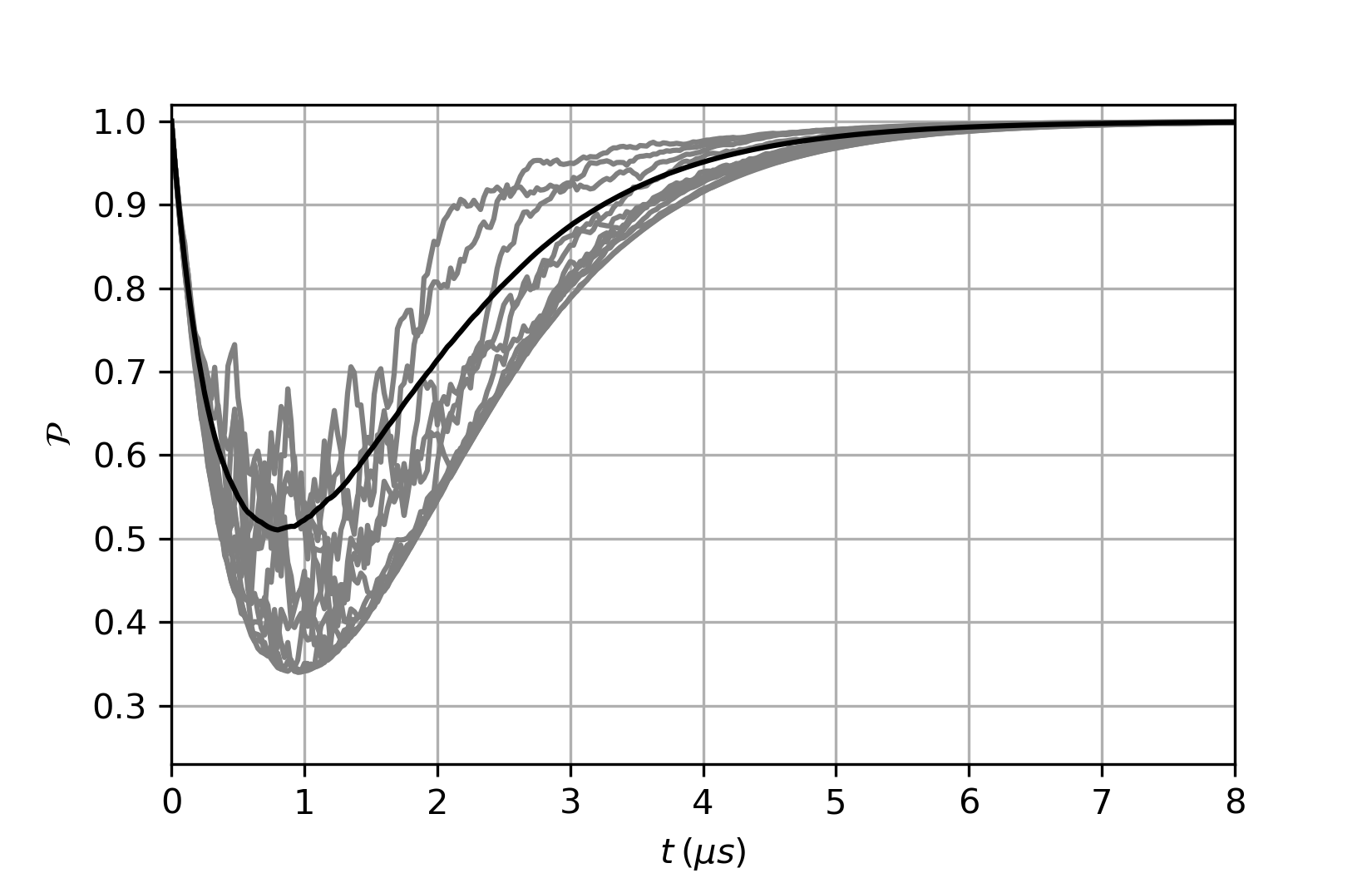}
    \caption{We initialize our two qubits in $\ket{ee}$, for the system diagrammed in Fig.~\ref{fig-MixedMeas} with the photodetector turned off. This scenario is modelled by the state update rule \eqref{stateup_MAdisc}. We plot the state purity $\mathcal{P}(t) = \text{tr}(\rho^2(t))$ as a function of time, showing both the purity of individual quantum trajectories in grey, and the average purity over an ensemble of such trajectories in black. The purity is 1 at the start and end, because both $\ket{ee}$ and $\ket{gg}$ are pure states, but the purity drops substantially during the dynamics moving between them, due to information being discarded at port 4 after the beamsplitter. Trajectories do not reach the maximally--mixed two--qubit state ($\mathcal{P} = 1/4$), and some stay well above, such that the average purity does not drop below $\mathcal{P} = 1/2$. We have $\gamma = 1~\mathrm{MHz}$, such that times in $\mathrm{\mu s}$ are also in units of both qubits' $T_1$. }
    \label{fig-het-disc-purity}
\end{figure}

Heterodyning one port and discarding the other does not lead to entanglement genesis. The purity of the system [given by $\text{tr}(\rho^2)$] drops considerably on its way from $\ket{ee}$ to $\ket{gg}$, a fact which substantially impedes the creation of entanglement on its own (i.e.~see the discussion of inefficient measurements), even without the other problematic properties of heterodyne detection with respect to generating concurrence (recall e.g.~the argument in and around \eqref{het-sep}). The purity recovers as the system decays, since $\ket{gg}$ is technically a pure state. We show the purity in Fig.~\ref{fig-het-disc-purity}.

\par We proceed to the case where the (ideal) photodetector in Fig.~\ref{fig-MixedMeas} is turned on, such that the click record at that port is available, and the two--qubit state update goes like \eqref{stateup_MAJ}.
The operation \eqref{tMa0}, which describes the diffusive dynamics due to heterodyning between click events, does nominally generate some correlations between $\ket{eg}$ and $\ket{ge}$, according to the matrix elements highlighted in {\color{huntergreen} green}, but does not generate concurrence. 
We can again attribute this to the argument in and around \eqref{het-sep}, although many of the other points we have mentioned above apply as well.
Any concurrence generated by this mixed detection scheme is generated by the click detector at port 4, \emph{not} the diffusive dynamics from the heterodyne measurement at port 3.

\subsection{Simulation Procedures}

{We describe our simulation procedures for the situation above, in the interests of completeness. The reader more interested in the behavior that arises from the present scheme should jump to the next sub-section.}

\par We describe how the operators \eqref{tMa0}, \eqref{tMa1}, \eqref{tMa2}, and state update \eqref{stateup_MAJ} are implemented numerically to simulate the stochastic trajectory dynamics, and then show results for a few revealing initial states. The case here is in many ways more numerically complex than the cases in the main text, and many statements about the general strategy employed here apply across all of our simulations. As in the main text, we expand the denominator of the state update equation to approximate the probability distribution describing possible measurement outcomes at each step. This leads us to define $\mathcal{G}$, such that 
\be
\text{tr}\left( \hat{\mathcal{M}}_{\alpha j} \rho(t) \hat{\mathcal{M}}_{\alpha j}^\dag \right) \approx e^{ C_j+\mathcal{G}_j dt + O(dt^2)}.
\ee
It turns out that when we do this kind of expansion, we will find some Gaussian terms with some additional state--dependent coefficients attached, i.e. we find
\be 
e^{\mathcal{G}_0 dt} = w_0 g_0,
\quad
e^{C_1}e^{\mathcal{G}_1 dt} = w_1 g_1, \quad \text{and}\quad
e^{C_2}e^{\mathcal{G}_2 dt} = w_2 g_2,
\ee
where the $g$ terms are Gaussians in $r_I$ and $r_Q$ (recall e.g.~\eqref{hetd_readouts}) with variance $1/dt$, and the remaining terms which survive the expansion are collected into the weight factors $w$. As in the case with photodetectors we considered earlier, the $w_j$ are state dependent and used to make a multinomial choice about whether (and how many times) the photodetector registers an event in a given timestep. This then also determines which Gaussian $g_j$ the heterodyne readout result is drawn from. 
Using the full set of available information to draw the measurement records is necessary, to ensure that we do not create records which are inconsistent with one another (even if we were to suppose we have multiple observers with incomplete information, the records would still have to be generated on the basis of a ``super observer's'' record(s), which account for all available measurement information); see \cite{Dziarmaga2004, Harrington2019, PL-Dissertation} for further comments in this vein.

The particulars of the weights and Gaussians are described here, starting with the terms in the no--click case. We have 
\begin{subequations} 
\be \begin{split} \label{G0HetMix}
\mathcal{G}_0 =& - \tfrac{1}{2}\left(r_I - \chi_I \sqrt{\tfrac{\gamma}{2}} \right)^2 - \tfrac{1}{2}\left( r_Q - \chi_Q \sqrt{\tfrac{\gamma}{2}} \right)^2 - \gamma \: \Xi + \tfrac{\gamma}{4}\left( \chi_I^2 + \chi_Q^2 \right), \quad \text{with}
\end{split} \ee
\be \begin{split}
 \chi_I \equiv & \cos\theta (q_5 + q_6 +q_8 +q_9 ) - \sin\theta (q_{11} + q_{12} +q_{13} + q_{15}),
\end{split}  \ee \be \begin{split} 
 \chi_Q \equiv & - \sin\theta (q_5 + q_6 +q_8 +q_9 ) - \cos\theta (q_{11} + q_{12} +q_{14} + q_{15}).
\end{split} \ee \end{subequations}
The coordinates $\mathbf{q}$ parameterizing the two--qubit density matrix are described in \ref{sec-bloch-coord}, and we defined $\Xi$ in \eqref{XiTheta}. We thus have Gaussians with variance $1/dt$ and means $\chi_I\sqrt{\gamma/2}$ and $\chi_Q\sqrt{\gamma/2}$ for $r_I$ and $r_Q$, respectively. The remaining terms are included to the weight factor used to determine the correct statistics for the no--click event, which is
\be \label{w0mixed}
w_0 = \mathcal{N}\exp\left[ - \gamma\: \Xi + \tfrac{\gamma}{4}\left( \chi_I^2 + \chi_Q^2 \right)  \right].
\ee
The term $\mathcal{N}$ is a normalization for the click probabilities, used to make $\sum_j w_j = 1$. The remaining Gaussians in $r_I$ and $r_Q$, for the one--click and two--click terms, both have mean zero, and the same variance $1/dt$. This indicates that in the event of a click, we know the photon went to port 4, and therefore did not go to the heterodyne detection at port 3; the heterodyne readouts then contain no signal in the requisite timestep (there is no information at port 3 without the possibility of a photon having arrived there), and only pure noise from the LO. 
These conditions are key to enforcing that correlations between the optical modes are properly reflected in simulations (again, this is necesary to generate mutually--consistent measurement records).
The probabilites associated with these jump events are given by 
\be \label{w1mixed}
w_1 = \mathcal{N} \frac{\gamma\: dt \: \varsigma}{6 \sqrt{2}} \exp\left[ - \frac{\gamma \: \kappa \: dt}{\varsigma \sqrt{2}} \right], \text{ and}
\ee
\be \label{w2mixed}
w_2 = \mathcal{N}\frac{dt^2 \gamma^2}{24} \left(3 + 6 \sqrt{2} q_1 + 2 \sqrt{6} q_2 + 2 \sqrt{3} q_3 \right),
\ee
where we have shorthanded some expressions
\begin{subequations} \label{vska}
\be \label{vsig}
\varsigma \equiv 3 \sqrt{2} + 3 q_1 + \sqrt{3} q_2 + 2 \sqrt{6} q_3 - 6 q_{4}, \ee \be \label{kapp}
\kappa \equiv \frac{3}{\sqrt{2}} + 6q_1 + 2 \sqrt{3} q_2 + \sqrt{6} q_3, \ee
\end{subequations}
for ease of notation.

\par The simulation procedure in each timestep can then be summarized as:
\begin{enumerate}
    \item Use the state--dependent $w_j$ as probabilities in a multinomial distribution; draw an outcome for the number of clicks at the detector at port 4 accordingly.
    \item Given the outcome at the click detector, draw $r_I$ and $r_Q$ from the appropriate Gaussian distributions, with variance $1/dt$ and state--dependent means, to simulate the heterodyne measurement at port 3.
    \item Choose the appropriate operator \eqref{tMa0}, \eqref{tMa1}, or \eqref{tMa2}, according to the jump outcome, put in the stochastic readouts $r_I$ and $r_Q$, and then update the state with \eqref{stateup_MAJ}. Repeat until desired evolution time is reached.
\end{enumerate}
The procedure for the codes in the main body of the text is quite similar {(and is simpler than that shown here, as it is more straightforward to draw readouts all of one type than it is to combine diffusive and jump dynamics)}. 
We have here shown, however, that our methods generalize to cases in which we mix diffusive and jump trajectories. 
Few works in the quantum trajectories literature have studied the dynamics arising under simultaneous different \emph{types} of continuous measurements (i.e.~jumps and diffusion) at all \cite{Kuramochi2013, Chantasri2019, Zhang2019Heralded}.

\subsection{Interference Effects}

\begin{figure}[t]
    \centering
    \includegraphics[width = \textwidth]{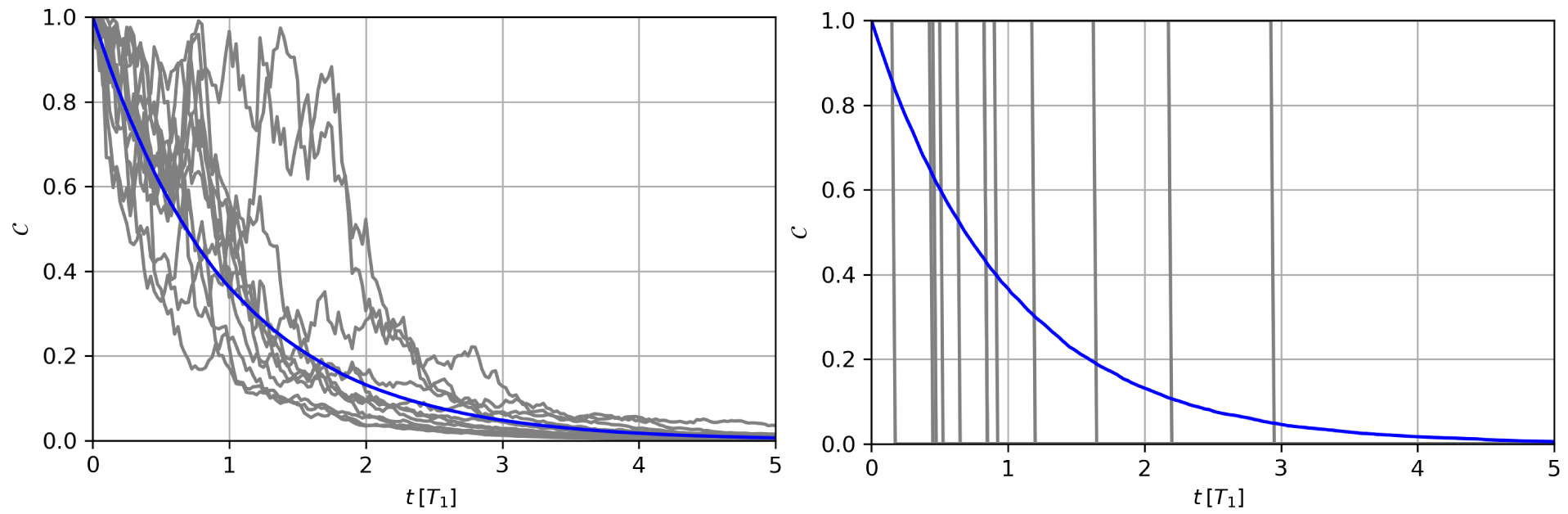}
    \caption{We plot concurrence as a function of time, originating either from $\ket{\Psi^+}$ (left) or $\ket{\Psi^-}$ (right) \eqref{psi-pm}, in the setup of Fig.~\ref{fig-MixedMeas} which combines heterodyne detection at one port with photodetection at the other. The effect of the interference at the beamsplitter created by the correlation or anti--correlation between $\ket{eg}$ and $\ket{ge}$ is clearly visible here, because determining the output port determines the type of measurement backaction; dynamics originating from $\ket{\Psi^+}$ only interact with the heterodyne device, resulting in diffusive quantum trajectories of the two--qubit state, whereas only jump dynamics arise from $\ket{\Psi^-}$, since all of the output goes to the photodetector in that case. The average concurrence is in good quantitative agreement with $\bar{\mathcal{C}}(t) = e^{-\gamma t}$ in both cases shown above, consistent with the photodetection case in Fig.~\ref{fig-photodect-conc}(b) and the homodyne case of Fig.~\ref{fig-Homx2-BellDecay}(a). No post--selection is used in the simulations above, because the interference conditions for the states in question perfectly select one output or the other.}
    \label{fig-Het1x-BareConcDecay}
\end{figure}

\par We put the numerical strategies we just described to work. In Fig.~\ref{fig-Het1x-BareConcDecay}, we plot the state evolution originating from $\ket{\Psi^\pm}$ \eqref{psi-pm}. We immediately see that one Bell state allows for trajectories which only experience diffusion ($\ket{\Psi^+}$ sends all its output to the heterodyne detector at output 3), and the other allows only jumps ($\ket{\Psi^-}$ sends all its output to the photodetector at output 4). 
The pure states which generate these interference conditions are, more generally, of the form
\be \label{interference-state-form}
\ket{\psi} = \mathcal{N} \left( \begin{array}{r} 
0 \\ \mathsf{a} \\ \pm\mathsf{a} \\ \mathsf{b}
\end{array} \right).
\ee
It is apparent from simulation that the relevant measurement dynamics preserve the correlations and coherence in the $\ket{eg}$ and $\ket{ge}$ terms, even as the population is shifted to $\ket{gg}$ over time by the natural decay process. Therefore the conditions for complete interference do not only appear at the beginning of the simulation, but are preserved for its duration once they are established. We also understand that the interference effect is naturally reproduced by the weight factors \eqref{w0mixed}, \eqref{w1mixed}, and \eqref{w2mixed}; we have applied them in the codes without adding any further constraints. 
{We may thus appreciate an interesting feature of our model, which stems from the assumption of indistinguishable photons: We have a case where the phase of the entangled two--qubit state may select a detector output with certainty; if different detector types are used at the outputs, then completely different types of measurement backaction are effectively selected by interference}. 

The concurrence in Fig.~\ref{fig-Het1x-BareConcDecay} decays on average at the rate $\gamma$ at which the individual qubits relax, despite exhibiting very different trajectories for individual realizations. 
We also see that the concurrence among the diffusive trajectories originating from $\ket{\Psi^+}$ does not decrease monotonically in individual realizations; although we cannot generate entanglement from simple separable states using heterodyne detection, certain trajectories do still exhibit partial decay and regrowth of concurrence.

\section{More on Measurement Inefficiency \label{app-inefficient}}
Recall from Fig.~\ref{fig-inefficient}, and the main text, that measurement inefficiency is effectively modeled by {imagining} the signal to be diverted into a lost mode with some probability, rather than arriving at an ideal detector with certainty. 
{In practice, no experimental detection scheme achieves perfect measurement efficiencies. 
In fact, a wider range of detector imperfections are commonplace (see e.g.~\cite{wein2020analyzing} for some details and characterization of the effect of such imperfections on similar entanglement schemes)}. 
The relevant optical transformation was summarized by \eqref{aeta}, and allows us to get a matrix $\mathcal{M}_\eta$ from the $\mathcal{M}$ we defined in \eqref{2Q_stateUp1}.

\par We have mentioned that our state update must be modified for the case of inefficient detection, by tracing out over possible outcomes in the ``lost'' modes that we do not have access to. 
For example, for inefficient photodetection with the outcome $\lbrace n_3,n_4 \rbrace = \lbrace 0,0 \rbrace$ at the signal ports (where $n_3$ and $n_4$ are a number of photons received in the requisite interval $dt$), we would have a group of Kraus operators
\be 
\hat{\mathcal{M}}_{0 0 n_3^\ell n_4^\ell} = \bra{0_{3}^s 0_{4}^s n_3^\ell n_4^\ell}\mathcal{M}_{\eta} \ket{0000},
\ee 
and the state update equation $\rho(t+dt) =$
\be \label{m00-eta-stateup}
 \frac{\hat{\mathcal{M}}_{0000}\rho\hat{\mathcal{M}}^\dag_{0000}+ \hat{\mathcal{M}}_{0010}\rho\hat{\mathcal{M}}^\dag_{0010}+\hat{\mathcal{M}}_{0001}\rho\hat{\mathcal{M}}^\dag_{0001}+\hat{\mathcal{M}}_{0020}\rho\hat{\mathcal{M}}^\dag_{0020}+\hat{\mathcal{M}}_{0002}\rho\hat{\mathcal{M}}^\dag_{0002}}{\text{tr}\left( \hat{\mathcal{M}}_{0000}\rho\hat{\mathcal{M}}^\dag_{0000}+ \hat{\mathcal{M}}_{0010}\rho\hat{\mathcal{M}}^\dag_{0010}+\hat{\mathcal{M}}_{0001}\rho\hat{\mathcal{M}}^\dag_{0001}+\hat{\mathcal{M}}_{0020}\rho\hat{\mathcal{M}}^\dag_{0020}+\hat{\mathcal{M}}_{0002}\rho\hat{\mathcal{M}}^\dag_{0002}\right)}.
\ee
for $\rho = \rho(t)$.
For such an update with finite measurement efficiency, the basis in which we do the trace over the outcomes in the lost mode does not matter, as long as it represents a complete set of outcomes.
By that token, inefficient homodyne detection is {most--straightforwardly modeled} by an operator
\be  
\hat{\mathcal{M}}_{X n_3^\ell n_4^\ell} = \bra{X_3 X_4 n_3^\ell n_4^\ell}\mathcal{M}_{\eta} \ket{0000},
\ee 
where the signal modes are projected into a quadrature $X_3$ or $X_4$, as in the ideal case, but the lost modes are projected into the Fock basis.
Such operators can be used with the state update $\rho(t+dt) = $
\be \label{hom-eta-stateup}
\frac{\hat{\mathcal{M}}_{X00}\rho\hat{\mathcal{M}}^\dag_{X00}+ \hat{\mathcal{M}}_{X10}\rho\hat{\mathcal{M}}^\dag_{X10}+\hat{\mathcal{M}}_{X01}\rho\hat{\mathcal{M}}^\dag_{X01}+\hat{\mathcal{M}}_{X20}\rho\hat{\mathcal{M}}^\dag_{X20}+\hat{\mathcal{M}}_{X02}\rho\hat{\mathcal{M}}^\dag_{X02}}{\text{tr}\left( \hat{\mathcal{M}}_{X00}\rho\hat{\mathcal{M}}^\dag_{X00}+ \hat{\mathcal{M}}_{X10}\rho\hat{\mathcal{M}}^\dag_{X10}+\hat{\mathcal{M}}_{X01}\rho\hat{\mathcal{M}}^\dag_{X01}+\hat{\mathcal{M}}_{X20}\rho\hat{\mathcal{M}}^\dag_{X20}+\hat{\mathcal{M}}_{X02}\rho\hat{\mathcal{M}}^\dag_{X02}\right)}.
\ee
Summing over the lost modes in the discrete Fock basis is computationally simpler than integrating out another pair of continuous--valued homodyne (quadrature basis) outcomes, although the latter would also be correct.
We continue into a more detailed discussion of each type of inefficient measurement.
The comparable exploration of the one qubit case can be found in \cite{FlorTeach2019,PL-Dissertation}.

\subsection{Methods for Inefficient Photodetection \label{sec-app-PDeta}}

We now summarize all possible outcomes and Kraus operators used to model inefficient photodetection. All Kraus operators here are notated according to $\mathcal{M}_{n_3^s n_4^s n_3^\ell n_4^\ell}$, where the $n$ are outcomes in the Fock basis at their respective ports.
The phases $\theta$ and $\vartheta$ in \eqref{bs} may be set to zero when we monitor the system via photodetection, without loss of generality.

\par When neither detector clicks, the following options are valid:
\begin{subequations} \label{M00-eta}
\be \label{M0000}
\hat{\mathcal{M}}_{0000} = \bra{0000}\mathcal{M}_\eta\ket{0000},
\ee \be \label{M0010}
\hat{\mathcal{M}}_{0010} = \bra{0010}\mathcal{M}_\eta\ket{0000},
\ee \be \label{M0001}
\hat{\mathcal{M}}_{0001} = \bra{0001}\mathcal{M}_\eta\ket{0000},
\ee \be \label{M0020}
\hat{\mathcal{M}}_{0020} = \bra{0020}\mathcal{M}_\eta\ket{0000},
\ee \be \label{M0002}
\hat{\mathcal{M}}_{0002} = \bra{0002}\mathcal{M}_\eta\ket{0000}.
\ee
\end{subequations}
The first represents the ideal option, in which we registered no photons because there were, in fact, no photons emitted. The next operators however account for the possibility that one or two photons were emitted, but were routed to one of the loss channels and weren't measured.
The probability of the outcome $\lbrace n_3^s, n_4^s \rbrace = \lbrace 0,0 \rbrace$ is given by
\be \begin{split}
w_{00} &= \text{tr}\left( \sum_j \hat{\mathcal{M}}_j \rho \hat{\mathcal{M}}_j^\dag \right) = 1 - \tfrac{\epsilon}{2}(\eta_3+\eta_4)\Xi + \tfrac{\epsilon}{\sqrt{2}}(\eta_4-\eta_3)q_4 + \tfrac{\epsilon^2}{2}(\eta_3^2+\eta_4^2)\Theta,
\end{split} \ee
where $j$ indexes the five operators in \eqref{M00-eta}, and we have defined $\Xi$ and $\Theta$ in \eqref{XiTheta}.

\par If the detector at port 3 registers the arrival of a single photon, the following options are at play: 
\begin{subequations} \label{M10-eta}
\be \label{M1000}
\hat{\mathcal{M}}_{1000} = \bra{1000}\mathcal{M}_\eta\ket{0000},
\ee \be \label{M1010}
\hat{\mathcal{M}}_{1010} = \bra{1010}\mathcal{M}_\eta\ket{0000}.
\ee 
\end{subequations}
(Either one photon was emitted, and we caught it, or two--photons were emitted, in which case they both had to go to the same output {under the assumption of photon indistinguishability}, in this case 3, but the second photon was lost on account of $\eta_3\neq1$).
The probability of the outcome $\lbrace n_3^s, n_4^s \rbrace = \lbrace 1,0 \rbrace$ is given by
\be \begin{split}
w_{10} &= \text{tr}\left( \hat{\mathcal{M}}_{1000} \rho \hat{\mathcal{M}}_{1000}^\dag + \hat{\mathcal{M}}_{1010} \rho \hat{\mathcal{M}}_{1010}^\dag  \right) = \frac{\epsilon \: \eta_3}{2} \Xi + \frac{\epsilon \: \eta_3}{\sqrt{2}}q_4 - \epsilon^2 \eta_3^2 \Theta.
\end{split} \ee
Likewise, if the detector at port 4 registers a single photon, we have the operators
\begin{subequations} \label{M01-eta}
\be \label{M0100}
\hat{\mathcal{M}}_{0100} = \bra{0100}\mathcal{M}_\eta\ket{0000},
\ee \be \label{M0101}
\hat{\mathcal{M}}_{0101} = \bra{0101}\mathcal{M}_\eta\ket{0000},
\ee 
\end{subequations}
and the probability of the outcome $\lbrace n_3^s, n_4^s \rbrace = \lbrace 0,1 \rbrace$ is given by
\be \begin{split}
w_{01} &= \text{tr}\left( \hat{\mathcal{M}}_{0100} \rho \hat{\mathcal{M}}_{0100}^\dag + \hat{\mathcal{M}}_{0101} \rho \hat{\mathcal{M}}_{0101}^\dag  \right) = \frac{\epsilon \: \eta_4}{2} \Xi - \frac{\epsilon \: \eta_4}{\sqrt{2}}q_4 - \epsilon^2 \eta_4^2 \Theta.
\end{split} \ee

If the detectors at port 3 or 4 register a double click, we know that we have received all of the photons, because it wasn't possible for more to be emitted in a single timestep. This situation involves the operators 
\begin{subequations} \label{M2-eta} 
\be \label{M2000}
\hat{\mathcal{M}}_{2000} = \bra{2000}\mathcal{M}_\eta\ket{0000},
\ee or \be \label{M0200}
\hat{\mathcal{M}}_{0200} = \bra{0200}\mathcal{M}_\eta\ket{0000},
\ee
\end{subequations}
which are invoked with the probabilities
\begin{subequations}
\be 
w_{20} = \text{tr}\left(\hat{\mathcal{M}}_{2000} \rho \hat{\mathcal{M}}_{2000}^\dag \right) = \frac{\epsilon^2\eta_3^2}{2} \Theta,
\ee or \be 
w_{02} =\text{tr}\left(\hat{\mathcal{M}_{0200}} \rho \hat{\mathcal{M}}_{0200}^\dag \right) = \frac{\epsilon^2\eta_4^2}{2} \Theta,
\ee
\end{subequations}
respectively.

\par We can verify that we have exhausted the set of possible Kraus operators for the Fock basis which can arise from $\mathcal{M}_\eta$, in that they form {elements of a POVM}, i.e.
\be 
\sum_j \hat{\mathcal{M}}_j^\dag \hat{\mathcal{M}}_j = \hat{\mathbb{I}},
\ee
where the sum here is over all of the operators \eqref{M00-eta}, \eqref{M10-eta}, \eqref{M01-eta}, and \eqref{M2-eta}. The probabilities $w$ for the five possible outcomes we have listed are also already properly normalized, in that they too sum to 1.

\subsection{Methods for Inefficient Homodyne Detection}

The same principles can be used to generate the corresponding model for inefficient homodyne detection. 
We project the output signal modes (corresponding to $\hat{a}_{3s}^\dag$ and $\hat{a}_{4s}^\dag$) in to a quadrature, as in the ideal case, and we will project the lost modes into the Fock basis as we just did for inefficient photodetection.
A total of five Kraus operators arise from such an analysis, and they are
\begin{subequations} \label{M-hom-eta} \be \begin{split}
&\hat{\mathcal{M}}_{X00} = \bra{X_3 X_4 00} \mathcal{M}_\eta \ket{0000} = \frac{e^{-(X_3^2+X_4^2)/2}}{\sqrt{\pi}} \times  \\
&{\footnotesize  \left(\begin{array}{cccc}
1-\epsilon & 0 & 0 & 0 \\
\sqrt{\epsilon(1-\epsilon)}(\sqrt{\eta_3} e^{i\theta}X_3-\sqrt{\eta_4}e^{i\vartheta}X_4) & \sqrt{1-\epsilon} & 0 & 0 \\
\sqrt{\epsilon(1-\epsilon)}(\sqrt{\eta_3} e^{i\theta}X_3+\sqrt{\eta_4}e^{i\vartheta}X_4) & 0 & \sqrt{1-\epsilon} & 0 \\
\epsilon [e^{2i\theta} \eta_3 (X_3^2-\tfrac{1}{2})-e^{2i\vartheta} \eta_4 (X_4^2-\tfrac{1}{2})] & \sqrt{\epsilon}(\sqrt{\eta_3} e^{i\theta}X_3+\sqrt{\eta_4}e^{i\vartheta}X_4) & \sqrt{\epsilon}(\sqrt{\eta_3} e^{i\theta}X_3-\sqrt{\eta_4}e^{i\vartheta}X_4) & 1
\end{array}\right)},
\end{split} \ee \be \begin{split}
\hat{\mathcal{M}}_{X10} &= \bra{X_3 X_4 10} \mathcal{M}_\eta \ket{0000} \\
&= \frac{e^{-(X_3^2+X_4^2)/2}}{\sqrt{\pi}} \left(\begin{array}{cccc}
0 & 0 & 0 & 0 \\
\sqrt{\frac{\epsilon(1-\epsilon)}{2} (1-\eta_3)} & 0 & 0 & 0 \\
\sqrt{\frac{\epsilon(1-\epsilon)}{2} (1-\eta_3)} & 0 & 0 & 0 \\
\epsilon \:e^{i\theta} \: X_3 \sqrt{2\eta_3(1-\eta_3)} & \sqrt{\frac{\epsilon}{2} (1-\eta_3)} & \sqrt{\frac{\epsilon}{2} (1-\eta_3)} & 0
\end{array} \right),
\end{split} \ee \be \begin{split}
\hat{\mathcal{M}}_{X01} &= \bra{X_3 X_4 01} \mathcal{M}_\eta \ket{0000} \\
&= \frac{e^{-(X_3^2+X_4^2)/2}}{\sqrt{\pi}} \left(\begin{array}{cccc}
0 & 0 & 0 & 0 \\
-\sqrt{\frac{\epsilon(1-\epsilon)}{2} (1-\eta_4)} & 0 & 0 & 0 \\
\sqrt{\frac{\epsilon(1-\epsilon)}{2} (1-\eta_4)} & 0 & 0 & 0 \\
-\epsilon \:e^{i\vartheta} \: X_4 \sqrt{2\eta_4(1-\eta_4)} & \sqrt{\frac{\epsilon}{2} (1-\eta_4)} & -\sqrt{\frac{\epsilon}{2} (1-\eta_4)} & 0
\end{array} \right),
\end{split} \ee 
 \be 
\begin{split}
\hat{\mathcal{M}}_{X20} &= \bra{X_3 X_4 20} \mathcal{M}_\eta \ket{0000} \\
&= \frac{e^{-(X_3^2+X_4^2)/2}}{\sqrt{\pi}} \left(\begin{array}{cccc}
0 & 0 & 0 & 0 \\
0 & 0 & 0 & 0 \\
0 & 0 & 0 & 0 \\
\tfrac{\epsilon}{\sqrt{2}}(1-\eta_3) & 0 & 0 & 0
\end{array} \right),
\end{split}
\ee \be 
\begin{split}
\hat{\mathcal{M}}_{X02} &= \bra{X_3 X_4 02} \mathcal{M}_\eta \ket{0000} \\
&= \frac{e^{-(X_3^2+X_4^2)/2}}{\sqrt{\pi}} \left(\begin{array}{cccc}
0 & 0 & 0 & 0 \\
0 & 0 & 0 & 0 \\
0 & 0 & 0 & 0 \\
-\tfrac{\epsilon}{\sqrt{2}}(1-\eta_4) & 0 & 0 & 0
\end{array} \right).
\end{split}
\ee\end{subequations}

\par We reiterate that the state update is given, in terms of the above operators, by \eqref{hom-eta-stateup}.
These operators again form a complete set, in that 
\be 
\iint dX_3 \: dX_4 \: \sum_j \hat{\mathcal{M}}_{Xj}^\dag \hat{\mathcal{M}}_{Xj} = \hat{\mathbb{I}},
\ee
where $j$ indexes all five operators in \eqref{M-hom-eta}.
The statistics of the measurement record they imply are summarized by 
\be \begin{split}
\mathcal{G}_{34}^\eta =&  -\tfrac{1}{2}\left(r_3 - \sqrt{\gamma \eta_3}\, \chi_3 \right)^2 - \tfrac{1}{2}\left(r_4 - \sqrt{\gamma \eta_4}\, \chi_4 \right)^2 \\    
& + \tfrac{\gamma}{2}\left( \eta_3\chi_3^2 + \eta_4\chi_4^2 \right) - \tfrac{\gamma}{2}(\eta_3+\eta_4) \: \Xi +\tfrac{\gamma}{2}q_4 (\eta_4-\eta_3) \\ & + \tfrac{\gamma}{\sqrt{2}}  q_7 (\eta_4 \cos(2\vartheta)-\eta_3 \cos(2\theta))  + \gamma \sqrt{2} \:q_{13}(\eta_4 \cos\vartheta \sin\vartheta - \eta_3\cos\theta \sin\theta),
\end{split} \ee
which should be compared with \eqref{G34}, and where $\chi_3$ and $\chi_4$ are still defined as in \eqref{chi34}.
In summary then, inefficient homodyne detection may be simulated using the rule \eqref{hom-eta-stateup}, and stochastic readouts $r_3$ and $r_4$, drawn from Gaussian distributions of variance $1/dt$ and mean $\sqrt{\gamma \eta_3}\, \chi_3(\theta)$ and $\sqrt{\gamma \eta_4}\, \chi_4(\vartheta)$, respectively.
Comparing to the ideal case, we can understand that we have the same noise, but the means $\chi_3$ and $\chi_4$ (which set the signal content of the readout) are now attenuated by a factor $\sqrt{\eta}$ relative to the noise. 
{Thus inefficient measurements, which only partially collect the information the optical degree of freedom ``knows'' about the qubits, result in a worsened signal to noise ratio, along with some decoherence due to averaging over lost information. }

\section{Review of Two--Qubit Density Matrices and Entanglement}
A two--qubit density matrix $\rho$ can be any $4\times 4$ matrix which satisfies the properties $\rho_{ij} = \rho_{ji}^\ast$, and $\text{tr}(\rho) = 1$, as demanded by quantum mechanics and normalization of probability densities. A subset of such density matrices describes separable systems. We discuss two--qubit entanglement in section \ref{sec-concurrence}, 
and then go through some details of a general coordinate parameterization of the two--qubit density matrix in Sec.~\ref{sec-bloch-coord}. 
{Inefficient measurements and mixed states are ubiquitous in laboratory situations, and we consequently use a density matrix rather than a pure state in many parts of the main text.}

\subsection{Two--qubit entanglement and concurrence \label{sec-concurrence}}
\par 
A commonly--accepted measure of entanglement between two qubits is concurrence \cite{wooters1998}. The concurrence $\mathcal{C}$ of a two--qubit density matrix is given by
\be 
\mathcal{C}[\rho] = \text{max} \lbrace 0 ,\lambda_1 - \lambda_2 - \lambda_3 - \lambda_4 \rbrace
\ee
where the $\lambda_i$ are the eigenvalues of the Hermitian matrix
\be 
\hat{R} = \sqrt{\sqrt{\rho} (\hat{\sigma}_y \otimes \hat{\sigma}_y) \rho^\ast (\hat{\sigma}_y \otimes \hat{\sigma}_y) \sqrt{\rho}}
\ee 
listed in \emph{decreasing} order. In practice, it is often easier to compute the eigenvalues of 
\be 
\rho (\hat{\sigma}_y \otimes \hat{\sigma}_y) \rho^\ast (\hat{\sigma}_y \otimes \hat{\sigma}_y), \quad\text{for}\quad \hat{\sigma}_y = \left( \begin{array}{cc} 0 & -i \\ i & 0 \end{array} \right),
\ee which give $\lambda_i^2$ instead. 
The qubits described by $\rho$ are said to be entangled iff $\mathcal{C}>0$, and are unentangled (separable) when $\mathcal{C} = 0$. 

\par Note that this simplifies nicely for the case of a pure two--qubit state 
\be \label{2Qpsi-gen-pure}
\bra{\psi} = \left( \mathsf{a}^\ast, \mathsf{b}^\ast, \mathsf{c}^\ast, \mathsf{d}^\ast \right),
\ee
(where we are still assuming a basis $\lbrace \ket{ee}$,$\ket{eg}$,$\ket{ge}$,$\ket{gg} \rbrace$).
The concurrence reduces in this case to 
\be 
\mathcal{C} = 2 |\mathsf{a} \mathsf{d} - \mathsf{b}\mathsf{c}|.
\ee
The concurrence $\mathcal{C}$ may range from $0$ to $1$, where $\mathcal{C}=0$ denotes a separable state, and $\mathcal{C} =1$ denotes a maximally entangled state, e.g.,~any of the standard Bell states
\begin{subequations}
\be \label{phi-pm}
\ket{\Phi^\pm} \equiv \tfrac{1}{\sqrt{2}} \ket{ee} \pm \tfrac{1}{\sqrt{2}} \ket{gg},
\ee \be \label{psi-pm}
\ket{\Psi^\pm} \equiv \tfrac{1}{\sqrt{2}} \ket{eg} \pm \tfrac{1}{\sqrt{2}} \ket{ge}.
\ee 
\end{subequations}
If we instead express our generic two--qubit pure state in the Bell basis
\be
\ket{\psi} = \mathsf{A} \ket{\Phi^+} + \mathsf{B} \ket{\Phi^-} + \mathsf{C} \ket{\Psi^+} + \mathsf{D} \ket{\Psi^-},
\ee
the concurrence reads
\be 
\mathcal{C} = |\mathsf{A}^2 - \mathsf{B}^2 - \mathsf{C}^2 + \mathsf{D}^2|. 
\ee
The mapping between the two bases listed here is given by the unitary
\be 
\mathcal{U} = \frac{1}{\sqrt{2}}\left( \begin{array}{cccc}
1 & 0 & 0 & 1 \\
1 & 0 & 0 & -1 \\
0 & 1 & 1 & 0 \\
0 & 1 & -1 & 0
\end{array}\right),
\ee
such that, if $\ket{\psi}$ is in the standard $\lbrace \mathsf{a},\mathsf{b},\mathsf{c},\mathsf{d}\rbrace$ basis, $\mathcal{U}\ket{\psi}$ is the same state expressed in the Bell basis $\lbrace \mathsf{A},\mathsf{B},\mathsf{C},\mathsf{D} \rbrace$.

\subsection{Generalized Gell--Mann matrices and effective system coordinates \label{sec-bloch-coord}}
We here describe the coordinate parameterization of the two--qubit density matrix which we use in the main text, and throughout our simulations.
It is always possible to decompose an $n\times n$ density matrix according to 
\be \label{rho-decomposeGM}
\rho = \frac{\hat{\mathbb{I}}_n}{n} + \mathbf{q} \cdot \boldsymbol{\hat{\Gamma}}.
\ee
Here $\mathbf{q}$ is a generalized Bloch vector, and $\boldsymbol{\hat{\Gamma}}$ is the vector of generalized Gell--Mann matrices. There are $n^2-1 = \text{dim}(\mathbf{q})$ coordinates and matrices. In the two--dimensional case, $\mathbf{q}$ is the usual Bloch coordinates, and $\boldsymbol{\hat{\Gamma}}$ are the Pauli matrices. Parameterizing a $4\times4$ density matrix requires 15 coordinates in the most general case. 

\par We adapt the matrices from \cite{Bertlmann2008} to define some coordinates for our two--qubit system, beginning with the three diagonal matrices
 \be \label{diagmat}
\hat{\Gamma}_1 = \frac{1}{\sqrt{2}} \left( \begin{array}{cccc}  
 1 & 0 & 0 & 0 \\
 0 & -1 & 0 & 0 \\
 0 & 0 & 0 & 0 \\
 0 & 0 & 0 & 0
\end{array} \right),
\quad
\hat{\Gamma}_2 = \frac{1}{\sqrt{6}} \left( \begin{array}{cccc}  
 1 & 0 & 0 & 0 \\
 0 & 1 & 0 & 0 \\
 0 & 0 & -2 & 0 \\
 0 & 0 & 0 & 0
\end{array} \right),
\quad
\hat{\Gamma}_3 = \frac{1}{\sqrt{12}} \left( \begin{array}{cccc}  
 1 & 0 & 0 & 0 \\
 0 & 1 & 0 & 0 \\
 0 & 0 & 1 & 0 \\
 0 & 0 & 0 & -3
\end{array} \right).
\ee
Next we list the six symmetric matrices of the set
\be \begin{split} \label{symmat}
& \hat{\Gamma}_4 = \frac{1}{\sqrt{2}} \left( \begin{array}{cccc}  
0 & 0 & 0 & 0 \\
0 & 0 & 1 & 0 \\
0 & 1 & 0 & 0 \\
0 & 0 & 0 & 0
\end{array} \right),
\quad
\hat{\Gamma}_5 = \frac{1}{\sqrt{2}} \left( \begin{array}{cccc}  
0 & 1 & 0 & 0 \\
1 & 0 & 0 & 0 \\
0 & 0 & 0 & 0 \\
0 & 0 & 0 & 0
\end{array} \right),
\quad 
\hat{\Gamma}_6 = \frac{1}{\sqrt{2}} \left( \begin{array}{cccc}  
0 & 0 & 1 & 0 \\
0 & 0 & 0 & 0 \\
1 & 0 & 0 & 0 \\
0 & 0 & 0 & 0
\end{array} \right),
\\ &
\hat{\Gamma}_7 =  \frac{1}{\sqrt{2}}\left( \begin{array}{cccc}  
0 & 0 & 0 & 1 \\
0 & 0 & 0 & 0 \\
0 & 0 & 0 & 0 \\
1 & 0 & 0 & 0
\end{array} \right),
\quad 
\hat{\Gamma}_8 =  \frac{1}{\sqrt{2}} \left( \begin{array}{cccc}  
0 & 0 & 0 & 0 \\
0 & 0 & 0 & 1 \\
0 & 0 & 0 & 0 \\
0 & 1 & 0 & 0
\end{array} \right),
\quad
\hat{\Gamma}_9 = \frac{1}{\sqrt{2}} \left( \begin{array}{cccc}  
0 & 0 & 0 & 0 \\
0 & 0 & 0 & 0 \\
0 & 0 & 0 & 1 \\
0 & 0 & 1 & 0
\end{array} \right).
\end{split} \ee
We conclude with the remaining six anti--symmetric matrices of the set
\be \begin{split} \label{asymmat}
&\hat{\Gamma}_{10} = \frac{1}{\sqrt{2}} \left( \begin{array}{cccc}  
0 & 0 & 0 & 0 \\
0 & 0 & -i & 0 \\
0 & i & 0 & 0 \\
0 & 0 & 0 & 0
\end{array} \right),
\quad 
\hat{\Gamma}_{11} = \frac{1}{\sqrt{2}} \left( \begin{array}{cccc}  
0 & -i & 0 & 0 \\
i & 0 & 0 & 0 \\
0 & 0 & 0 & 0 \\
0 & 0 & 0 & 0
\end{array} \right),
\quad
\hat{\Gamma}_{12} = \frac{1}{\sqrt{2}} \left( \begin{array}{cccc}  
0 & 0 & -i & 0 \\
0 & 0 & 0 & 0 \\
i & 0 & 0 & 0 \\
0 & 0 & 0 & 0
\end{array} \right),
\\ &
\hat{\Gamma}_{13} = \frac{1}{\sqrt{2}} \left( \begin{array}{cccc}  
0 & 0 & 0 & -i \\
0 & 0 & 0 & 0 \\
0 & 0 & 0 & 0 \\
i & 0 & 0 & 0
\end{array} \right),
\quad 
\hat{\Gamma}_{14} = \frac{1}{\sqrt{2}} \left( \begin{array}{cccc}  
0 & 0 & 0 & 0 \\
0 & 0 & 0 & -i \\
0 & 0 & 0 & 0 \\
0 & i & 0 & 0
\end{array} \right),
\quad
\hat{\Gamma}_{15} = \frac{1}{\sqrt{2}} \left( \begin{array}{cccc}  
0 & 0 & 0 & 0 \\
0 & 0 & 0 & 0 \\
0 & 0 & 0 & -i \\
0 & 0 & i & 0
\end{array} \right).
\end{split} \ee
Using \eqref{rho-decomposeGM}, we may write an arbitrary $4\times4$ density matrix in terms of the 15 generalized Bloch coordinates $\mathbf{q}$. This yields $\rho = $
\be \label{densmat_from_q}
 \frac{1}{\sqrt{2}} \left(\begin{array}{cccc}
\tfrac{\sqrt{2}}{4} + q_1 + \tfrac{1}{\sqrt{3}} q_2 + \tfrac{1}{\sqrt{6}}q_3 & q_5-i q_{11} & q_6 -i q_{12} & q_7-i q_{13} \\
q_5 + i q_{11} & \tfrac{\sqrt{2}}{4} + \tfrac{1}{\sqrt{3}} q_2 + \tfrac{1}{\sqrt{6}}q_3 - q_1 & q_4 - i q_{10} & q_8 - i q_{14} \\
q_6 + i q_{12} & q_4 + i q_{10} & \tfrac{\sqrt{2}}{4} + \tfrac{1}{\sqrt{6}} q_3 - \tfrac{2}{\sqrt{3}} q_2 & q_9 - i q_{15} \\
q_7 + i q_{13} & q_8 + i q_{14} & q_9 + i q_{15} & \tfrac{\sqrt{2}}{4} - \tfrac{3}{\sqrt{6}} q_3
\end{array}\right).
\ee 
We see that the populations are described by coordinates 1--3 (corresponding to matrices \eqref{diagmat}), and that the coherences are described by the remaining coordinates, with real parts corresponding to \eqref{symmat} and the imaginary parts to \eqref{asymmat}.
We can also codify this information visually, as it relates to the arrays of plots in Figs.~\ref{fig-SQT_table-HomEnt} through \ref{fig-DENS_table-HomFail}, by
\be \label{colorrho}
\rho = \left(\begin{array}{cccc} {\color{coquelicot} \blacktriangle} & 
{\color{electricgreen} \blacksquare}-\text{\colorbox{black}{${\color{white}i}{\color{electricgreen}\blacklozenge}$}} &  
{\color{mint} \blacksquare}-\text{\colorbox{black}{${\color{white}i}{\color{mint}\blacklozenge}$}}
& {\color{dodgerblue} \blacksquare}-\text{\colorbox{black}{${\color{white}i}{\color{dodgerblue}\blacklozenge}$}} \\ 
{\color{electricgreen} \blacksquare}+\text{\colorbox{black}{${\color{white}i}{\color{electricgreen}\blacklozenge}$}} & 
{\color{crimsonglory} \blacktriangleleft} & 
{\color{lincolngreen} \blacksquare}-\text{\colorbox{black}{${\color{white}i}{\color{lincolngreen}\blacklozenge}$}} & 
{\color{persianblue} \blacksquare}-\text{\colorbox{black}{${\color{white}i}{\color{persianblue}\blacklozenge}$}}\\ 
{\color{mint} \blacksquare}+\text{\colorbox{black}{${\color{white}i}{\color{mint}\blacklozenge}$}} & 
{\color{lincolngreen} \blacksquare}+\text{\colorbox{black}{${\color{white}i}{\color{lincolngreen}\blacklozenge}$}}& 
{\color{deeppink} \blacktriangleright}& 
 {\color{patriarch} \blacksquare}-\text{\colorbox{black}{${\color{white}i}{\color{patriarch}\blacklozenge}$}}\\ 
{\color{dodgerblue} \blacksquare}+\text{\colorbox{black}{${\color{white}i}{\color{dodgerblue}\blacklozenge}$}}
& {\color{persianblue} \blacksquare}+\text{\colorbox{black}{${\color{white}i}{\color{persianblue}\blacklozenge}$}}
& {\color{patriarch} \blacksquare}+\text{\colorbox{black}{${\color{white}i}{\color{patriarch}\blacklozenge}$}}
& {\color{electricviolet} \blacktriangledown}\end{array} \right).
\ee

\par In terms of the above coordinates, the purity of the state is described by
\be 
\mathcal{P}(\rho) = \text{tr}(\rho^2) = \tfrac{1}{4}+ \sum_i q_i^2.
\ee

\section{Mapping Equations of Motion from Kraus Operators to the SME \label{sec-KrausVSME}}

Our expectation, based on the one--qubit cases \cite{Jordan2015flor,FlorTeach2019,PL-Dissertation} and other {continuous measurement schemes}, is that the equations of motion derived from the stochastic master equation (SME) \eqref{SME} can be taken as It\^{o} {Langevin} equations, converted to their Stratonovich form, and will then be found to be identical to the equations of motion derived by expanding the state update with our Kraus operators to $O(dt)$ (where the latter is performed without treating the readout variables in any special way, and using regular calculus).
{T}he It\^{o} and Stratonovich conventions basically concern which Riemann sum is used to integrate a stochastic differential equation (SDE): {While different Riemann sums will give equivalent results when integrating ordinary differential equations, they give different results when applied to diffusive stochastic equations, because the latter are non-differentiable at all timescales.} 
The details of either convention, and the rules for converting between them are well understood (see e.g.~\cite{BookGardiner2}). 
We perform the requisite computations and conversions for our two--qubit homodyne detection and heterodyne detection models in turn; {further pertinent details appear in e.g.~Ref.~\cite{PL-Dissertation}.} 

\subsection{Homodyne Detection}

\subsubsection{Ideal Case}

Let us start with our Kraus operator methods. We will use double homodyne detection, in the entangling case $\theta = 0$ and $\vartheta = 90^\circ$. We may expand our operator \eqref{M-hom} according to
\be \label{expandhomo}
\hat{\mathcal{M}}_{34} e^{X_3^2/2+X_4^2/2} \approx \hat{\mathbb{I}} +\hat{\mathfrak{Z}} dt + O(dt^2) ,
\ee
where we can eliminate the Gaussian pre--factor and any other constants which appear in every matrix element of the operator, because they will cancel off from the state update normalization momentarily anyway. 
Then the state update can be be approximated by
\be \label{2QKraus-Eqmo-Odt} \begin{split}
\rho(t+dt) &\approx \frac{(\hat{\mathbb{I}} + \hat{\mathfrak{Z}} dt) \rho(t) (\hat{\mathbb{I}} + \hat{\mathfrak{Z}}^\dag dt)}{\text{tr}\left((\hat{\mathbb{I}} + \hat{\mathfrak{Z}} dt) \rho(t) (\hat{\mathbb{I}} + \hat{\mathfrak{Z}}^\dag dt) \right)} \\
&\approx \rho + dt \left(  \hat{\mathfrak{Z}} \rho + \rho \hat{\mathfrak{Z}}^\dag - \rho\: \text{tr}\left( \hat{\mathfrak{Z}} \rho + \rho \hat{\mathfrak{Z}}^\dag\right) \right),
\end{split} \ee
which can then be rearranged according to $\rho(t+dt)-\rho(t) \approx dt \: \dot{\rho}$, such that 
\be \label{2QKraus-Eqmo-Odt2}
\dot{\rho} \approx \hat{\mathfrak{Z}} \rho + \rho \hat{\mathfrak{Z}}^\dag - \rho\: \text{tr}\left( \hat{\mathfrak{Z}} \rho + \rho \hat{\mathfrak{Z}}^\dag\right).
\ee
The non--trivial part of the Kraus operator, to $O(dt)$, is
\be \label{kraus-homo-odt}
\hat{\mathfrak{Z}} \equiv \left( \begin{array}{cccc}
-\gamma & 0 & 0 & 0 \\
\sqrt{\frac{\gamma}{2}}(r_3-ir_4) & -\gamma/2 & 0 & 0 \\
\sqrt{\frac{\gamma}{2}}(r_3+ir_4) & 0 & -\gamma/2 & 0 \\
-\gamma & \sqrt{\frac{\gamma}{2}}(r_3+ir_4) & \sqrt{\frac{\gamma}{2}}(r_3-ir_4) & 0
\end{array} \right).
\ee
The equation of motion \eqref{2QKraus-Eqmo-Odt2} with \eqref{kraus-homo-odt} is best represented as fifteen coupled equations in the coordinates $\mathbf{q}$, obtained as $\dot{\mathbf{q}} = \text{tr}(\boldsymbol{\hat{\Gamma}} \dot{\rho})$; for the sake of brevity we do not list them out here.

\par The SME \eqref{SME} is derived in a similar spirit, using It\^{o} calculus \cite{Jacobs2006}. Recall that the corresponding Lindblad operators for the SME \eqref{homodyne-SME-ops} are
\be 
\hat{L}_3 = \sqrt{\tfrac{\gamma}{2}}\left(\hat{\sigma}_-^A+\hat{\sigma}_-^B\right), \quad \hat{L}_4= i\sqrt{\tfrac{\gamma}{2}}\left(\hat{\sigma}_-^A-\hat{\sigma}_-^B\right),
\ee
which denote the observables in channels 3 and 4, {which should} be used in \eqref{SME} with $c = 3,4$. We are able to obtain another fifteen--dimensional system of equations (represented in $\mathbf{q}$) from this SME--derived expression for $\dot{\rho}$, expressed in terms of the white noise terms $\xi_3 \sim dW_3/dt$ and $\xi_4 \sim dW_4/dt$ in channels 3 and 4, respectively. If the system of SDEs from the SME are a set of It\^{o} equations
\be
\dot{\mathbf{q}} = \mathbf{a}(\mathbf{q}) + \mathbf{b}_3(\mathbf{q}) \: \xi_3 + \mathbf{b}_4(\mathbf{q})\: \xi_4,
\ee
then they can be converted to the corresponding Stratonovich form 
\be \label{strato-generalform-homodynex2}
\dot{\mathbf{q}} = \mathbf{A}(\mathbf{q}) + \mathbf{b}_3(\mathbf{q}) \: \xi_3 + \mathbf{b}_4(\mathbf{q}) \: \xi_4
\ee
according to the transformation \cite{BookGardiner2}
\be 
\mathbf{A} = \mathbf{a} - \tfrac{1}{2}(\mathbf{b}_3\cdot\nabla)\mathbf{b}_3 - \tfrac{1}{2}(\mathbf{b}_4 \cdot \nabla)\mathbf{b}_4,
\ee
where $\nabla$ is here the vector derivative in all fifteen coordinates $\mathbf{q}$. Again without listing out the fifteen equations or details of the transformation, we find for the example at hand that the Stratonovich version of the SME equations \eqref{strato-generalform-homodynex2} are exactly equivalent to the equations \eqref{2QKraus-Eqmo-Odt2}, with the relationship
\begin{subequations}\be 
r_3 = \sqrt{\gamma}(q_5+q_6+q_8+q_9) + \xi_3, \ee \be
r_4 = \sqrt{\gamma}(-q_{11}+q_{12}+q_{14}-q_{15}) + \xi_4,
\ee \end{subequations}
between the readouts and white noise (valid to $O(dt)$). Thus the two approaches we have described are equivalent, provided we account for the fact that we have carried each of them out using a different stochastic calculus. 

\subsubsection{Inefficient Measurement \label{sec-hometa-Cmaxderiv}}

We are able to repeat the same procedure for the case of $\eta_3 \neq 1 \neq \eta_4$, and thereby check that our equations for inefficient measurement correctly translate to the SME as well. We verify that the correspondence between our Kraus operators (yielding a Stratonovich--like equation, expressed in terms of the records $r_3$ and $r_4$) and It\^{o} SME (yielding an It\^{o} equation expressed in terms of $dW_3$ and $dW_4$) works for the general homodyne case characterized by the SME operators
\be 
\hat{L}_3 = \sqrt{\tfrac{\gamma}{2}}e^{i\theta}\left(\hat{\sigma}_-^A+\hat{\sigma}_-^B\right), \quad \hat{L}_4= \sqrt{\tfrac{\gamma}{2}}e^{i\vartheta}\left(\hat{\sigma}_-^A-\hat{\sigma}_-^B\right),
\ee
for any $\eta_3 \in [0,1]$, $\eta_4 \in [0,1]$, and any combination of $\theta$ and $\vartheta$, corresponding to the most general update rule \eqref{hom-eta-stateup} with \eqref{M-hom-eta}.
{The procedure is similar to that above, and the details are not repeated here (but more appear in Ref.~\cite{PL-Dissertation}).}

{Instead, we show some of the $r_3 = 0 = r_4$ equations (derived using regular calculus / using the Stratonovich approach) that underpin the results of Sec.~\ref{sec-hom-eta}.}
{In particular,} the case characterized by $q_1 = \sqrt{2}\: q_2$, $q_5 = q_6$, $q_8 = q_9$, $q_{11}=-q_{12}$, $q_{13} = 0$, and $q_{14} = -q_{15}$ {is especially useful}. 
The condition $q_1 = \sqrt{2} q_2$ ensures that the element $\ket{eg}\bra{eg}$ and $\ket{ge}\bra{ge}$ are identical. This leads us to define two special coordinates for this case, according to 
\begin{subequations}
\label{qab} \be 
q_a \equiv \tfrac{1}{4} + \tfrac{1}{\sqrt{2}} q_1 + \tfrac{1}{\sqrt{6}} q_2 + \tfrac{1}{\sqrt{12}}q_3
\ee \be 
q_b \equiv \tfrac{1}{4} + \tfrac{1}{\sqrt{6}} q_2+ \tfrac{1}{\sqrt{12}}q_3 - \tfrac{1}{\sqrt{2}} q_1,
\ee
\end{subequations}
which, combined with $q_1 = \sqrt{2} \,q_2$, set the diagonal terms of the density matrix to be defined only in terms of $q_a$ and $q_b$. Specifically, $q_a$ is the population in $\ket{ee}$, $q_b$ is the correlated population on \emph{both} $\ket{eg}$ and $\ket{ge}$, and $1-q_a-2 q_b$ is the population on $\ket{gg}$.
We then list the system of nine differential equations which define the state dynamics \emph{under the condition} $r_3 = 0 = r_4$ (i.e.~we list the Stratonovich--like dynamics corresponding to the first--order expansion of the rule \eqref{hom-eta-stateup}, for the special case $X_3 = 0 = X_4$ and $\eta_3 = \eta = \eta_4$):
\begin{subequations} \label{2xHOM-eta-r0}
\be 
\dot{q}_a = -2 \gamma q_a + \eta \gamma \sqrt{2}\: q_a q_7 + 2\eta\gamma q_a (q_a+q_b),
\ee \be 
\dot{q}_b = \gamma (q_a(1-\eta) - q_b) +\eta \gamma \sqrt{2}\: q_b q_7 + 2\eta\gamma q_b (q_a+q_b),
\ee \be 
\dot{q}_4 = -\gamma q_4 + \eta \gamma \sqrt{2}\: q_4 q_7 + 2\eta\gamma q_4 (q_a+q_b),
\ee \be 
\dot{q}_5 = \tfrac{3}{2}\gamma q_5 + \eta \gamma \sqrt{2}\: q_5 q_7 + 2\eta\gamma q_5 (q_a+q_b),
\ee \be 
\dot{q}_7 = -\gamma q_7 +\eta \gamma \sqrt{2}(q_7^2-q_a) + 2\eta\gamma q_7 (q_a+q_b),
\ee \vspace{3pt} \be \begin{split}
\dot{q}_8 =& \gamma q_5 - \tfrac{1}{2}\gamma q_8 - 2 \eta \gamma q_5 + \eta \gamma \sqrt{2}\: q_8 q_7 + 2\eta\gamma q_8 (q_a+q_b),
\end{split} \ee \be 
\dot{q}_{10} = -\gamma q_{10} + \eta \gamma \sqrt{2}\: q_{10} q_7 + 2\eta\gamma q_{10} (q_a+q_b),
\ee \be 
\dot{q}_{11} = -\tfrac{3}{2}\gamma q_{11} + \eta \gamma \sqrt{2}\: q_{11} q_7 + 2\eta\gamma q_{11} (q_a+q_b),
\ee \be \begin{split}
\dot{q}_{14} =& -\gamma q_{11} - \tfrac{1}{2}\gamma q_{14} + 2\eta \gamma q_{11} +  \eta \gamma \sqrt{2}\: q_{14} q_7 + 2\eta\gamma q_{14} (q_a+q_b).
\end{split} \ee
\end{subequations}
This system of equations is somewhat messy, and these nine coordinates do not reduce nicely into a simple expression for the concurrence. We can however understand this system of equations as the generalized case of \eqref{dotCbound}; by numerically integrating the system \eqref{2xHOM-eta-r0} for the initial condition $q_a = 1$ and $\mathbf{q} = 0$ otherwise, we obtain the bounds on concurrence creation, e.g.~as shown in Fig.~\ref{fig-HOMeta}. {While the full set of equations are needed for more general initial conditions, several equations may be eliminated for the particular initial conditions above}, leaving only
\begin{subequations} 
\be 
\dot{q}_a = -2 \,\gamma \,q_a + \eta \,\gamma \,\sqrt{2}\: q_a\, q_7 + 2\,\eta\,\gamma\, q_a (q_a+q_b),
\ee \be 
\dot{q}_b = \gamma (q_a(1-\eta) - q_b) +\eta\, \gamma\, \sqrt{2}\: q_b\, q_7 + 2\,\eta\,\gamma\, q_b (q_a+q_b),
\ee \be 
\dot{q}_7 = -\gamma \,q_7 +\eta\, \gamma \,\sqrt{2}(q_7^2-q_a) + 2\,\eta\,\gamma\, q_7 (q_a+q_b).
\ee
\end{subequations}
Note also that by integrating this system for the case $\eta \leq \tfrac{1}{2}$, we find $\mathcal{C} = 0$ for all time, demonstrating that the upper bound forces the concurrence yield down to zero for homodyne efficiencies below $50\%$, as shown in simulations and discussed in the main text.

\subsection{Heterodyne Detection \label{sec-het-sme}}
We repeat the analysis above for the cases of interest involving heterodyne detection, again for the case $\theta = 0$ and $\vartheta = 90^\circ$ emphasized in the main text. Recall that the full operator is \eqref{M-het}, which leads to an approximate form 
\be 
\hat{\mathfrak{A}} = \left(\begin{array}{cccc} 
- \gamma & 0 & 0 & 0 \\
\tfrac{\sqrt{\gamma}}{2}(r_I - i r_Q - i r_X - r_Y) & - \gamma/2 & 0 & 0 \\
\tfrac{\sqrt{\gamma}}{2}(r_I - i r_Q + i r_X + r_Y) & 0 & -\gamma/2 & 0 \\
0 & \tfrac{\sqrt{\gamma}}{2}(r_I - i r_Q + i r_X + r_Y) & \tfrac{\sqrt{\gamma}}{2}(r_I - i r_Q - i r_X - r_Y) & 0
\end{array} \right),
\ee 
which is defined according to 
\be 
\hat{\mathcal{M}}_{\alpha \beta} e^{|\alpha|^2/2+|\beta|^2/2} \approx \hat{\mathbb{I}} + \hat{\mathfrak{A}} dt + O(dt^2).
\ee
As above, we can then write some first--order state update rule 
\be \label{2xHT-eqmo}
\dot{\rho} \approx \hat{\mathfrak{A}} \rho + \rho \hat{\mathfrak{A}}^\dag - \rho\: \text{tr}\left( \hat{\mathfrak{A}} \rho + \rho \hat{\mathfrak{A}}^\dag\right)
\ee
which uses the approximate measurement operator $\hat{\mathfrak{A}}$.

\par The readouts have means $\sqrt{\tfrac{\gamma}{2}}(q_5 + q_6 + q_8 + q_9)$ (for $r_I$), $-\sqrt{\tfrac{\gamma}{2}}(q_{11} + q_{12} + q_{14} + q_{15})$ (for $r_Q$), $\sqrt{\tfrac{\gamma}{2}}(-q_{11} + q_{12} + q_{14} - q_{15})$ (for $r_X$), and $\sqrt{\tfrac{\gamma}{2}}(-q_5 + q_6 + q_8 -q_9)$ (for $r_Y$). This is emininently sensible as compared with the homodyne case, as we have {the signal portion of the readout attenuated by a factor $\sqrt{2}$, such that the signal to noise ratio of e.g.~$r_I$ is reduced compared to that of $r_3$ on account of our having now split our attention between both (non-commuting) quadratures.} We can thus infer the corresponding Lindblad operators, which are 
\be \begin{split} \label{2xHet_Lops2}
    L_I &= \tfrac{\sqrt{\gamma}}{2} \left(\hat{\sigma}_-^A+\hat{\sigma}_-^B\right), \quad L_Q = -i\tfrac{\sqrt{\gamma}}{2} \left(\hat{\sigma}_-^A+\hat{\sigma}_-^B\right), \\
    L_X &= i\tfrac{\sqrt{\gamma}}{2} \left(\hat{\sigma}_-^A-\hat{\sigma}_-^B\right), \quad L_Y = \tfrac{\sqrt{\gamma}}{2} \left(\hat{\sigma}_-^A-\hat{\sigma}_-^B\right).
\end{split} \ee
As in the other cases, the It\^{o} equation \eqref{SME} and Stratonovich--like equation \eqref{2xHT-eqmo} are equivalent, in that either one is derived exactly from the other by using $r_j = \left\langle\hat{L}_j+\hat{L}_j^\dag\right\rangle + \xi_j$ and the appropriate transformation dictated by a consistent stochastic calculus \cite{BookGardiner2}.

\bibliography{refs}

\newpage
\section*{The Authors}
\noindent
\parbox[c][][c]{.16\textwidth}{\includegraphics[width = .16\textwidth, trim = {210 90 190 40}, clip]{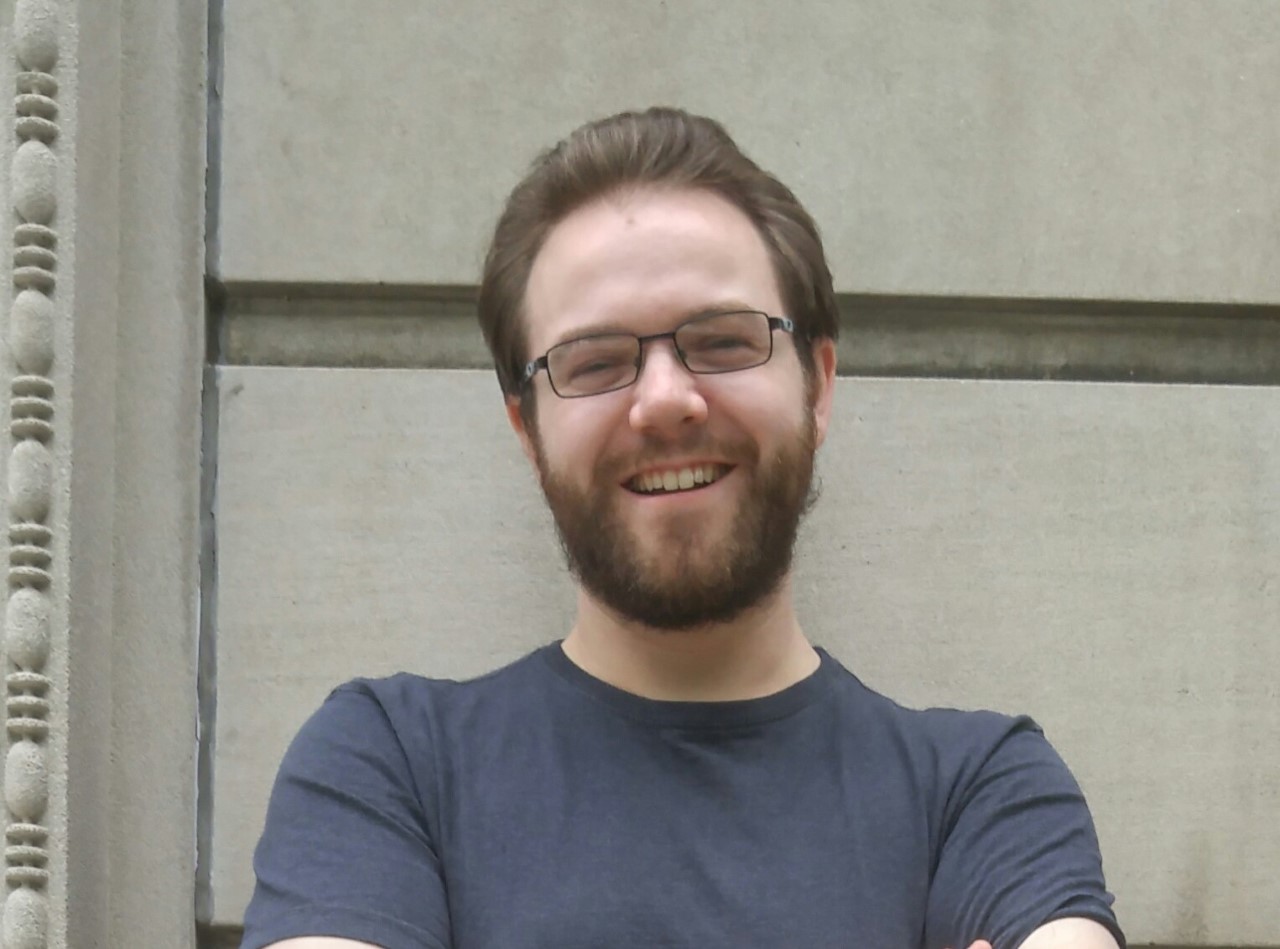}}
\hfill \parbox[c][][c]{.82\textwidth}{
{\bf\sc Philippe Lewalle} completed his undergraduate studies (B.S.~in Physics, and B.A.~in Music) at the University of Rochester in 2014, and continued on to complete his Ph.D.~at the University of Rochester under the direction of Prof.~Andrew N.~Jordan in 2021. He will soon join the group of Prof.~K.~Birgitta Whaley at UC Berkeley as a postdoc. His research interests include quantum information, open systems, measurement, and control. 
}\\\vspace{-3pt}\\ ~ \\
\parbox[c][][c]{.16\textwidth}{\includegraphics[width = .16\textwidth, trim = {7 20 10 2}, clip]{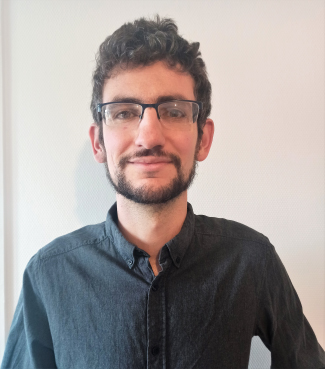}}
\hfill \parbox[c][][c]{.82\textwidth}{
{\bf\sc Cyril Elouard} completed a Ph.D.~in theoretical physics at the University Grenoble-Alpes (France). After three years of postdoc in the group of Prof. Andrew N. Jordan at the University of Rochester, New York, USA, he joined as a postdoc the QUANTIC Lab at Inria, Paris, France. His research activities span quantum thermodynamics, quantum measurement theory and quantum control, with applications in superconducting circuits, opto-mechanical devices and nanoelectronic devices.
}\\\vspace{-3pt}\\ ~ \\
\parbox[c][][c]{.16\textwidth}{\includegraphics[width = .16\textwidth, trim = {150 100 100 100}, clip]{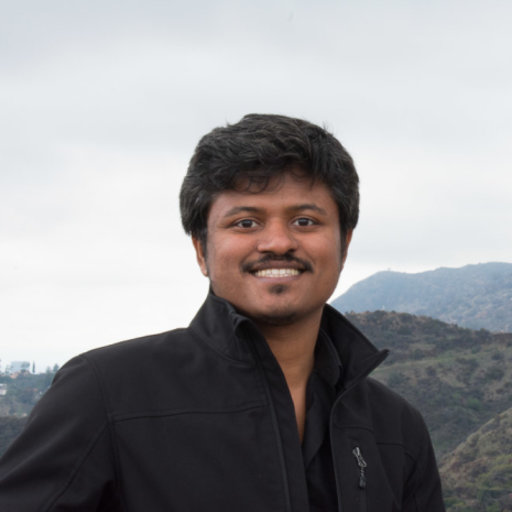}}
\hfill \parbox[c][][c]{.82\textwidth}{
{\bf\sc Sreenath K.~Manikandan} completed his integrated Bachelor of Science--Master of Science degree with a major in physics and minor in mathematics (2015) from the Indian Institute of Science Education and Research Thiruvananthapuram, India, and his Ph.D.~in physics from the University of Rochester, New York, USA, supervised by Prof. Andrew N. Jordan. His research interests are in quantum/condensed matter physics: quantum thermodynamics, quantum measurements, and applications to mesoscopic quantum heat engines and quantum refrigerators.
}\\\vspace{-3pt}\\ ~ \\
\parbox[c][][c]{.16\textwidth}{\includegraphics[width = .16\textwidth, trim = {0 40 0 0}, clip]{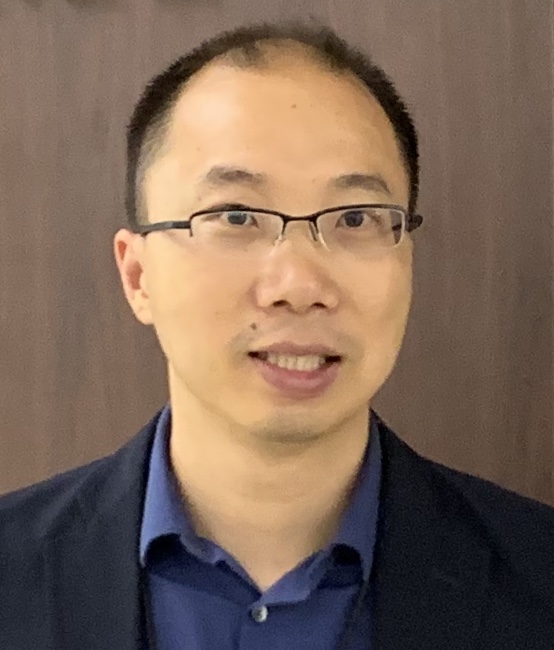}}
\hfill \parbox[c][][c]{.82\textwidth}{
Dr.~{\bf\sc Xiao-Feng Qian} received his Ph.D.~in 2014 from the Physics Department of University of Rochester, supervised by Dr.~Joseph H.~Eberly. He then became a Research Associate at the Institute of Optics at Rochester. He joined the Physics Faculty of Stevens Institute of Technology since 2019. His research focuses on experimental and theoretical investigations of coherence and entanglement in both quantum and classical wave systems.
}\\\vspace{-3pt}\\ ~ \\
\parbox[c][][c]{.16\textwidth}{\includegraphics[width = .16\textwidth]{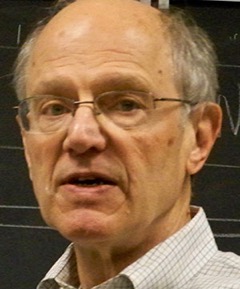}}
\hfill \parbox[c][][c]{.82\textwidth}{
{\bf\sc Joseph Eberly} is the Carnegie Professor of Physics and Professor of Optics at the University of Rochester, where his ongoing research interests are focused on the quantum behavior of optical radiation.  He is a Ph.D.~graduate of Stanford University where E.T.~Jaynes was his research supervisor. He has served as Chair of the Division of Laser Science for APS, was President of OSA in 2007, and is the Founding Editor of the journal Optics Express. 
}\\\vspace{-3pt}\\ ~ \\
\parbox[c][][c]{.16\textwidth}{\includegraphics[width = .16\textwidth]{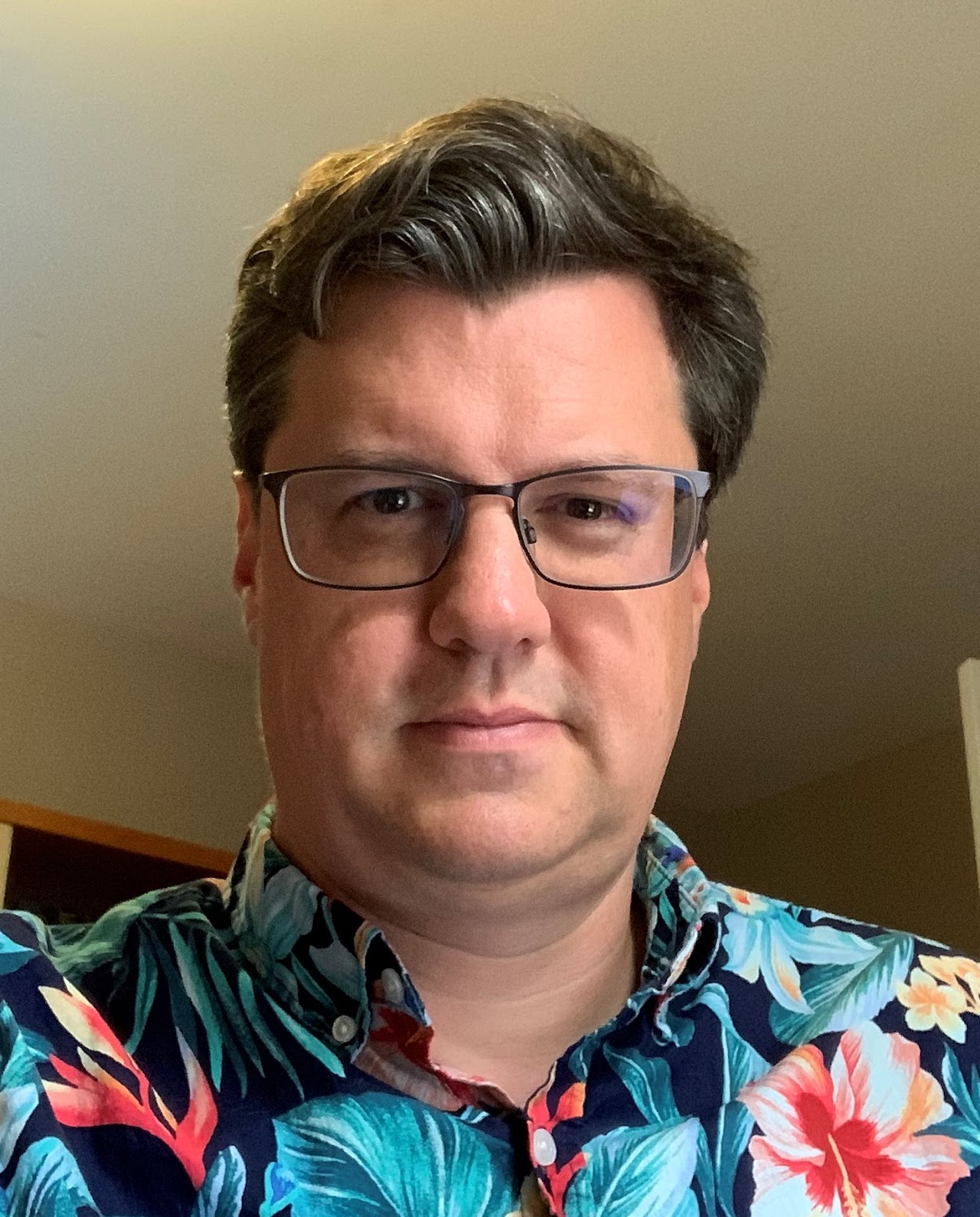}}
\hfill \parbox[c][][c]{.82\textwidth}{
Prof.~{\bf\sc Andrew N.~Jordan} received his B.S.~in Physics and Mathematics (1997) from Texas A\&M University and his Ph.D.~in Theoretical Physics (2002) from the University of California, Santa Barbara, supervised by Prof.~Mark Srednicki. He was a postdoctoral fellow at the University of Geneva (2002-2005) with Prof.~Markus B\"{u}ttiker, and a research scientist at Texas A\&M (2005-2006) with Prof.~Marlan Scully. He joined the University of Rochester as Assistant Professor of Physics in 2006, was promoted to Associate Professor with Tenure in 2012, and full Professor in 2015. 
}

\end{document}